\documentclass{aa} 

\usepackage{graphicx}
\usepackage{txfonts}
\usepackage{subfigure}
\usepackage{hyperref}
\usepackage[rightcaption]{sidecap}
\def\ms{\hbox{\,m\,s$^{-1}$}}         
\newcommand{\mearth}{M_\oplus}
\newcommand{\rearth}{R_{\rm \oplus}}
\newcommand{\msun}{M_\odot}

\begin{document}

	\title{The GAPS Programme at TNG}
	
	\subtitle{XLII. A characterisation study of the multi-planet system around the 400 Myr-old star HD\,63433 (TOI-1726)\thanks{Based on observations made with the Italian {\it Telescopio Nazionale
Galileo} (TNG) operated by the Fundaci\'on Galileo Galilei (FGG) of the
Istituto Nazionale di Astrofisica (INAF) at the
Observatorio del Roque de los Muchachos (La Palma, Canary Islands, Spain).}}
	
	 \titlerunning{The GAPS Programme at TNG XLVI. The exoplanetary system around HD\,63433}
	
	\author{M.~Damasso,\inst{1}
	    D.~Locci, \inst{2}
        S.~Benatti,\inst{2}
	    A.~Maggio, \inst{2}
        D.~Nardiello, \inst{3,4}
        M.~Baratella, \inst{5}
        K.~Biazzo, \inst{6}
	    A.~S.~Bonomo, \inst{1}
	    S.~Desidera, \inst{3}
	    V.~D'Orazi, \inst{3}  
	    M.~Mallonn, \inst{5} 
	    A.~F.~Lanza, \inst{7}
	    A.~Sozzetti, \inst{1}
        F.~Marzari, \inst{3}
        F.~Borsa, \inst{8} %
        J.~Maldonado, \inst{2}
        L.~Mancini, \inst{9,1,10} %
        E.~Poretti, \inst{11}
        G.~Scandariato, \inst{7}
        A.~Bignamini, \inst{12} 
        L.~Borsato, \inst{3}
        R.~Capuzzo~Dolcetta,\inst{13} %
        M.~Cecconi, \inst{11}
        R.~Claudi, \inst{3}
        R.~Cosentino, \inst{11} 
        E.~Covino, \inst{14} %
        A.~Fiorenzano,  \inst{11} 
        A.~Harutyunyan,  \inst{11}
        A.~W.~Mann, \inst{15} 
        G.~Micela, \inst{2}
        E.~Molinari, \inst{16} %
        M.~Molinaro, \inst{12} 
        I.~Pagano, \inst{7}
        M.~Pedani,  \inst{11} 
        M.~Pinamonti, \inst{1}
        G.~Piotto, \inst{17} %
        H.~Stoev \inst{11}
	}


	\institute{INAF -- Osservatorio Astrofisico di Torino, Via Osservatorio 20, I-10025 Pino Torinese, Italy\\
		\email{mario.damasso@inaf.it}
		\and INAF -- Osservatorio Astronomico di Palermo, Piazza del Parlamento 1, I-90134, Palermo, Italy
		\and INAF -- Osservatorio Astronomico di Padova, Vicolo dell'Osservatorio 5, I-35122, Padova, Italy
        \and Aix Marseille Univ, CNRS, CNES, LAM, Marseille, France
        \and Leibniz-Institut f\"{u}r Astrophysik Potsdam (AIP), An der Sternwarte 16, 14482 Potsdam, Germany
        \and INAF -- Osservatorio Astronomico di Roma, Via Frascati 33, 00078 -- Monte Porzio Catone (Roma), Italy
        \and INAF -- Osservatorio Astrofisico di Catania, Via S.~Sofia,78 - 95123 Catania, Italy
        \and INAF -- Osservatorio Astronomico di Brera, Via E. Bianchi 46, 23807 -- Merate (LC), Italy 
        \and Department of Physics, University of Rome ``Tor Vergata'', Via della Ricerca Scientifica 1, 00133, Rome, Italy       
        \and Max Planck Institute for Astronomy, K\"{o}nigstuhl 17, 69117 -- Heidelberg, Germany   
        \and Fundación Galileo Galilei - INAF, Rambla Jos\'e Ana Fernandez P\'erez 7, E-38712, Bre\~na Baja, TF - Spain
	    \and INAF -- Osservatorio Astronomico di Trieste, via Tiepolo 11, 34143 Trieste	
	    \and Dipartimento di Fisica - Universit\`{a} di Roma La Sapienza, P.le A.~Moro 5, 00185 -- Roma, Italy
	    \and INAF -- Osservatorio Astronomico di Capodimonte, Salita Moiariello 16, I-80131, Napoli, Italy 
	    \and Department of Physics and Astronomy, The University of North Carolina at Chapel Hill, Chapel Hill, NC 27599, USA 
	    \and INAF -- Osservatorio Astronomico di Cagliari, Via della Scienza 5, I-09047, Selargius (CA), Italy     
	    \and Dipartimento di Fisica e Astronomia "G. Galilei"-- Universt\`a degli Studi di Padova, Vicolo dell'Osservatorio 3, I-35122 Padova
	}
	
	\date{Received Accepted}
	
	
	\abstract
	{The GAPS collaboration is carrying out a spectroscopic and photometric follow-up of a sample of young stars with planets (age $\lesssim$600 Myr) to characterise planetary systems at the early stages of their evolution.}
	{For more than two years, we monitored with the HARPS-N spectrograph the 400 Myr-old star HD\,63433, which hosts two close-in (orbital periods $P_b\sim7.1$ and $P_c\sim20.5$ days) sub-Neptunes detected by the TESS space telescope, and it was announced in 2020. Using  radial velocities and additional TESS photometry, we aim to provide the first measurement of their masses, improve the measure of their size and orbital parameters, and study the evolution of the atmospheric mass-loss rate due to photoevaporation. } 
	{We tested state-of-the-art analysis techniques and different models to mitigate the dominant signals due to stellar activity that are detected in the radial velocity time series. We used a hydro-based analytical description of the atmospheric mass-loss rate, coupled with a core-envelope model and stellar evolutionary tracks, to study the past and future evolution of the planetary masses and radii. }
	{We derived new measurements of the planetary orbital periods and radii ($P_b=7.10794\pm0.000009$ d, $r_b=2.02^{+0.06}_{-0.05}$ $R_{\oplus}$; $P_c=20.54379\pm0.00002$ d, $r_c=2.44\pm0.07$ $R_{\oplus}$), and determined mass upper limits ($m_b\lesssim$11 $M_{\oplus}$; $m_c\lesssim$31 $M_{\oplus}$; 95$\%$ confidence level), with evidence at a 2.1--2.7$\sigma$ significance level that HD\,63433\,c might be a dense mini-Neptune with a Neptune-like mass. For a grid of test masses below our derived dynamical upper limits, we found that HD\,63433\,b has very likely lost any gaseous H-He envelope, supporting HST-based observations that are indicative of there being no ongoing atmospheric evaporation. HD\,63433\,c will keep evaporating over the next $\sim$5 Gyr if its current mass is $m_c\lesssim$15 $M_{\oplus}$, while it should be hydrodynamically stable for higher masses.} 
	{}
	
	
	\titlerunning{The planetary system around HD\,63433}
    \authorrunning{Damasso et al.}
	\maketitle
	%
	
	\section{Introduction} \label{sec:intro}
	
	Thanks to the constantly increasing number of characterised extrasolar planets, there is a growing number of studies related to the exoplanet demographics, defined by \cite{gaudi2021} as a research area mainly dedicated to study planets' distribution as a function of the physical parameters that may influence planet formation and evolution, exploring a broad range of the parameter space. Among the many physical parameters involved, a particularly crucial one is the age of the exoplanet systems. Drawing a relationship between the observed properties of the systems and their ages is a necessary step to take in order to build a general picture of how planets form and evolve with time, and to constrain the timescales of the main processes at work within the first hundreds of millions years from planet formation. That is why it is crucial to detect planets at different stages of their evolution (infant: age $<$20 Myr; young: 20$<$ age $<$100 Myr; intermediate-age: up to 800 Myr), and determine their orbital and main physical parameters. Among several key aspects, characterising young and close-in super-Earths and sub-Neptunes spotted at different ages offers the opportunity of theoretically studying how the atmospheric mass-loss rate evolves with time, in response to the high stellar high-energy irradiation, after dissipation of the protoplanetary disk. 
	
	Currently, the statistical sample of exoplanet systems younger than $\sim$800 Myr is still small if compared to the total number of discoveries achieved so far. This prevents drawing results from comparative analyses with more evolved systems. Detecting infant and young planets is particularly challenging, especially for blind searches carried out with the radial velocity (RV) method because of the very high level of scatter (up to several hundreds of \ms) mostly due to stellar magnetic activity, which heavily hampers the detection of periodic signals produced by planetary companions. The photometric transit method proved to be a more fruitful detection technique in this case, in particular thanks to the observations of the \textit{Kepler/K2} and the Transiting Exoplanet Survey Satellite (TESS) space telescopes. They yielded discoveries of planetary-sized companions, which triggered RV follow-up campaigns to measure their masses and bulk densities. Among these, the long-term Italian project Global Architecture of Planetary Systems (GAPS; \citealt{Covino2013,Poretti2016}) is investing a good fraction of the allocated observing time to follow-up stars with an age between $\sim$2--600 Myr using the High Accuracy Radial velocity Planet Searcher (HARPS-N) spectrograph \citep{Cosentino2012} located in the Northern Hemisphere at Telescopio Nazionale Galileo (TNG). After an initial effort in revising previously announced and debated planet detections (e.g. \citealt{carleo2018,carleo2020,damasso_v830tau}), the main focus of the GAPS young planets sub-programme shifted to the measurement of the fundamental properties of transiting planets. GAPS has observed and characterised several systems so far (e.g. \citealt{carleo2021,Mascareno2021,Nardiello2022}), and further characterisation studies have been carried out in parallel in the Southern Hemisphere with the HARPS spectrograph (e.g. \citealt{benatti2021,2022arXiv221007933D}). The measurement of the planetary dynamical masses remains crucial to investigate the evolution of exo-atmospheres even when it is only possible to provide upper limits, and the current escape rates can be assessed through atmospheric evaporation models (e.g. \citealt{benatti2021,carleo2021,poppen2021,maggio2022}). 
	
	In this paper, we present a characterisation study of the two-planet system detected by the TESS space telescope \citep{ricker2016} around the bright and 400-Myr old Sun-like star HD\,63433, a member of the Ursa Major moving group, also known as TESS object of interest (TOI) 1726. The system was first validated and partly characterised by \cite{mannetal2020}. In particular, they revealed that the orbit of the innermost planet is prograde, and emphasised that HD\,63433 represents a benchmark system for investigating the planet's evolution after the first hundreds millions years. The first results for this system led the GAPS collaboration to select HD\,63433 as a promising target for an intense RV follow-up for further characterisation. The architecture of HD\,63433 was also investigated by \cite{2020AJ....160..193D}, who measured the sky-projected spin-orbit angle of planet c, consistent with an aligned and prograde orbit. Results from \cite{mannetal2020} and \cite{2020AJ....160..193D} indicate that the system apparently did not experience catastrophic events capable of heavily influencing the dynamical evolution. The presence of a cold Jupiter in the system was examined by \cite{hirsch2021}, who analysed sparse RVs with a long time baseline, and found evidence only for a signal clearly ascribable to stellar activity.
	
	The main purpose of the GAPS follow-up campaign of HD\,63433 was measuring the masses of the two known planets. Accomplishing this goal becomes even more compelling to interpret the recent observational results of \cite{2022AJ....163...68Z}. They found evidence of Ly$\alpha$ absorption during a transit of HD\,63433\,c, suggesting ongoing atmospheric evaporation from this planet, while they suggest that HD\,63433\,b has already lost its primordial atmosphere: these observations have to be necessarily reconciled with mass measurements within the framework of current theoretical models. The study by \cite{2022AJ....163...68Z} emphasises a key aspect: the two planets represent a great opportunity for a comparative study within the same system, to examine how their different orbital configurations led to the apparently diverse atmospheric mass-loss timescales.
	
	In this work, we focus on the planetary mass measurements through the exploitation of RVs calculated from HARPS-N spectra, and we investigate the time evolution of the planetary atmospheres through evaporation models and updated stellar evolutionary tracks. 
	The paper is organised as follows. In Sect. \ref{sec:dataset} we describe the photometric and spectroscopic datasets, and in Sect. \ref{sec:stellarparameters} we present our measured fundamental stellar parameters. In Sect. \ref{sec:rvactivityfreqcontent} we present a frequency content analysis of the spectroscopic dataset, as a preparatory step to a more sophisticated analysis discussed in Sect. \ref{sec:rvlcanalysis}. In Sect. \ref{sec:mrdiag} we use the mass-radius diagram for a general comment on the possible planetary structures constrained by our results, especially in the context of known young exoplanets. Our results for planets b and c are then used for investigating the atmospheric evolutionary history in Sect. \ref{sec:atmphotoev}. Conclusions are summarised in Sect. \ref{sec:conclusions}.     
	
	\section{Observations and data reduction}
	\label{sec:dataset}
	
	\subsection{TESS photometry}
	\label{sec:tessphotometry}
	
\begin{figure}
		\centering
		\includegraphics[width=0.5\textwidth]{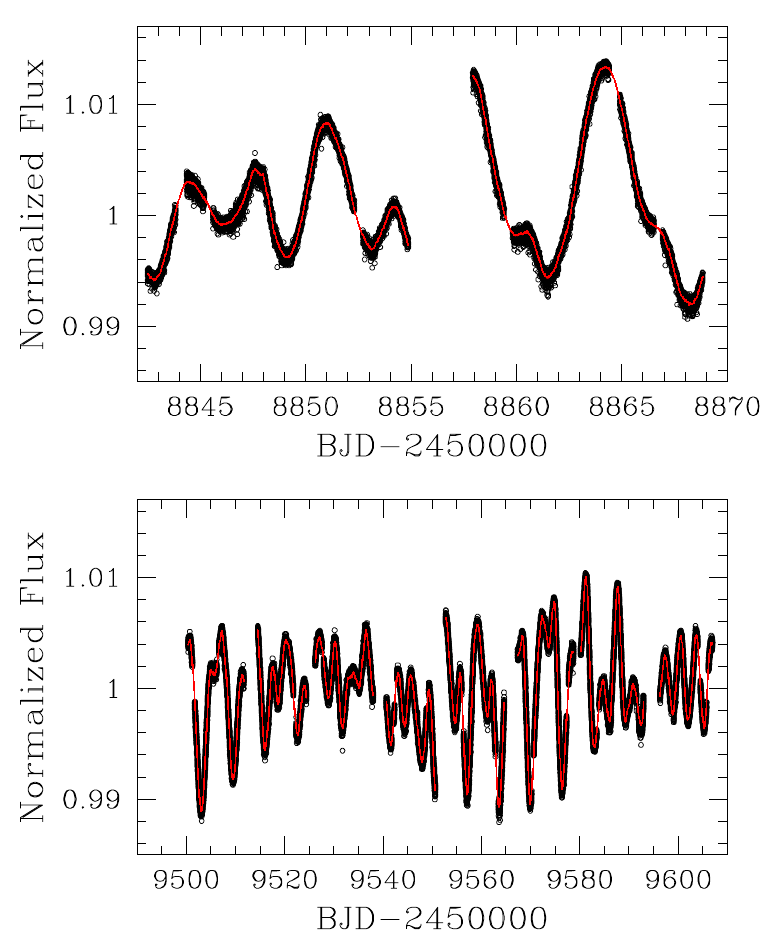}
		\caption{TESS light curve of HD\,63433 for Sector 20 (upper panel), and Sectors from 44 to 47 (lower panel). Planetary transits and bad-quality points have been removed from the dataset. The red curve represents the best-fit GP model.}
		\label{fig:tess_lc}
	\end{figure}

	TESS observed HD\,63433 in Sector 20 (from 24 December 2019 to 21 January 2020) during Cycle 2 (GO22038, PI: Roettenbacher; GO22203, PI: Ge; GO22032, PI: Metcalfe), and again in Sectors 44--47 (from 12 October 2021 to 28 January 2022) during Cycle 4 (GO4104, PI: Huber; GO4242, PI: Mayo; GO4191, PI: Burt; GO4060, PI: ; GO4039, PI: Mann). In our study, we used the short-cadence (2-minutes) light curves. We did not make use of the official Presearch Data Conditioning Simple Aperture Photometry (PDCSAP, \citealt{2012PASP..124.1000S,2012PASP..124..985S,2014PASP..126..100S}) light curves, because they are affected by systematics due to over-corrections and/or injection of spurious signals. We corrected the Simple Aperture Photometry (SAP) data by using cotrending basis vectors obtained with the tools developed by \cite{Nardielloetal2020}. 
	The extracted light curve for all Sectors is shown in Fig. \ref{fig:tess_lc}.
	In order to model and remove the clearly visible variability due to the stellar activity, we used the package \texttt{PyORBIT} \citep{2016ascl.soft12008M,2016A&A...588A.118M,2018AJ....155..107M}. First, we masked all the transits from the light curve, then we modelled the stellar activity by using Gaussian rocess (GP) regression performed with the package \texttt{celerite2}
	\citep{2017AJ....154..220F,2018RNAAS...2...31F}. We interpolated each excluded point within the masked transits using the best-fit GP solution, and we corrected the light curve variability by dividing the observed flux for the best-fit model. The final result is a flattened light curve with all the transits preserved, which we used in the combined light curve-RV analysis (Sect. \ref{sec:rvlcanalysis}). We already verified in previous works (see, e.g. \citealt{2020MNRAS.498.5972N, 2021MNRAS.505.3767N}) that this procedure for detrending the light curve does not alter the transit signal, by affecting its shape and depth.
	At the time of writing, according to the Web TESS Viewing Tool HD\,63433 won't be observed by TESS anymore up to September 2023 (Cycle 5).
	
	\subsection{STELLA photometry} \label{sec:stella}
	HD 63433 was observed with the STELLA telescope and its
	wide-field imager WiFSIP \citep{strassmeier2004} between 29 September 2020 and 17 March 2021 in Johnson \textit{V} and Cousin \textit{I} bands (82 and 81 epochs, respectively). The follow-up was intended to monitor the photometric evolution in response to a changing activity during one of the spectroscopic observing seasons. Each nightly observing block consisted of seven exposures with an exposure time of 1 second for each band. To avoid saturation of this bright target ($V_T\sim 7$ mag), the telescope was defocussed to get about 5 arcseconds for the full width at half maximum (FWHM) of the object point spread function (PSF). The data reduction followed that detailed description in \cite{mallonn2016}. We averaged the nightly exposures for each filter. The light curves are shown in Fig. \ref{fig:stelladata} (left column). The scatter and median uncertainty for the \textit{V} and \textit{I}-band light curve are 0.009 mag and 0.01 mag, respectively.

	\subsection{HARPS-N spectroscopic data}
	\label{sec:harpndata}
	The spectroscopic follow-up of HD\,63433 was carried out with the HARPS-N spectrograph \citep{Cosentino2012} mounted at the Telescopio Nazionale Galileo (TNG) on the island of La Palma (Canary Islands, Spain). The observations consist of 103 spectra collected between 26 February 2020 and 17 April 2022 (781 days), with a typical exposure time of 600 s, and a median S/N of 203 measured at a wavelength of $\sim$ 550 nm. The spectra have been reduced with the standard Data Reduction Software (\texttt{DRS}) pipeline (version 3.7.1) through the \texttt{YABI} workflow interface \citep{YABI}, which is maintained by the Italian centre for Astronomical Archive (IA2)\footnote{\url{ https://ia2.inaf.it}}. The RVs and activity diagnostics such as the FWHM and bisector velocity span (BIS) have been derived from the \texttt{DRS} cross-correlation function (CCF), which was calculated by adopting a reference mask for a star of spectral type G2, and a half-window width of 200 km/s. 
	The RVs have a median internal error $\sigma_{\rm RV,\,DRS}=0.81 \ms$, and an rms of 23.3 $\ms$.
	
	As a sanity check, we measured the RVs also using the Template Enhanced Radial velocity Reanalysis Application (\texttt{TERRA}) pipeline (v1.8) \citep{2012ApJS..200...15A}, which, for young and active stars, is in principle more effective in providing RVs less affected by stellar activity than those calculated from the CCF. \texttt{TERRA} dataset is characterised by a median internal error $\sigma_{\rm RV,\,TERRA}=1.25 \ms$, and an rms of 24.1 $\ms$. In Sect. \ref{sec:rvlcanalysis} we investigate the results obtained using both \texttt{DRS} and \texttt{TERRA} RVs. 
	
	We calculated two additional chromospheric activity diagnostics from the HARPS-N spectra, namely the $\log R^{\prime}_{\rm HK}$ from the $\mathrm{CaII}$ H$\&$K lines (provided by the \texttt{DRS}), and the index based on the H-alpha line calculated with the tool \texttt{ACTIN} (v1.3.9; \citealt{gomesdasilva18}). RVs and activity diagnostics are listed in Tables \ref{tab:dataset1} and \ref{tab:dataset2}.
	
	We note that there is very little overlap between the beginning of the last season of the RVs and the final part of the \textsc{TESS} light curve of Sector 47. The lack of simultaneous photometric monitoring with the RVs makes it impossible to use the light curve as an ancillary dataset to filter out the activity term in the RVs. That is especially true for a star such as HD\,63433, which shows rapidly changing photospheric active regions in the light curve that will generate RV variations independent of any planet. 
	The light curve observed by STELLA is characterised by sparse sampling, and overlaps only partially with the second season covered by HARPS-N. Therefore, even the STELLA dataset is not of practical use to effectively correct stellar activity in the RVs.  
	
	\section{Fundamental stellar parameters} \label{sec:stellarparameters}
	
	HD\,63433 is a dwarf star that belongs to the Ursa Major moving group, with an estimated age of roughly 400 Myr (see \citealt{mannetal2020}). Given its relatively young age, the spectroscopic analysis might be hampered by the intense magnetic fields that alter the structure of the stellar photosphere, in particular the upper layers. This alteration affects the formation of spectral lines in these layers for which abundances show a trend with optical depths (see \citealt{baratellaetal2020b} for further details). As a consequence, the standard spectroscopic approach, that is the equivalent width (EW) method, based on the use of iron (Fe) lines, could fail. In particular, when deriving the microturbulence velocity ($\xi$) by imposing that weak and strong iron lines have the same abundance, $\xi$ could be over-estimated, which leads to an under-estimation of the iron abundance ($[\rm Fe/H]$).
	
	Following the same strategy and the same line list as in \cite{baratellaetal2020a}, we applied a new method that consists of using a combination of Fe and titanium (Ti) lines to derive $T_{\rm eff}$ (through excitation equilibrium), and using only Ti lines to derive $\log g$ (through ionisation equilibrium) and $\xi$ (by zeroing the trend between individual abundances and strength, or EW, of the lines). The code MOOG (\citealt{sneden1973}) was used to analyse the HARPS-N co-added spectrum. We measured line EWs with the ARES v2 code (\citealt{sousaetal2015}) and discarded lines with errors larger than 10\% and with $EW>120$\,m\AA. The 1D LTE model atmospheres linearly interpolated from the ATLAS9 grid of \cite{castellikurucz2003}, with new opacities (ODFNEW), were adopted. 
	
	From the spectral type of the star (G2V), we estimated an input $T_{\rm eff}$ of 5660\,K from the relation by \cite{PecautMamajek2013}. Using different colour indexes (the reddening is negligible as the star is 22\,pc distant from the Sun) and different calibrated relations (\citealt{Casagrandeetal2010}, \citealt{Mucciarellietal2021}), the photometric estimates of $T_{\rm eff}$ range between 5589\,K (from $B_p-R_p$), 5619\,K (from $V-K$) and 5762\,K (from $J-K$). The derived surface gravity based on the Gaia DR3 parallax (\citealt{Gaia2022}) is between 4.50 and 4.54 dex (error of the order of 0.05), depending on the photometric $T_{\rm eff}$ considered. Finally, using the \cite{DutraFerreiraetal2016} relation, we estimated an initial $\xi$ of $0.095\pm0.05$\,km/s.
	
	The final values obtained with the new spectroscopic approach are: $T_{\rm eff}$=5700$\pm$75 K, $\log g$=4.54$\pm$0.05 and $\xi$=1.03$\pm$0.10 km/s. The iron abundance is \ion{[Fe/H]}{i}=0.02$\pm$0.06, \ion{[Fe/H]}{ii}=0.05$\pm$0.05, while the titanium abundance is \ion{[Ti/H]}{i}=0.04$\pm$0.07, \ion{[Ti/H]}{ii}=0.05$\pm$0.04. The uncertainties reported are the quadratic sum of the scatter due to the EW measurements and the contribution of the parameter uncertainties to the final abundances. Our results are in perfect agreement with the initial guesses and with the recent analysis of \cite{mannetal2020}.
	
	Fixing $T_{\rm eff}$, $\log g$, $\xi$, and \ion{[Fe/H]}{i} to the values found above, we measured the stellar projected rotational velocity ($v\sin{i_{\star}}$) using the same MOOG code and applying the spectral synthesis of two regions around 6200\,and 6700\,$\AA$. We adopted the same grid of model atmosphere and, after fixing the macroturbulence velocity to the value of 3.2\,km/s from the relationship by \cite{Breweretal2016}, we find a $v\sin{i_{\star}}$ of $7.2\pm0.7$\,km/s, consistent with the result by \cite{mannetal2020}. 
	
	Finally, we also derived the lithium abundance $\log A$(Li)$_{\rm NLTE}$ from the measured lithium EW (=84.5$\pm$2.0\,m\AA) and considering our stellar parameters previously derived together with the NLTE corrections by \cite{lindetal2009}. The values of the lithium abundance is $2.56\pm0.06$ and the position of the target in a $\log A$(Li)$_{\rm NLTE}$-$T_{\rm eff}$ diagram is compatible with the UMa group, as expected (see \citealt{mannetal2020}).
	
\begin{figure}
		\centering
		\includegraphics[width=0.5\textwidth]{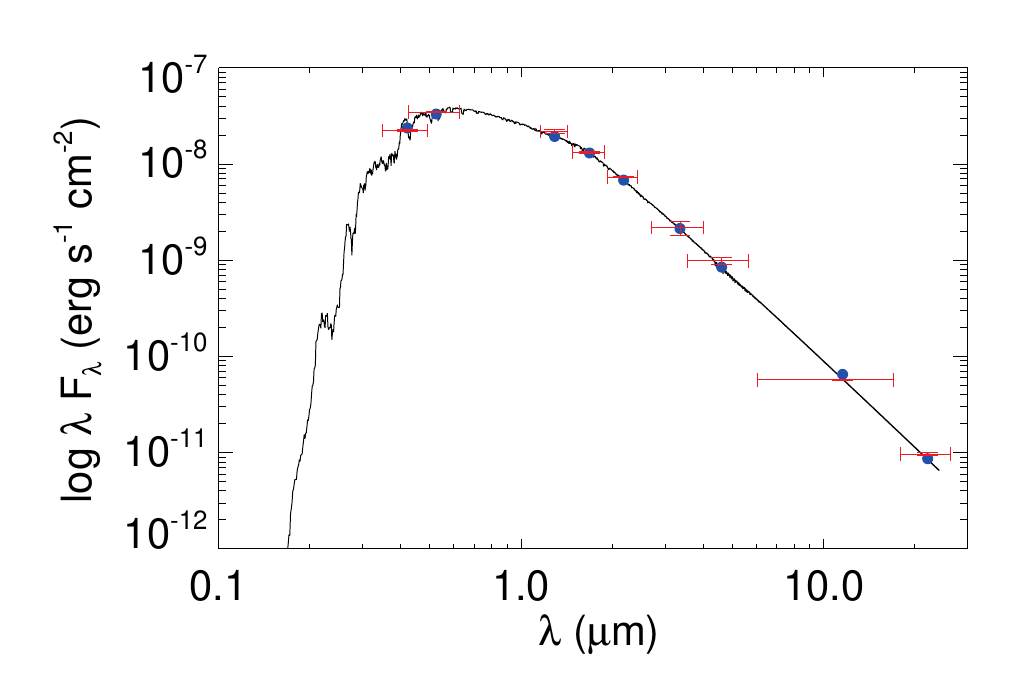}
		\caption{Spectral energy distribution of the host star HD\,63433 with the best-fit model overplotted (solid line). Red and blue points correspond to the observed and predicted values, respectively. }
		\label{fig:sed}
	\end{figure}

	We calculated the stellar radius and mass with the EXOFASTv2 tool \citep{Eastmanetal2019}, by fitting the stellar Spectral Energy Distribution (SED) and 
	providing the stellar luminosity derived from the SED as input to the MIST stellar evolutionary tracks \citep{Dotter2016}. 
	For the SED we considered the Tycho $B$ and $V$ magnitudes \citep{hog2000}, the 2MASS near-IR $J$, $H$ and $K$ magnitudes \citep{cutri2003}, and the WISE mid-IR $W1$, $W2$, $W3$ and $W4$ magnitudes \citep{cutri2013}. 
	We imposed Gaussian priors on 
	i) the stellar effective temperature and metallicity from our analysis of the HARPS-N spectra; 
	ii) the parallax $44.6848\pm0.0228$~mas from the Gaia EDR3 \citep{gaia2016,gaia2021} and 
	iii) the stellar age $413\pm23$~Myr from the most updated value for the Ursa Major association \citep{jones2015}. 
	We used uninformative priors for all the other parameters. 
	The best fit of the SED is shown in Fig.~\ref{fig:sed}. 
	We found $R_\star=0.897 \pm 0.019~\rm R_\odot$ and $M_\star=0.994 \pm 0.028~\rm M_\odot$, in excellent agreement 
	with the previous determination by \cite{mannetal2020}.
	
	From modelling the activity of the TESS light curve (Fig. \ref{fig:tess_lc}) with a GP regression, we obtained a rotation period $P=6.48 \pm 0.08$~d. This result is in agreement with the measurement by \cite{mannetal2020}, and it is confirmed by looking at the periodogram of the \textit{V}-band STELLA light curve calculated with the generalised Lomb-Scargle (\texttt{GLS}, \citealt{zechmeister2009}) tool (Fig. \ref{fig:stelladata}), which shows the main and significant peak at 6.4 d. 	
	The fundamental stellar properties of HD\,63433 are summarised in Tab. \ref{tab:stellarparam}. 
	
	\begin{table}
		\caption[]{Fundamental parameters of HD\,63433 (TOI-1726)}
		\label{tab:stellarparam}
		\centering
		\begin{tabular}{lcc}
			\hline
			\noalign{\smallskip}
			Parameter      &  Value &  Ref. \\
			\noalign{\smallskip}
			\hline
			\noalign{\smallskip}
			$B_T$ [mag]  & 7.749 $\pm$ 0.016 & [1]  \\ 
			$V_T$ [mag]  & 6.987 $\pm$ 0.010 & [1] \\
			$J$ [mag]    & 5.624 $\pm$ 0.043 & [2] \\
			$H$  [mag]   & 5.359 $\pm$ 0.026 & [2] \\
			$K_s$ [mag] & 5.258 $\pm$ 0.016 & [2]  \\
			WISE1   [mag] &  5.246 $\pm$ 0.178 & [3] \\
			WISE2   [mag] & 5.129 $\pm$ 0.087  & [3] \\
			WISE3   [mag] & 5.297 $\pm$ 0.016  & [3] \\
			WISE4  [mag]  & 5.163 $\pm$ 0.031  & [3] \\
			parallax, $\varpi$ [mas] & 44.685 $\pm$ 0.023 & [4]\\
			$T_{\rm eff}$ [K] & 5700 $\pm$ 75 & [5] \\
			$\log g$ [dex]  & 4.54 $\pm$ 0.05 & [5] \\
			$v_t$ [km s$^{-1}$] & 1.03 $\pm$ 0.10 & [5] \\
			$\ion{[\rm Fe/H]}{i}$ [dex] & 0.02 $\pm$ 0.06 & [5] \\
			A(Li) & 2.56 $\pm$ 0.06 & [5] \\
			$v\sin i_{\star}$ [km s$^{-1}$] & 7.2 $\pm$ 0.7 & [5] \\ 
			$P_{\rm rot,\,\star}$ [days] & $6.48 \pm 0.08$ & [5] \\
			Mass, $M_{\star}$ [M$_{\odot}$] & $0.994^{+0.027}_{-0.028}$  & [5] \\
			Radius, $R_{\star}$ [R$_{\odot}$] & $0.897 \pm 0.019$ & [5] \\
			Density, $\rho_{\star}$ [$\rho_{\odot}$] & 1.376 $\pm$ 0.078 & [5] \\
			$L_{\star}$ [$L_{\odot}$] & $0.765^{+0.053}_{-0.050}$ & [5] \\
			$L_{\star,\, \rm X}$ [erg s$^{-1}$] & $7.5 \times 10^{28}$ & [6] \\
			Age [Myr] & 414 $\pm$ 23 & [7] \\
			\noalign{\smallskip}
			\hline
		\end{tabular}
		\tablebib{[1] Tycho catalogue, \cite{hog2000}; [2] 2MASS, \cite{cutri2003}; [3] AllWISE, \cite{cutri2013}; [4] Gaia eDR3, \cite{gaia2016,gaia2021,Gaia2022}; 
		[5] This work; [6] \cite{2022AJ....163...68Z}; [7] \cite{jones2015}.}
	\end{table}

	\section{Frequency content analysis of radial velocities and activity diagnostics}
	\label{sec:rvactivityfreqcontent}
	
\begin{figure*}
		\centering
		\includegraphics[width=\textwidth]{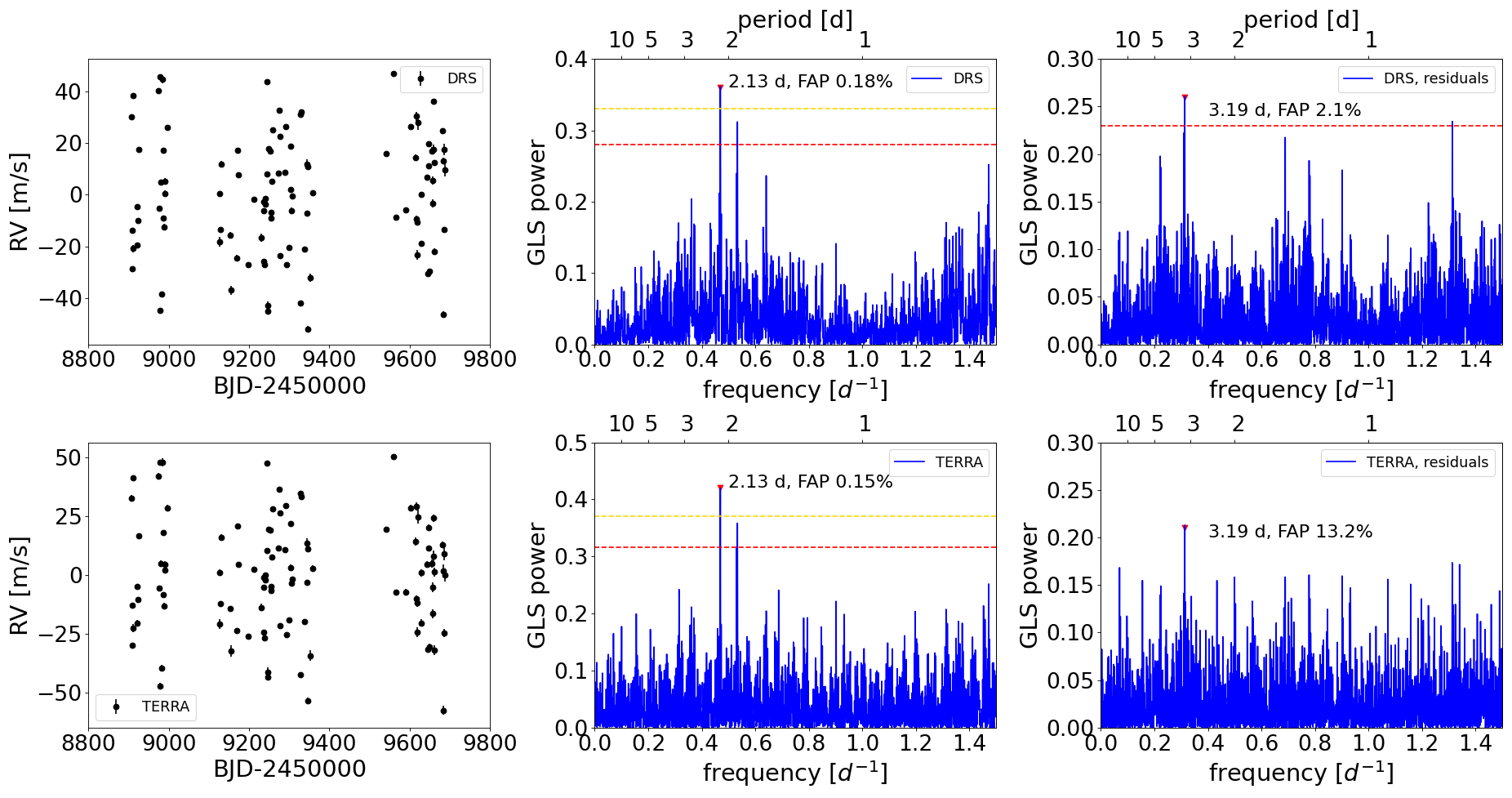}
		\caption{Time series of the RVs calculated from HARPS-N spectra with the \texttt{DRS} and \texttt{TERRA} pipelines, and their GLS periodograms. \textit{First column.} RV timeseries. The mean value has been subtracted from the original \texttt{DRS} data. \textit{Second column.} The corresponding GLS periodograms of the two RV dataset. FAP levels are of 1\% and 10\% are indicated by yellow and red horizontal lines, respectively. They are calculated through a bootstrap analysis. \textit{Third column.} GLS periodograms of the pre-whitened RV data.}
		\label{fig:rv_gls_drs_terra}
	\end{figure*}

	We analysed the frequency content of RVs and activity indicators to investigate the presence of sinusoidal periodic signals. The time series of the RVs (both \texttt{DRS} and \texttt{TERRA}), and the corresponding \texttt{GLS} periodograms are shown in the panels of the first two columns of Fig. \ref{fig:rv_gls_drs_terra}. The main peak is statistically significant (bootstrap false alarm probability FAP$<1\%$) and located at the second harmonic of the stellar rotation period. The periodograms of the residuals, after removing the best-fit sinusoid calculated by \texttt{GLS} (third column of Fig. \ref{fig:rv_gls_drs_terra}), show the main peak at the first harmonic of the rotation period, with low significance in the case of \texttt{TERRA} RVs. No other significant peaks are detected, especially at the orbital frequencies of the planets, indicating that the RVs scatter is dominated by stellar activity.
	
\begin{figure*}
		\centering
		\includegraphics[width=\textwidth]{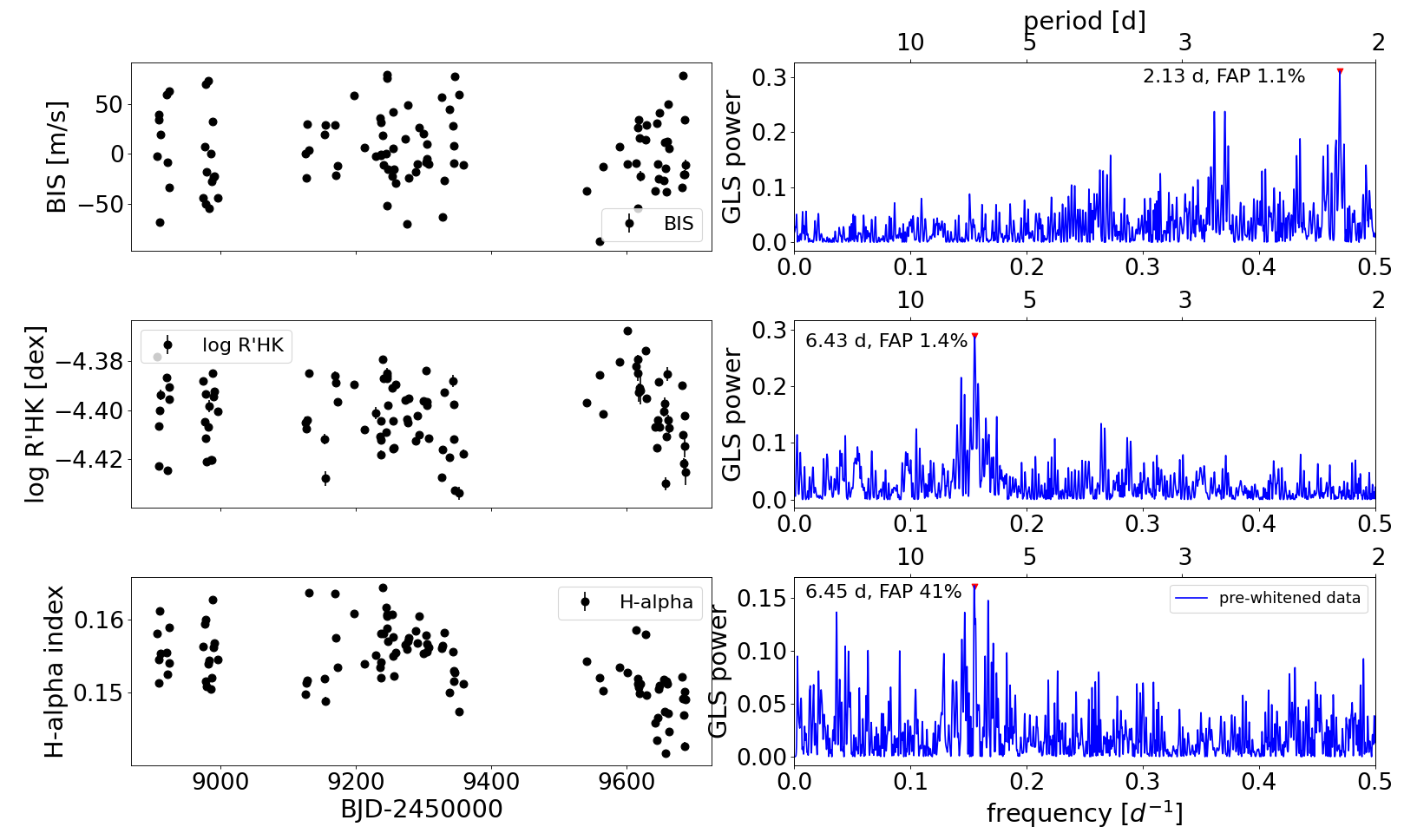}
		\caption{Time series (left column) and periodograms (right column) of spectroscopic activity diagnostics. The periodogram of the H-alpha index refers to pre-whitened data, after removing the long-term curvature visible in the time series. FAPs are calculated through a bootstrap analysis.}
		\label{fig:actindex}
	\end{figure*}

	The time series and \texttt{GLS} periodograms of the spectroscopic activity diagnostics are shown in Fig. \ref{fig:actindex}. The $\log R^{\prime}_{\rm HK}$ index and BIS contain a quite significant signal related to the stellar rotation period. The main peaks in the periodograms occur at a frequency corresponding to $P_{\rm rot,\,\star}$ and to the second harmonic of $P_{\rm rot,\,\star}$, respectively. A non-significant (FAP=41$\%$) peak at a frequency compatible with the rotational frequency is also detected in the residuals of the H-alpha index, after removing a long-term curvature clearly visible in the time series, for which we cannot assess a periodicity, if actually present. 
	The periodogram of the FWHM time series has the main peak located at $\sim$162 days (FAP=0.3$\%$), whose origin is not clear. After a pre-whitening, the periodogram of the FWHM residuals shows a peak at 2.15 days with FAP=51$\%$. Given the very low statistical significance of this signal, which corresponds to the second harmonic of $P_{\rm rot,\,\star}$, in the following analysis we do not use the FWHM to correct the RVs for the stellar activity term.
	
	We note that the RV and BIS data are significantly correlated. The Spearman’s rank correlation coefficient is $\rho=-0.88$, calculated using the \texttt{r$\_$correlate} function in \texttt{IDL}.

	\section{Joint analysis of transits and RVs}
	\label{sec:rvlcanalysis}
	
	We modelled the RV time series jointly with the detrended and flattened TESS light curve, which includes only the part of the data with the transits, and a sufficiently long out-of-transit baseline for a proper modelling of the planet ingress and egress.
	A key point in our analysis is the modelling of the stellar activity term which is the dominant signal in the RV time series. We took special care of it by testing different models, mostly based on the GP regression analysis. A schematic description of the test models is provided in Table \ref{tab:models}, with more details given in Appendix \ref{app:modeldetails}. 
	
	For all the tested models, we explored the full parameter space using the publicly available Monte Carlo (MC) nested sampler and Bayesian inference tool \texttt{MultiNest v3.10} (e.g. \citealt{Feroz2019}), through the \texttt{pyMultiNest} wrapper \citep{Buchner2014}. Our MC set-up included 300 live points, a sampling efficiency of 0.5, and a Bayesian tolerance of 0.3. We used the code \texttt{batman} \citep{Kreidberg2015} for modelling the photometric transits. In a few cases, when the same dataset has been used, we performed a Bayesian model selection by comparing the values of the Bayesian evidence $\ln\mathcal{Z}$ calculated by \texttt{MultiNest}, and using the empirical scale indicated in Table 1 of \cite{feroz2011} to assess their relative statistical significance\footnote{According to that scale for interpreting  model probabilities,  $(\log\mathcal{Z}_1-\log\mathcal{Z}_2)<1$ indicates inconclusive evidence in favour of model 1 over model 2; if $(\log\mathcal{Z}_1-\log\mathcal{Z}_2)\sim1$, there is weak evidence in favour of model 1; if $(\log\mathcal{Z}_1-\log\mathcal{Z}_2)\sim2.5$, there is moderate evidence in favour of model 1; if $(\log\mathcal{Z}_1-\log\mathcal{Z}_2)\geq5$, model 1 is strongly favoured over model 2.}. We assigned the same a-priori probability to each model.  
	
Concerning the light curve, we modelled the limb darkening with a quadratic law, and fitted the coefficients $LD_{\rm c1}$ and $LD_{\rm c2}$ using the formalism and uniform priors given by \cite{kipping2013} (see Eq. 15 and 16 therein). We also introduced constant jitters $\sigma_{\rm jit,\,TESS}$ added in quadrature to the nominal photometric uncertainties, one jitter term for Sector 20, and one for Sectors 44-47, to take into account the change in the TESS performance over nearly two years. 
The complete list of priors used for all the test models is provided in Table \ref{table:priors}.
	
The main goals of our analysis are the improvement of the orbital parameters for HD\,63433\,b and HD\,63433\,c, and the measure of the fundamental planetary parameters radius, mass, and average density. Based on the statistical results and mass measurements summarised in Table \ref{tab:models}, we notice that: 
	\begin{itemize}
		\item the GP quasi-periodic and quasi-periodic with cosine (QPC) kernels perform equally well (models M1 and M8), and better than the double simple harmonic oscillator (dSHO) kernel (model M7), when modelling only the RV time series. It is not statistically advantageous treating each observing season separately (model M2). The stellar rotation period is recovered with high precision (GP hyper-parameter $\theta=6.390^{+0.007}_{-0.005}$ days for M1), even with an adopted uniform and large prior $\mathcal{U}$(0,10) days; 
		\item results for all the test models show that the planetary orbits can be assumed circular. The data do not allow for significant non-zero eccentricities (see Table \ref{tab:resultmodel});
		\item we cannot constrain the mass $m_b$ of HD\,63433\,b, but only derive upper limits, with values that change depending on the model. We notice that the orbital period $P_b$ and the stellar rotation period $P_{\rm rot,\,\star}$ are not dissimilar, and this could make more difficult the identification of a planetary Doppler signal with a small semi-amplitude, despite the very precise transit ephemeris recovered;
		\item most of the models allow us to retrieve the mass $m_c$ of HD\,63433\,c with a statistical significance (defined as the ratio $m_c/\sigma^-_{\rm m_c}$) in the range 2.1--2.7$\sigma$ (with the exception of models M6, M7, and M9, for which the significance is lower);
		\item using the more complex multi-dimensional GP (quasi-periodic kernel), we find lower upper limits for the masses, with no improvement in their precision.
	\end{itemize} 
	
\begin{figure}
		\centering
		\includegraphics[width=0.45\textwidth]{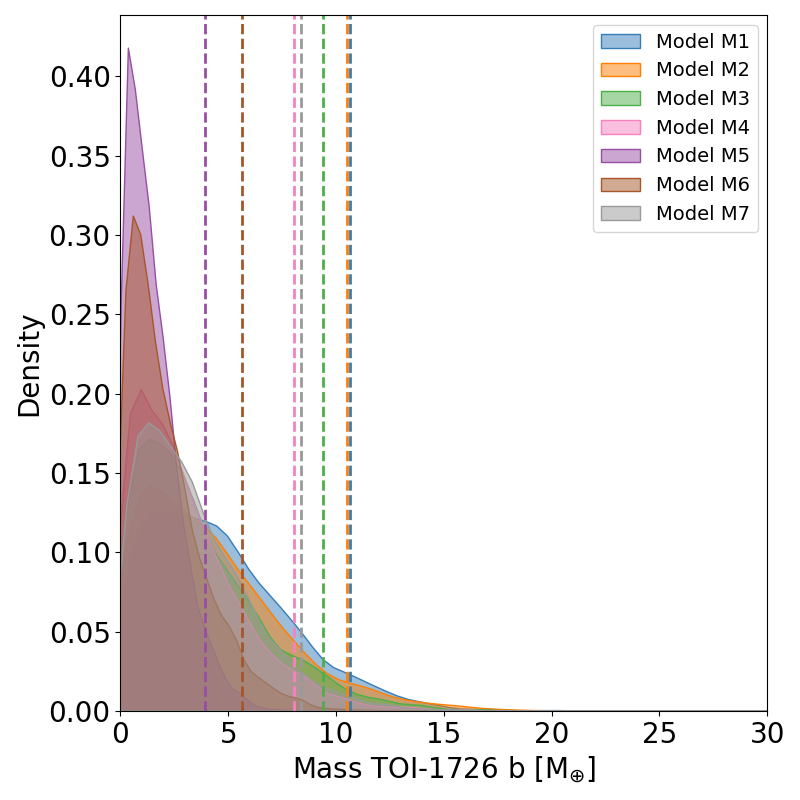}
		\includegraphics[width=0.45\textwidth]{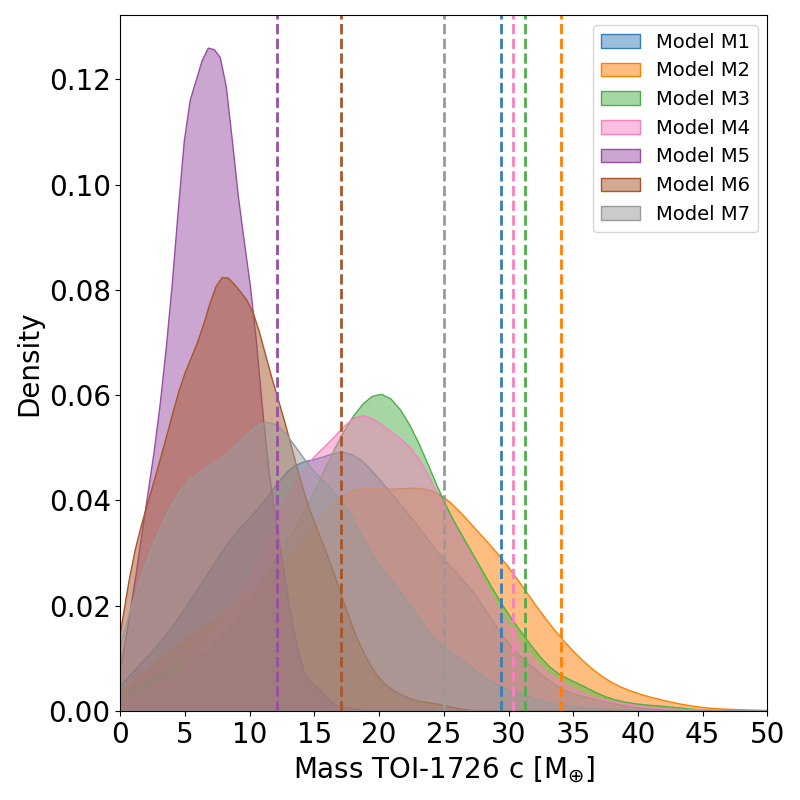}
		\caption{Posterior distributions for the planetary masses for some of the models tested in this work (Table \ref{tab:models}). Vertical dashed lines indicate the 95$\%$ percentiles.}
		\label{fig:posteriormasses}
	\end{figure}

	Despite the fact that we followed several pathways to model the stellar activity in the RV dataset, the data do not allow for a significant and accurate measurement of the mass $m_c$, but we can conclude with high confidence that the mass of HD\,63433\,c is less than twice the mass of Neptune. Nonetheless, we note that in a few cases the median values of Gaussian-like posterior distributions are very similar, and suggest that $m_c$ could be actually consistent with the mass of Neptune. Examples of posteriors for $m_b$ and $m_c$ are shown in Fig. \ref{fig:posteriormasses}, with reference to Table \ref{tab:models}. We also note that, in a few cases, masses of young planets have been claimed with a precision of the order of 3-3.5$\sigma$. For example, \cite{2022arXiv221007933D} found that TOI-179\,b (further discussed in Sect. \ref{sec:mrdiag}) has a mass of $24.1^{+7.1}_{-7.7}$ M$_{\oplus}$ (3.1$\sigma$ precision). The host TOI-179 has a similar mass, and an activity-induced RV scatter very similar to that measured for HD\,63433 ($\sim$ 24 m/s), while TOI-179\,b has an orbital period nearly five times shorter than HD\,63433\,c, making a more precise RV detection of the planet potentially easier. K2-100\,b \citep{barragan2019}, is a 750 Myr-old planet with an orbital period of 1.7-d, and a measured mass of 21.8$\pm$6.2 M$_{\oplus}$ (3.5$\sigma$ precision). Similarly, mass measurements have been claimed for TOI-560\,b (480 Myr; $m=10.2^{+3.4}_{-3.1}$ $\mearth$, i.e. 3.3$\sigma$ precision;  \citealt{barragan2022}), and for Kepler-411\,d (200 Myr; $m=15.2\pm5.1$, i.e. 3$\sigma$ precision, measured through transit timing variations; \citealt{sun2019}). Since we determined the mass of HD\,63433\,c with a significance of 2.7$\sigma$ at best, we do not consider this result significant enough for claiming a mass measurement, but indeed it deserves further attention to be confirmed with additional data.
	
\begin{figure*}
		\centering
		\includegraphics[width=0.8\textwidth]{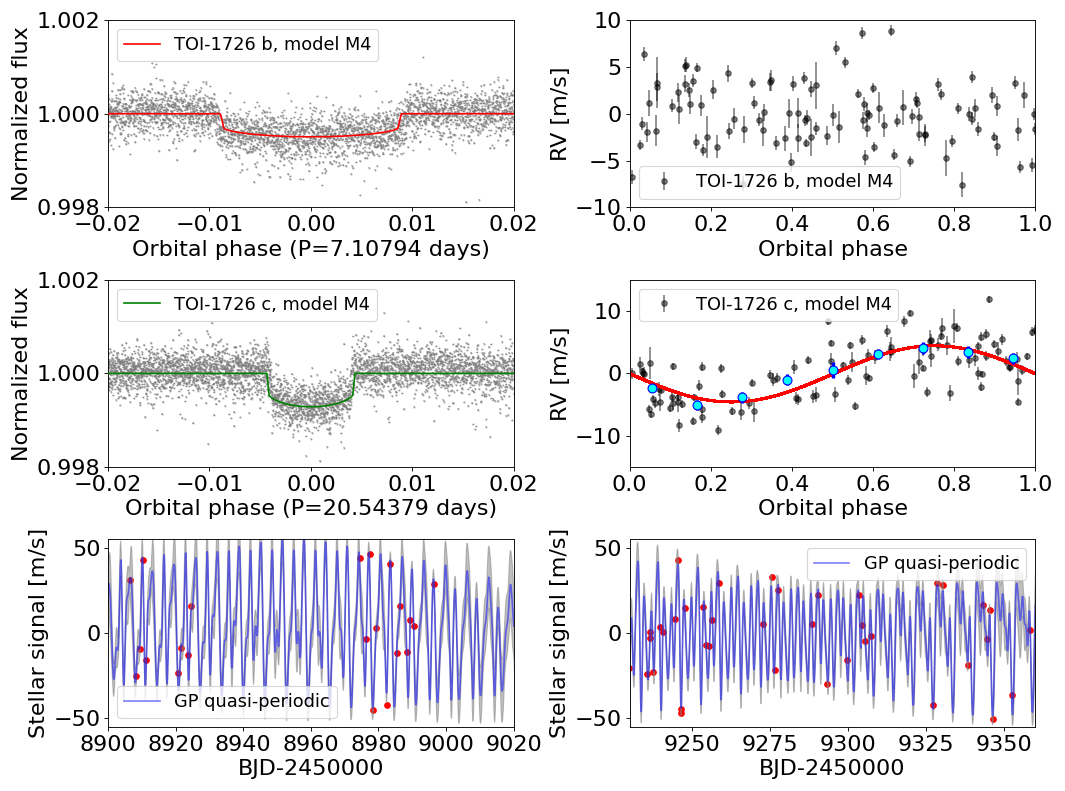}
		\caption{\textit{First two rows.} \textsc{TESS} light curve and best-fit transit models (left panels), and spectroscopic orbits (HARPS-N \texttt{DRS} RVs; right panels) related to planets HD\,63433\,b and HD\,63433\,c. Here, we show the solution obtained with model M4. The upper right plot shows the RV residuals, after removing the Doppler signal due to HD\,63433\,c and the stellar activity term from the original data, phased to the orbital period of 7.01794 d: it appears clear that the spectroscopic orbit corresponding to HD\,63433\,b remains not characterised. The phase-folded plot for HD\,63433\,c shows the RV residuals after removing the stellar activity term from the original data. \textit{Third row.} Zoomed view of the stellar activity term in the HARPS-N RVs, after removing the Doppler signal due to HD\,63433\,c from the original RVs, as modelled by a GP quasi-periodic kernel. In model M5, the GP quasi-periodic kernel was trained on the time series of the $\log R^{\prime}_{\rm HK}$ chromospheric activity diagnostic. }
		\label{fig:lc_rv_solution}
	\end{figure*}

	Concerning the transit ephemeris and other specific parameters of the transit model, we did not find significant differences among the results obtained for all the tested models (except for the models M5 and M6, where the light curve is not fitted). We provide in Table \ref{tab:resultmodel} the results of model M4, selected among the solutions available, which are all equivalent concerning the transit ephemeris. We underline that the election of one of the test models to our reference model is a tricky issue in this case, because there is not a representation of the activity term in the RVs which we can significantly define the best. Therefore, the choice of model M4 should be considered mainly for illustrative purposes. We found that for model M4 the fitted RV jitter $\sigma_{\rm jitt,\, HARPS-N}$ is nearly half that of model M1, for which we got $\sigma_{\rm jitt,\, HARPS-N}$ = 12.1$\pm$1.8 $\ms$, suggesting that, at some level, the activity modelling benefits from using the time series of the $\log R^{\prime}_{\rm HK}$ index to constrain the GP hyper-parameters. Nonetheless, we note that even the RV uncorrelated jitter is lower for model M5 ($\sigma_{\rm jitt,\, HARPS-N}=3.8^{+1.8}_{-2.1}$ $\ms$), but the masses of planet c are different: for model M4, it is similar to that of Neptune at a $\sim2.7\sigma$ level, while for model M5 we find 7 $\mearth$ with $\sim2\sigma$ significance. 
	We show the best-fit solution (model M4) for transits and spectroscopic orbits in Fig. \ref{fig:lc_rv_solution}.
		
With more data from new TESS Sectors, we improved the precision of the transit ephemeris and other transit-related parameters, like the planetary radii, with respect to the results of \cite{mannetal2020}, confirming that the planets have co-planar orbits. Nevertheless, some caution is required about the precision of our determined planetary radii, which, for each planet, is calculated by fitting a light curve obtained by combining all the available transits together, and it is likely optimistic. The unspotted level of HD\,63433 is unknown and this introduces a systematic error on the planetary radii. Considering that the flux in the TESS passband is modulated at the level of about 2\%, we estimate a systematic error at the level of 1\% on the planetary radii from transits close to light maxima and minima, respectively. This comes from the fact that the square of the ratio between the planetary and the stellar radii is proportional to the relative flux deficit at the centre of a transit. Therefore, a change in the out-of-transit flux by about 2\% between light maxima and minima will affect the planetary radius derived from the corresponding transits by about 1\%. This error estimate is indeed a lower limit because starspots uniformly distributed in longitude do not contribute to the amplitude of the photometric modulation, but affect the measurement of the radius ratio.
An additional systematic effect is produced by the occultations of starspots during transits, but its amplitude is probably lower by about 50\% than the effect produced by the changes in the out-of-transit light reference level, if we assume that spots in this G dwarf have a temperature deficit of about 1000 K with respect to the unperturbed photosphere and a filling factor of about 2\% in the latitude bands occulted by the planets \citep[e.g.][]{ballerini2012}. 
	We did not find a significant RV acceleration ($\dot{\gamma}=0.004\pm0.017 \ms d^{-1}$). 
	
	Limits on outer Jovian planets around HD\,63433 can be placed by the Hipparcos-Gaia proper motion anomaly technique \citep{brandt2021ApJS..254...42B,kervella2022A&A...657A...7K}. For example, the sensitivity limits based on this approach would be compatible with the presence of a $\gtrsim1.0$ M$_\mathrm{Jup}$ companion in the 'sweet spot' separation range $3-10$ au \citep{kervella2022A&A...657A...7K}. Interestingly, we note that the reported S/N of the proper motion anomaly for HD\,63433 is in the range 2--2.5 (with a growing trend from Gaia DR2 to DR3). While low, it is intriguing, as in this regime of orbital separations Jupiter- and super-Jupiter-mass companions might still be missed by existing RV datasets (see e.g. Fig. 9 of \citealt{hirsch2021}), particularly if they were to lie on significantly non-coplanar orbits. Given that the star is nearby, there are good prospects for improved sensitivity to such companions based on Gaia DR4 results, which are expected to be released in late 2025. 

	We did not measure different nor more precise results for the planetary masses when using the RVs extracted with \texttt{TERRA}. As an example, for the case of models M1, M4, and M6 we got $m_b=3.8^{+3.9}_{-2.6}$ and $m_c=14.8^{+8.0}_{-7.7}$ $\mearth$, $m_b=2.6^{+3.2}_{-1.8}$ and $m_c=15.1^{+7.5}_{-7.1}$ $\mearth$, and $m_b=1.8^{+2.3}_{-1.3}$ and $m_c=5.9^{+4.6}_{-3.8}$ $\mearth$ respectively. 
	
	\begin{table*}[htbp] 
		\tiny
		\centering
		\caption{Summary of the different RV (\texttt{DRS}) + transit photometry + activity diagnostics (when specified) models tested in this work. All include two Keplerians for planets b and c, except for models M5 and M6, for which we adopted circular orbits. }
   \label{tab:models}
		\begin{tabular}{ccccc}
			\noalign{\smallskip}
			\hline
			\textbf{Model ID} & \textbf{Description} & \textbf{Bayesian evidence}\tablefootmark{a} & \textbf{Planetary mass\tablefootmark{b}} & \textbf{Planetary density}\\
			& & ($\ln \mathcal{Z}$) & $\mearth$ & $g cm^{-3}$ \\
			\hline
			\noalign{\smallskip}
			M1   & GP quasi-periodic kernel applied to RVs & $103927.9\pm0.6$ & $m_b = 4.1^{+3.7}_{-2.7}$ (10.6) & $\rho_b=2.7^{+2.5}_{-1.9}$ (7.3) \\
			&  &                  & $m_c = 16.6^{+8.1}_{-7.9}$ (29.4)  & $\rho_c=6.0^{+3.1}_{-2.8}$ (11.0)      \\
			\noalign{\smallskip}
			M2   & GP quasi-periodic kernel applied to RVs, with a different set  & $103925.5\pm0.6$ & $m_b = 3.5^{+3.7}_{-2.5}$ (10.5) & $\rho_b=2.3^{+2.6}_{-1.7}$ (7.0) \\
			& of hyper-parameters (\textit{h,w,$\lambda$}) for each observing season &                  & $m_c = 20.5^{+8.7}_{-9.4}$ (34.1) & $\rho_c=7.7^{+3.3}_{-3.6}$ (13.2)  \\  
			\noalign{\smallskip}
			M3 & GP quasi-periodic kernel trained\tablefootmark{c} on BIS & $103417.5\pm0.6$ & $m_b = 2.9^{+3.5}_{-1.7}$ (9.4) & $\rho_b=1.9^{+2.3}_{-1.3}$ (6.2) \\
			&  &                  & $m_c = 19.7^{+6.8}_{-7.3}$ (31.3) & $\rho_c=7.4\pm2.8$ (12.0) \\ 
			\noalign{\smallskip}
			M4 & GP quasi-periodic trained\tablefootmark{c} on $\log R^{\prime}_{\rm HK}$ & $104237.8\pm0.6$ & $m_b = 2.4^{+3.0}_{-1.8}$ (8.1) & $\rho_b=1.6^{+1.9}_{-1.1}$ (5.4)\\
			&  &                  & $m_c = 18.9\pm7.0$ (30.4) & $\rho_c=7.1^{+2.8}_{-2.6}$ (11.7) \\ 
			\noalign{\smallskip}
			M5 & Multi-dimensional GP framework\tablefootmark{d}  
			& $836.6\pm0.3$ & $m_b = 1.2^{+1.4}_{-0.9}$ (3.9) & $\rho_b=0.8^{+1.0}_{-0.6}$ (2.6)\\
			& using $\log R^{\prime}_{\rm HK}$ and BIS, and the quasi-periodic kernel; circular orbits & & $m_c = 7.0^{+3.2}_{-3.0}$ (12.1) & $\rho_c=2.6^{+1.2}_{-1.1}$ (4.6) \\
			\noalign{\smallskip}
			M6 & Multi-dimensional GP framework\tablefootmark{d}  
			& $565.6\pm0.5$ & $m_b = 1.6^{+2.2}_{-1.2}$ (5.6) & $\rho_b=1.1^{+1.5}_{-0.8}$ (3.8) \\
			& using $\log R^{\prime}_{\rm HK}$ and the quasi-periodic kernel; circular orbits & & $m_c = 8.7^{+5.1}_{-4.6}$ (17.1) & $\rho_c=3.3^{+2.0}_{-1.7}$ (6.6) \\
			\noalign{\smallskip}
			M7 & GP dSHO kernel applied to RVs & $103923.6\pm0.5$ & $m_b = 2.9^{+3.0}_{-2.0}$ (8.4) & $\rho_b=1.9^{+2.1}_{-1.3}$ (5.8) \\
			& & & $m_c = 11.8^{+7.7}_{-6.8}$ (25.0) & $\rho_c=4.3^{+2.9}_{-2.5}$ (9.1)\\
			\noalign{\smallskip}  
			M8 & GP QPC kernel applied to RVs & $103928.1\pm0.6$ & $m_b = 4.3^{+4.2}_{-2.9}$ (11.4) & $\rho_b=2.8^{+2.8}_{-1.9}$ (7.5) \\
			& & &  $m_c = 16.6^{+8.7}_{-8.3}$ (30.5) & $\rho_c=6.2^{+3.3}_{-3.1}$ (11.7) \\
			\noalign{\smallskip}
			M9 & Activity corrected through a linear regression fit between RVs and BIS & $103991.2\pm0.5$ & $m_b = 2.1^{+2.1}_{-1.4}$ (6.2) & $\rho_b=1.4^{+1.4}_{-1.0}$ (4.2) \\ 
			&  &  & $m_c =8.4\pm4.9$ (16.7) & $\rho_c=3.1\pm1.8$ (6.3)\\
			\noalign{\smallskip}
			\hline
		\end{tabular}
		\label{tab:my_label}
		\tablefoot{\tablefoottext{a}{We note that the analysed datasets related to each model are not always the same, therefore model comparison based on the values of the Bayesian evidence $\ln\mathcal{Z}$ can be performed only when the same datasets are involved.}
		\tablefoottext{b}{The 95$^{th}$ percentile is provided in parenthesis.}
			\tablefoottext{c}{With ``trained'' we mean that the same GP kernel is used to model the RVs and the activity diagnostic. In the case of a quasi-periodic kernel, three hyper-parameters over four, namely \textit{w,$\theta$}, and \textit{$\lambda$}, are shared by the two datasets, while two distinct amplitudes \textit{h} are used.}
			\tablefoottext{d}{To speed up the analysis, we did not fit the photometric transits, and we assumed circular orbits (the results from other models show that the eccentricities are consistent with zero). The priors for the planet ephemeris are defined based on the results of the M4 model. We adopted planet radii calculated from M4 to derive the densities. }
		}
	\end{table*}

	\begin{table*}[htbp]
		\caption{Best-fit values of the free parameters of model M4 (RVs calculated by the \texttt{DRS}).}
		\label{tab:resultmodel}
		\tiny
		\begin{center}
			\begin{tabular}{ll}
				\hline
				\textbf{Parameter}   & \textbf{Best-fit value}\tablefootmark{a}\\
				\hline
				\noalign{\smallskip} 
				\textbf{Fitted} &  \\
				\textit{RV stellar activity GP term:} & \\
				\noalign{\smallskip}
				$h$ [\ms] & $26.0^{\rm +3.2}_{\rm -3.0}$\\
				\noalign{\smallskip}
				$\lambda$ [d] & $45.7^{\rm +14.0}_{\rm -9.9}$ \\
				\noalign{\smallskip}
				$w$ & $0.23\pm0.02$ \\ 
				\noalign{\smallskip}
				$\theta$ [d] & $6.39\pm0.01$ \\
				\noalign{\smallskip}
				\textit{Planet-related parameters:} &  \\
				\noalign{\smallskip}  
				$K_b$ [\ms] & $0.8^{\rm +1.0}_{\rm -0.6}$ (2.7)\\
				\noalign{\smallskip} 
				orbital period, $P_b$ [d] & $7.10794\pm0.000009$ \\
				\noalign{\smallskip}
				T$_{conj,\,b}$ [BJD-2450000] & $9584.5991\pm0.0004$ \\
				\noalign{\smallskip}
				$\sqrt{e_b}\cos{\omega_{\star,\,b}}$ & $-0.029^{+0.335}_{-0.303}$ \\
				\noalign{\smallskip}
				$\sqrt{e_b}\sin{\omega_{\star,\,b}}$ & $-0.133^{+0.092}_{-0.138}$ \\
				\noalign{\smallskip}
				$K_c$ [\ms] & $4.5\pm1.7$ (7.3) \\
				\noalign{\smallskip} 
				orbital period, $P_c$ [d] & $20.54379\pm0.00002$ \\
				\noalign{\smallskip}
				T$_{conj,\,c}$ [BJD-2450000] & $9583.6367\pm0.0004$ \\
				\noalign{\smallskip}
				$\sqrt{e_c}\cos{\omega_{\star,\,c}}$ & $-0.277^{+0.373}_{-0.255}$ \\
				\noalign{\smallskip}
				$\sqrt{e_c}\sin{\omega_{\star,\,c}}$ & $-0.060^{+0.128}_{-0.140}$ \\
				\noalign{\smallskip}
				acceleration, $\dot{\gamma}$ [$\ms d^{-1}$] &  $0.004\pm0.017$ \\
				\noalign{\smallskip} 
				$R_{\rm b}/R_{\star}$ & $0.02063^{\rm +0.00052}_{\rm -0.00028}$ \\
				\noalign{\smallskip}
				inclination, $i_b$ [deg] &  $88.96^{+0.70}_{-0.82}$  \\
				\noalign{\smallskip}
				$R_{\rm c}/R_{\star}$ & $0.0249^{+0.0004}_{-0.0003}$ \\
				\noalign{\smallskip}
				inclination, $i_c$ [deg] &  $89.41^{+0.33}_{-0.26}$  \\
				\noalign{\medskip}
				\textit{RV-related parameters} &  \\
				\noalign{\smallskip}          
				$\sigma_{\rm jit,\: HARPS-N}$ [\ms] & $6.6_{\rm -4.5}^{\rm +2.9}$ \\ 
				\noalign{\smallskip}
				$\gamma_{\rm HARPS-N}$ [\ms] & $-15806.1_{\rm -4.6}^{\rm +4.5}$ \\
				\noalign{\smallskip}
				\textit{Light curve-related parameters} &  \\  
				\noalign{\smallskip}
				$\sigma_{jit, \,TESS\,sect.\,20}$ & $0.000341\pm0.000004$ \\
				\noalign{\smallskip}
				$\sigma_{jit, \,TESS\,sect.\,44-47}$ & $0.000262\pm0.000002$ \\
				\noalign{\smallskip}
				$LD_{\rm c1}$ & $0.50^{+0.13}_{-0.14}$ \\
				\noalign{\smallskip}
				$LD_{\rm c2}$ &  $-0.05^{+0.22}_{-0.17}$\\
				\textbf{Derived} & \\
				eccentricity, $e_b$ & $0.09^{+0.14}_{-0.07}$ (0.37) \\
				\noalign{\smallskip}
				$a_b/R_{\star}$\tablefootmark{b} &  $16.8^{+0.4}_{-1.0}$ \\
				\noalign{\smallskip}
				$a_{b}$ [au] &  $0.0722\pm0.0007$\\
				\noalign{\smallskip}
				impact param., $b_b$ & $0.32_{\rm -0.21}^{\rm +0.28}$ \\
				\noalign{\smallskip}
				transit duration, $T_{b,\,1,4}$ [d] & $0.129^{+0.005}_{-0.011}$ \\
				\noalign{\smallskip}
				radius\tablefootmark{c}, $r_b$ [$\rearth$]  & $2.02^{+0.06}_{-0.05}$ \\
				\noalign{\smallskip}
				mass, $m_b$ [$\mearth$] & $2.4^{+3.0}_{-1.8}$ (8.1) \\
				\noalign{\smallskip}
				mean density, $\rho_b$ [g cm$^{\rm -3}$] & $1.6^{+2.0}_{-1.2}$ (5.4) \\
				\noalign{\smallskip}
				$T_{\rm eq.,\, b}$ [K] & $969\pm17$ \\
				\noalign{\smallskip}
				insolation, $S_b$ [$S_{\rm \oplus}$] & $146\pm10$ \\
				\noalign{\smallskip}
				eccentricity, $e_c$ & $0.11^{+0.18}_{-0.07}$ (0.43) \\
				\noalign{\smallskip}
				$a_c/R_{\star}$\tablefootmark{b} &  $36.4^{+1.0}_{-1.5}$ \\
				\noalign{\smallskip}
				$a_{c}$ [au] &  $0.147\pm0.001$\\
				\noalign{\smallskip}
				impact param., $b_c$ & $0.34^{+0.17}_{-0.19)}$ \\
				\noalign{\smallskip}
				transit duration, $T_{c,\,1,4}$ [d] & $0.173^{+0.006}_{-0.008}$ \\
				\noalign{\smallskip}
				radius\tablefootmark{c}, $r_c$ [$\rearth$] & $2.44\pm0.07$ \\
				\noalign{\smallskip}
				mass, $m_c$ [$\mearth$] & $18.9^{+7.0}_{-6.9}$ (30.3) \\
				\noalign{\smallskip}
				mean density, $\rho_c$ [g cm$^{\rm -3}$] & $7.1^{+2.8}_{-2.6}$ (11.7) \\
				\noalign{\smallskip}
				$T_{\rm eq.,\, c}$ [K] & $680\pm12$ \\
				\noalign{\smallskip}
				Insolation, $S_c$ [$S_{\rm \oplus}$] & $36\pm3$ \\
				\noalign{\smallskip}
				\hline
				\hline
			\end{tabular}
			\tablefoot{\tiny
				\tablefoottext{a}{The uncertainties are given as the $16^{\rm th}$ and $84^{\rm th}$ percentiles of the posterior distributions. For some of the parameters, we provide the $95^{\rm th}$ percentile in parenthesis.}
				\tablefoottext{b}{In place of $a_{\rm}/R_{\star}$, we used the stellar density $\rho_{*}$ as a free parameter with Gaussian prior $\mathcal{N}$(1.376,0.078) $\rho_{\odot}$, from which we derived the $a_{\rm p}/R_{\star}$ ratios at each step of the MC sampling.}
				\tablefoottext{c}{A systematic error of at least 1$\%$ of the best-fit radius should be added, to take into account effects due to stellar activity (Sect. \ref{sec:rvlcanalysis}).}
			}
		\end{center}
	\end{table*}
	
\section{Considerations on planetary structure based on the mass-radius diagram} \label{sec:mrdiag}
The more conservative conclusion that we can draw from the analysis described in Sect. \ref{sec:rvlcanalysis} is that we determined upper limits for the planet masses, $m_b\la$11 $\mearth$ and $m_c\la$31 $\mearth$. The mass of HD\,63433\,b remains undetected (and we cannot exclude the possibility that this result is mainly influenced by the choice of a specific model). The mass of HD\,63433\,c remains (very) uncertain, but the clue that it might be Neptune-like would imply a bulk density that, if confirmed, would be interestingly high for a planet with the size of a mini-Neptune. For this reason, in this Section we comment on this and other possibilities for the planets' physical structures which are compatible with our results, by inspecting the location of the two planets in a mass-radius diagram for exoplanets. Specifically, our considerations concern the solutions which correspond to models M4 and M5 of Table \ref{tab:models}).
	
\begin{figure*}
    \centering
    \includegraphics[width=0.8\textwidth]{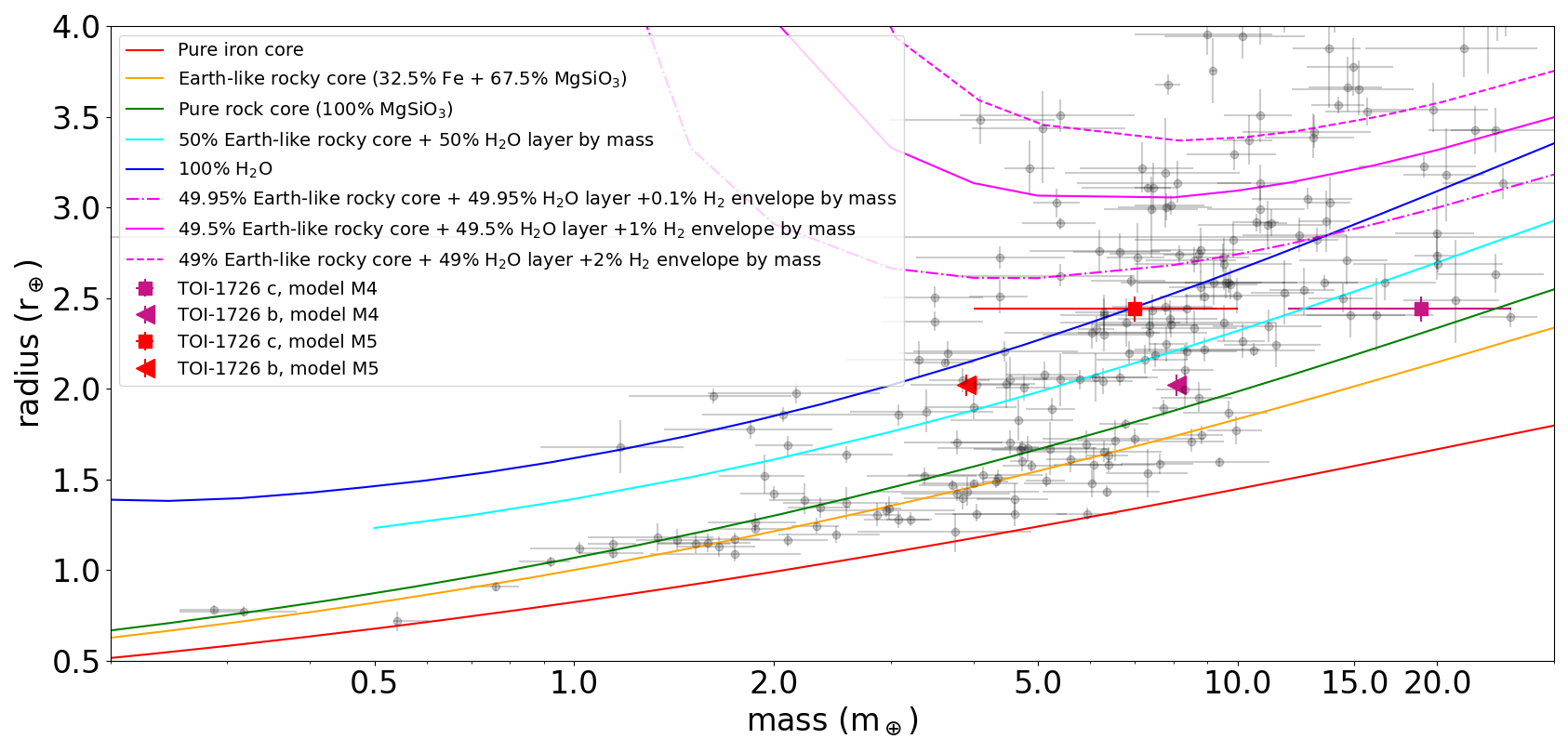}
    \includegraphics[width=0.8\textwidth]{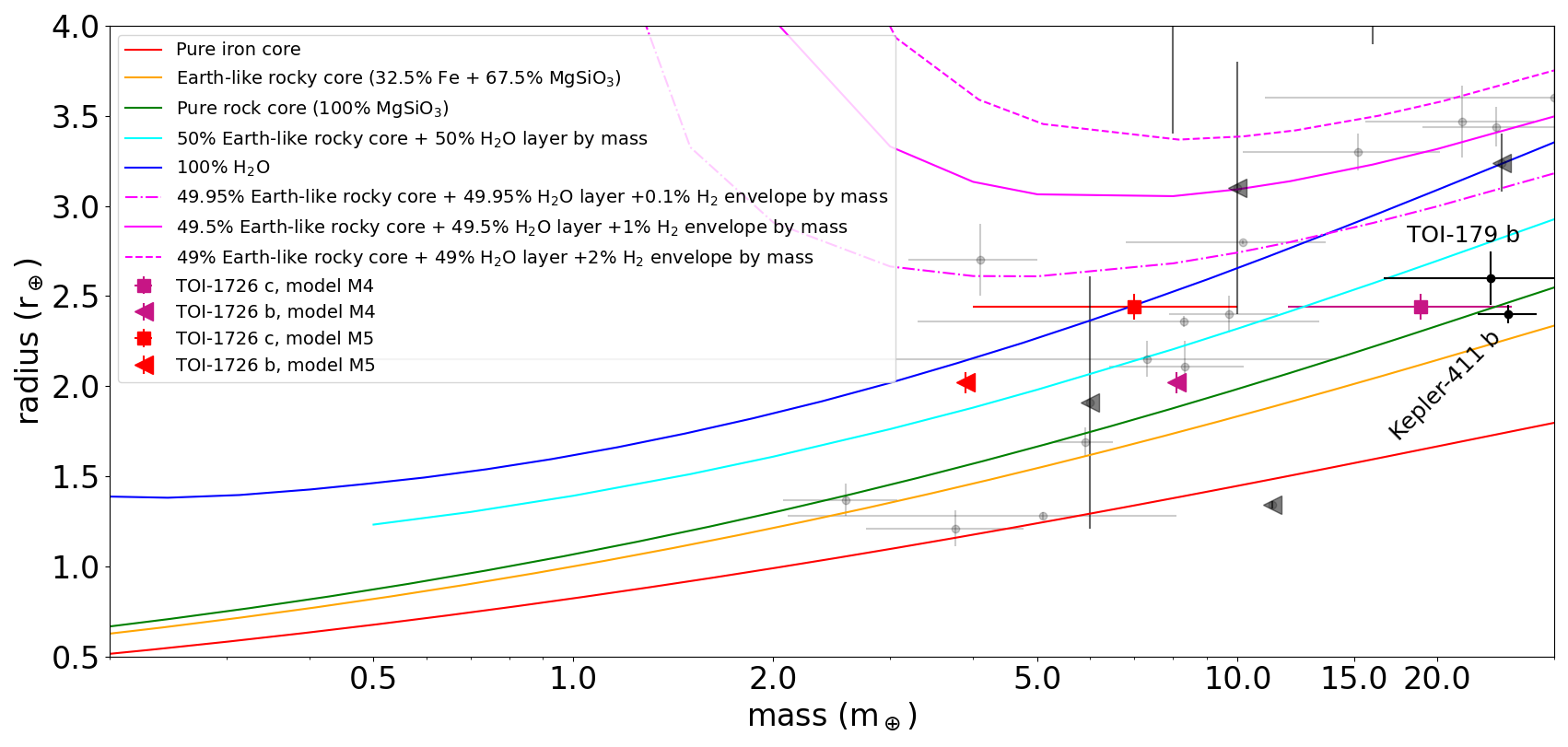}
    \caption{\textit{Upper panel.} Mass-radius diagram for exoplanets selected from the TEPCAT sample, available at \url{https://www.astro.keele.ac.uk/jkt/tepcat/} (updated to 7 October 2022; \citealt{2011MNRAS.417.2166S}). Black dots represent planets with mass and radius measured with a relative precision lower than 30$\%$ and 10$\%$, respectively. Planets of the HD\,63433 system are indicated by reddish triangles (planet b), which denote mass upper limits, and squares (planet c), with reference to the corresponding models listed in Table \ref{tab:models}. Theoretical curves for some planet compositions are overplotted, as calculated by \cite{2019PNAS..116.9723Z} assuming 1 milli-bar surface pressure. Models for a planet with an H$_2$O-gaseous atmosphere (50$\%$ and 100$\%$ H$_2$O by mass, cyan and blue curves, respectively), and for a planet with different percentages of an H$_2$ gaseous envelope over a 50\% water-rich layer (magenta curves) are calculated for an isothermal fluid/steam envelope equilibrium temperature of 700 K, similar to that of HD\,63433\,c (680$\pm12$ K).
    \textit{Lower panel.} Mass-radius diagram for planets with age $<$ 900 Myr. Triangles are used to identify planets for which only mass upper limits are available. No selection is made based on the precision of mass and radius measurements. We used \cite{2022arXiv221007933D} as a reference for the mass and radius of TOI-179\,b.}
    \label{fig:mass_radius_diag}
\end{figure*}

	We show in the upper panel of Fig. \ref{fig:mass_radius_diag} a mass-radius diagram which includes well-characterised planets (with masses and radii measured with a precision lower than 30\% and 10\%, respectively). Assuming the lower value for $m_c$ (model M5), HD\,63433\,c is located in a well populated region of the diagram, corresponding to water worlds, i.e. planets containing significant amounts of H$_2$O-dominated fluid/ice in addition to rock and gas. Instead, if the estimate of $m_c$ from model M4 is the one more representative of the real mass of HD\,63433\,c, the planet would occupy a position on the diagram which is less populated, but not empty, suggestive of a structure with a rocky core and a water layer $<50\%$ by mass. 
	The lower panel of Fig. \ref{fig:mass_radius_diag} shows a mass-radius diagram which includes only planets with age $<$ 900 Myr, with no selection based on the precision of their masses and radii. We note that, assuming the higher mass (density) estimate from model M4, HD\,63433\,c would have a similar structure of TOI-179\,b \citep{2022arXiv221007933D,vines2023} and Kepler-411\,b \citep{sun2019}. TOI-179 is a K2V star with an age similar to that of the system HD\,63433 (400$\pm$100 Myr; \citealt{2022arXiv221007933D}). TOI-179\,b is a dense mini-Neptune that moves on an eccentric orbit ($e=0.34^{+0.07}_{-0.09}$) with a shorter orbital semi-major axis ($a=0.0481\pm0.0004$ au) and higher equilibrium temperature ($T_{\rm eq}\sim990$ K, in the case of a circular orbit with same semi-major axis $a$) than HD\,63433\,c. \cite{2022arXiv221007933D} show that the typical composition of TOI-179\,b can be assumed 75\% rock + 25\% water, and conclude that the planet has likely lost most of its primordial atmosphere, and it is stable against hydrodynamic evaporation. Kepler-411\,b is a member of a fairly compact four-planet system orbiting a $\sim200$ Myr old K2V star with a period similar to that of TOI-179\,b.
	
	Based on our derived upper limits to the mass of HD\,63433\,b, we can conclude that its composition could be compatible with that of a water world, and we cannot exclude that it is similar to that of the outermost companion, especially if we consider the results of the M5 model. As for the case of TOI-179\,b, noticing that the two planets have similar equilibrium temperatures, HD\,63433\,b could have lost any primordial gaseous H-He envelope (as we discuss in detail in the next Section), and reached its definitive position on the mass-radius diagram.
 
	\section{Planetary atmosphere mass-loss due to photoevaporation }
	\label{sec:atmphotoev}
	
	We evaluated the mass-loss rate of the planetary atmosphere of planet b and c using the hydro-based approximation developed by \cite{kuby+2018a,kuby+2018b}, coupled with the planetary core-envelope model by \cite{LopFor14} and the MESA Stellar Tracks (MIST; \citealt{choi+2016}). For the stellar X-ray emission at different ages, we adopted the analytic description by \cite{Penz08a}, anchored to the current value of the X-ray luminosity, $L_{\star,\, X} = 7.5 \times 10^{28}$ erg s$^{-1}$ in the band 5--100 $\AA$, derived from XMM-Newton spectra \citep{2022AJ....163...68Z}. The stellar EUV luminosity (100--920 $\AA$) was computed by means of the new scaling law by \cite{SF22}. 
	More details on our modelling of atmospheric evaporation are provided in Appendix \ref{app:photoevap}. Following \citet{maggio2022}, we expect small but non negligible differences in the evaporation efficiency and timescales if a different X-ray to EUV scaling law is assumed (e.g. \citealt{King+Wheatley21}, \citealt{Johnstone+2021}), but a detailed comparison of results is not warranted at this stage, given the uncertainties on the planetary masses.
	
	We performed several simulations of the past and future evolution of the planetary atmospheres, assuming different possible values for the planetary masses at the current age. The grids of test masses are defined by taking into account the upper limits presented in Sect. \ref{sec:rvlcanalysis}. For planet b, we explored the mass range 2--12 $\mearth$, while for planet c we limited our analysis to masses in the range 5--15 $\mearth$, because we found that for $m_c>15$ $\mearth$ the planet is stable against photoevaporation. 
	
	\begin{figure*}[]
		\centering
		\subfigure[]{\includegraphics[width=0.45\textwidth, trim = 2cm 0cm 2cm 0.2cm]{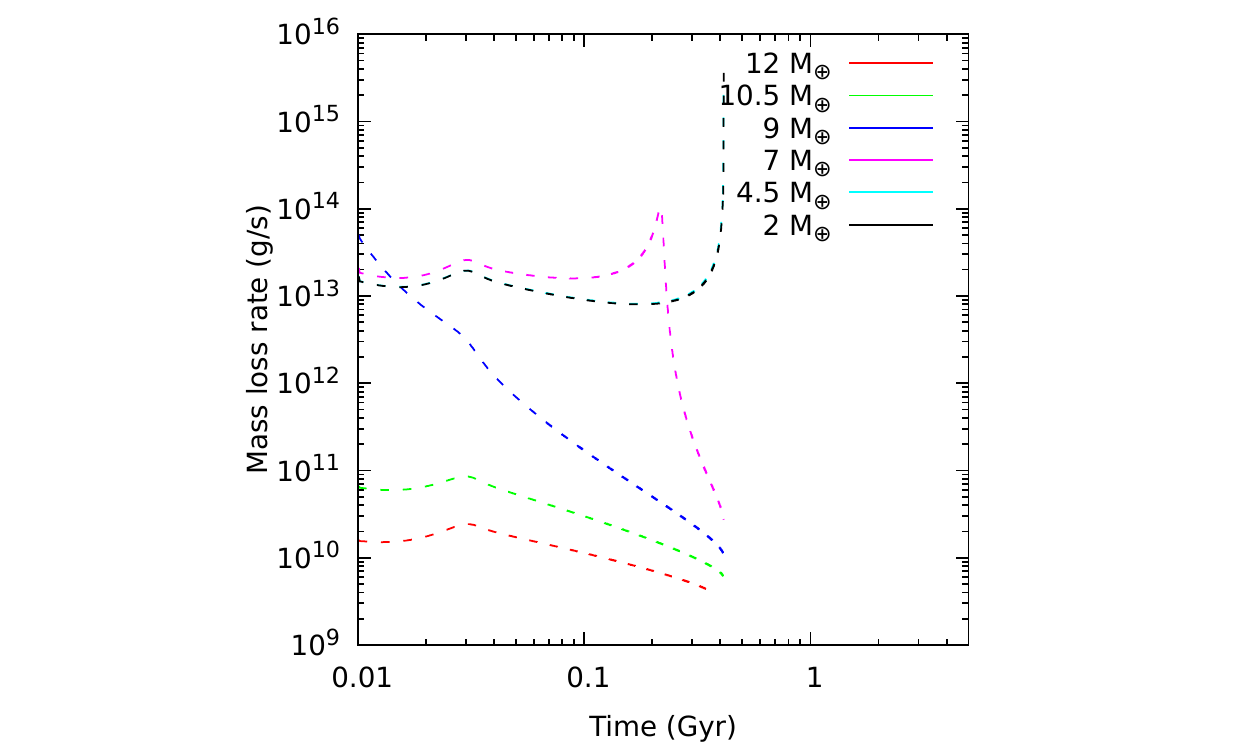}}
		\subfigure[]{\includegraphics[width=0.45\textwidth, trim = 2cm 0cm 2cm 0.2cm]{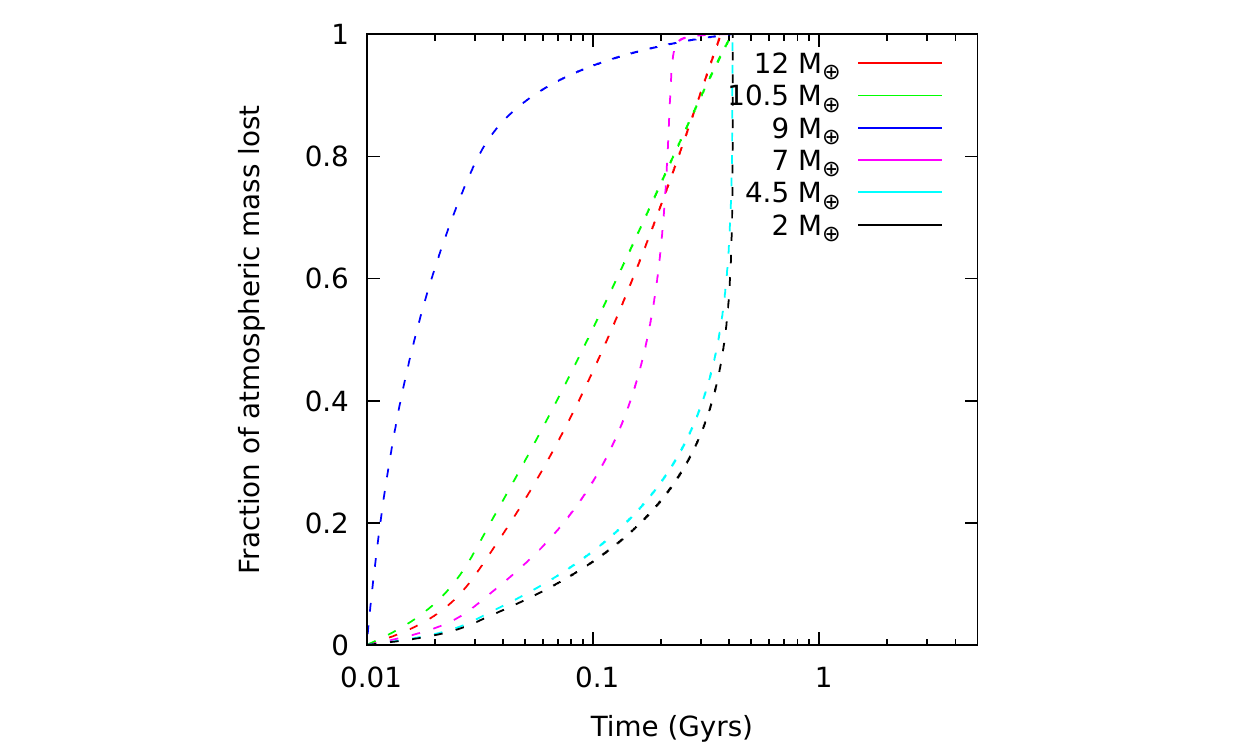}}
		\subfigure[]{\includegraphics[width=0.45\textwidth, trim = 2cm 0cm 2cm 0.2cm]{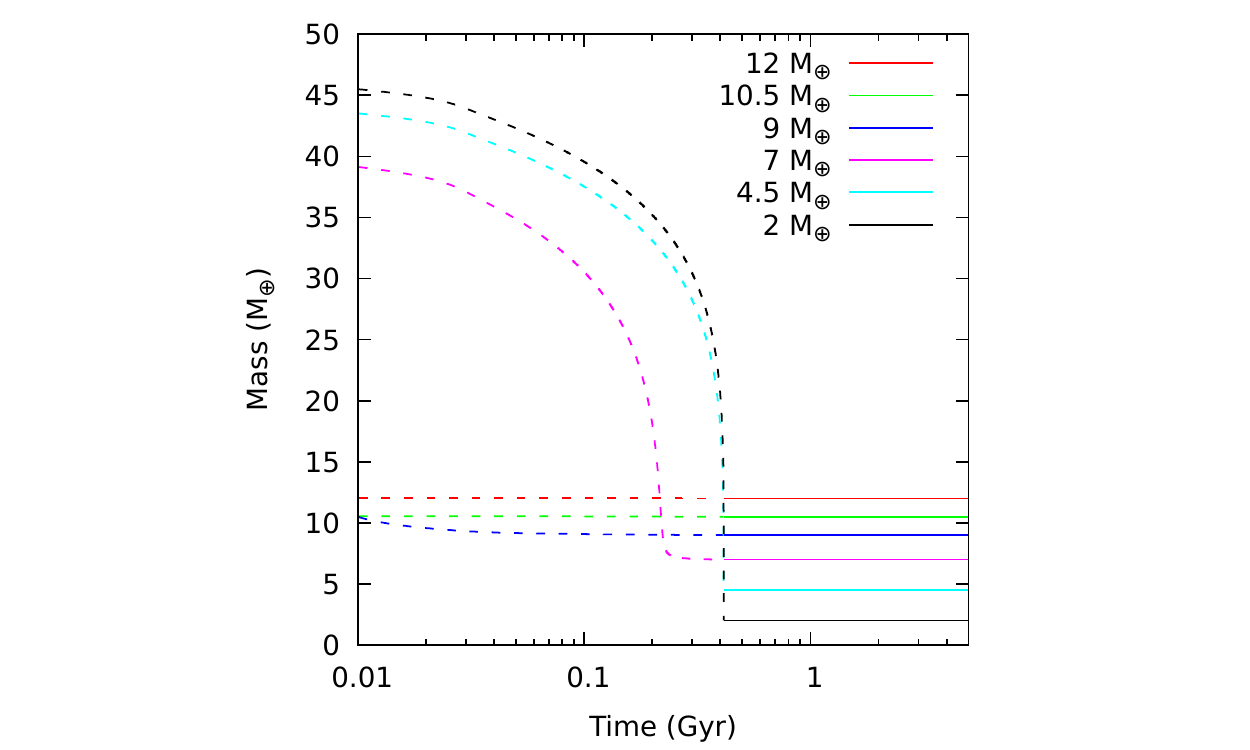}}
		\subfigure[]{\includegraphics[width=0.45\textwidth, trim = 2cm 0cm 2cm 0.2cm]{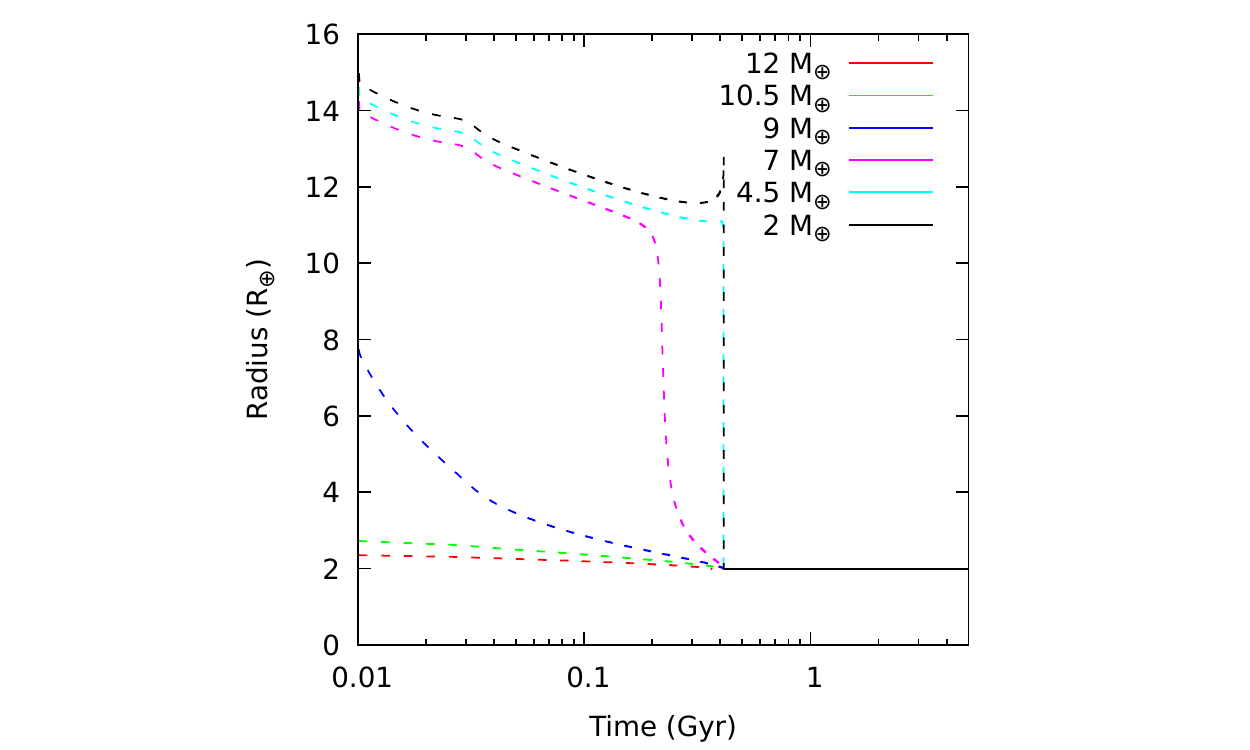}}
		\caption{Evolutionary history of planet HD 63433\,b. (a) Mass-loss rate vs. time for different values of the planetary mass at the present age. (b) Fraction of planetary mass lost vs. time. Panels (c) and (d) show the evolution of the planetary mass and radius, respectively. Dashed lines indicate backward time evolution from the current age, solid lines forward time evolution. Note that all the simulations were constructed with the boundary condition that the planet completely lost the atmosphere at the present age of 414\,Myr. } 
		\label{fig:evaporation_pl_b}
	\end{figure*}
	
	\begin{figure*}[]
		\centering
		\subfigure[]{\includegraphics[width=0.45\textwidth, trim = 2cm 0cm 2cm 0.2cm]{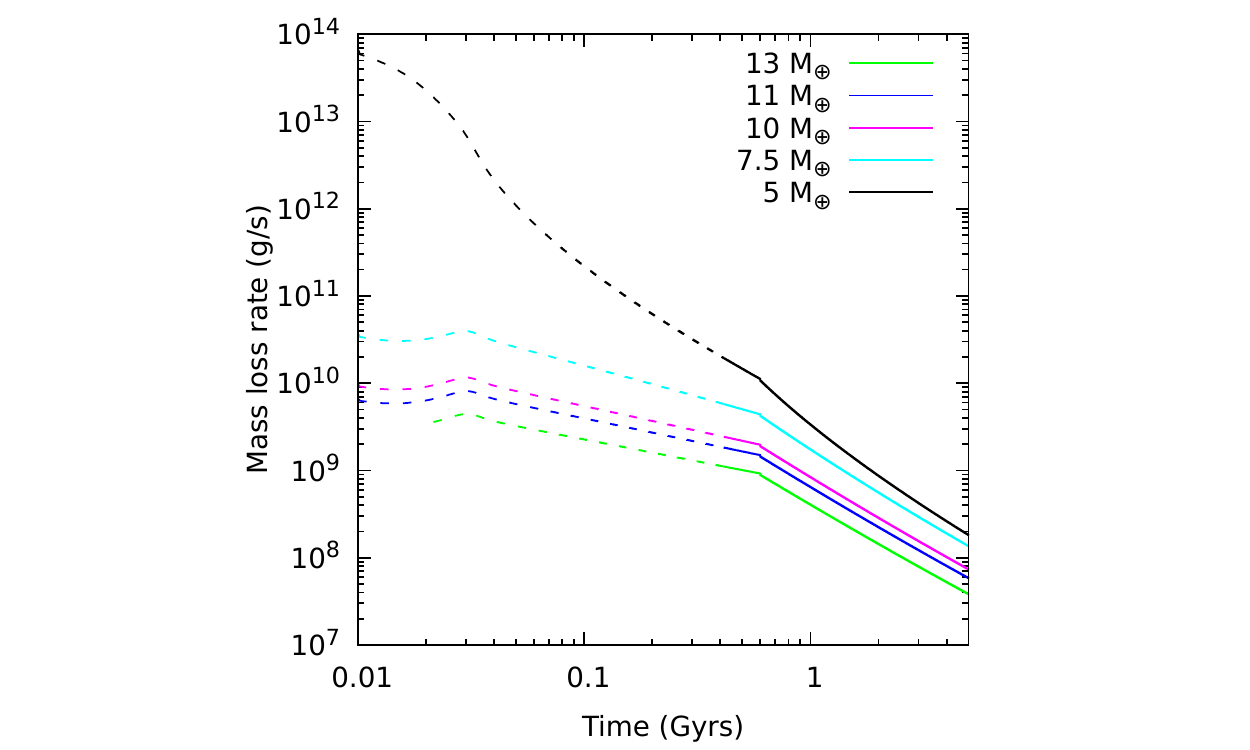}}
		\subfigure[]{\includegraphics[width=0.45\textwidth, trim = 2cm 0cm 2cm 0.2cm]{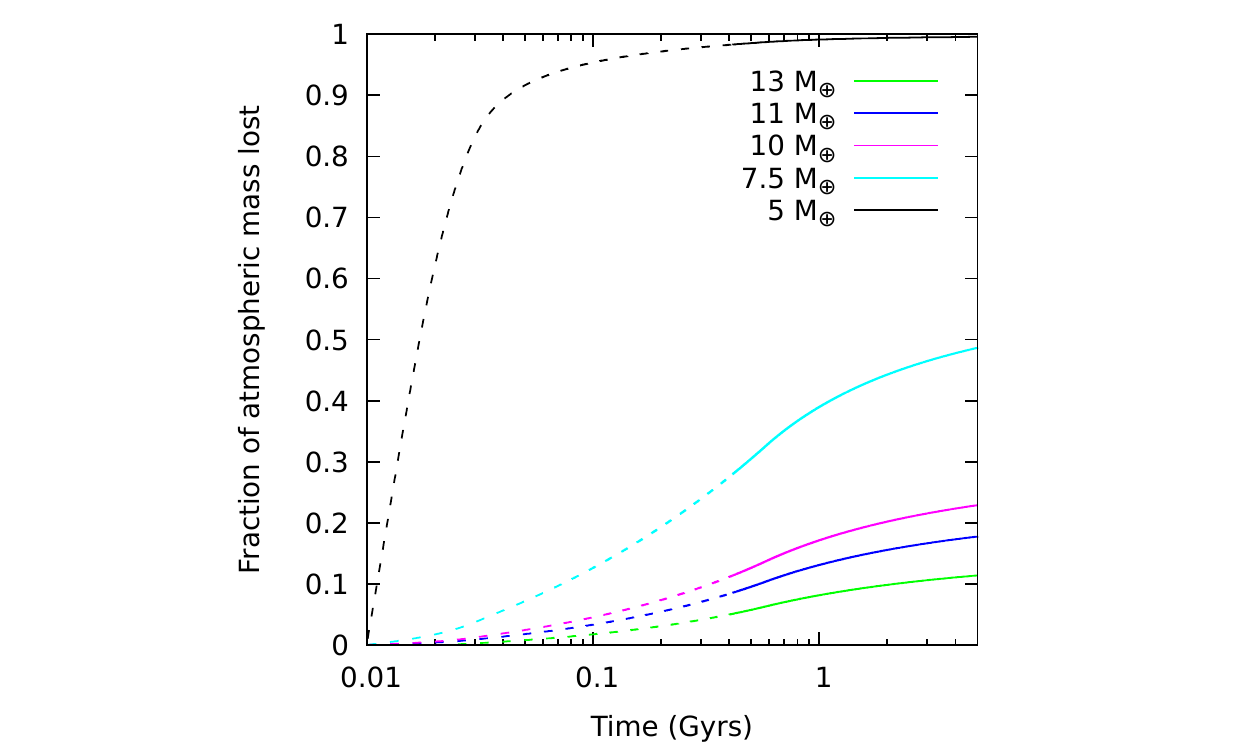}}
		\subfigure[]{\includegraphics[width=0.45\textwidth, trim = 2cm 0cm 2cm 0.2cm]{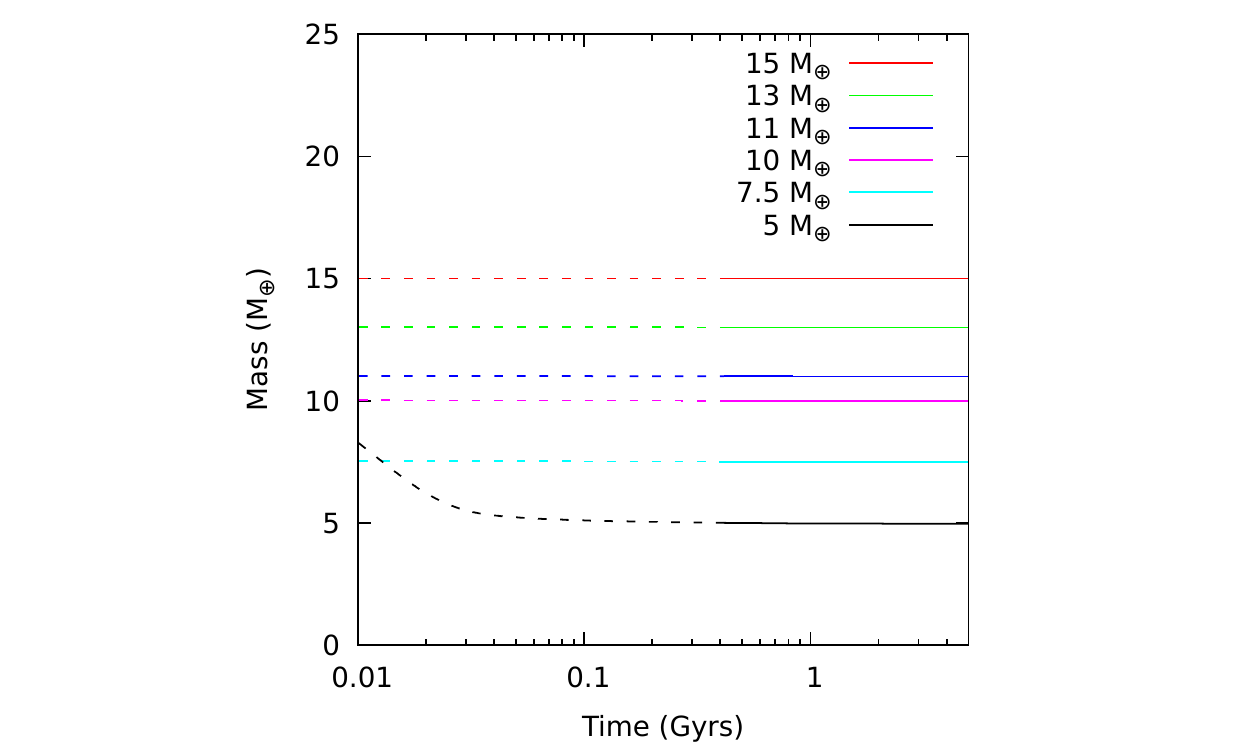}}
		\subfigure[]{\includegraphics[width=0.45\textwidth, trim = 2cm 0cm 2cm 0.2cm]{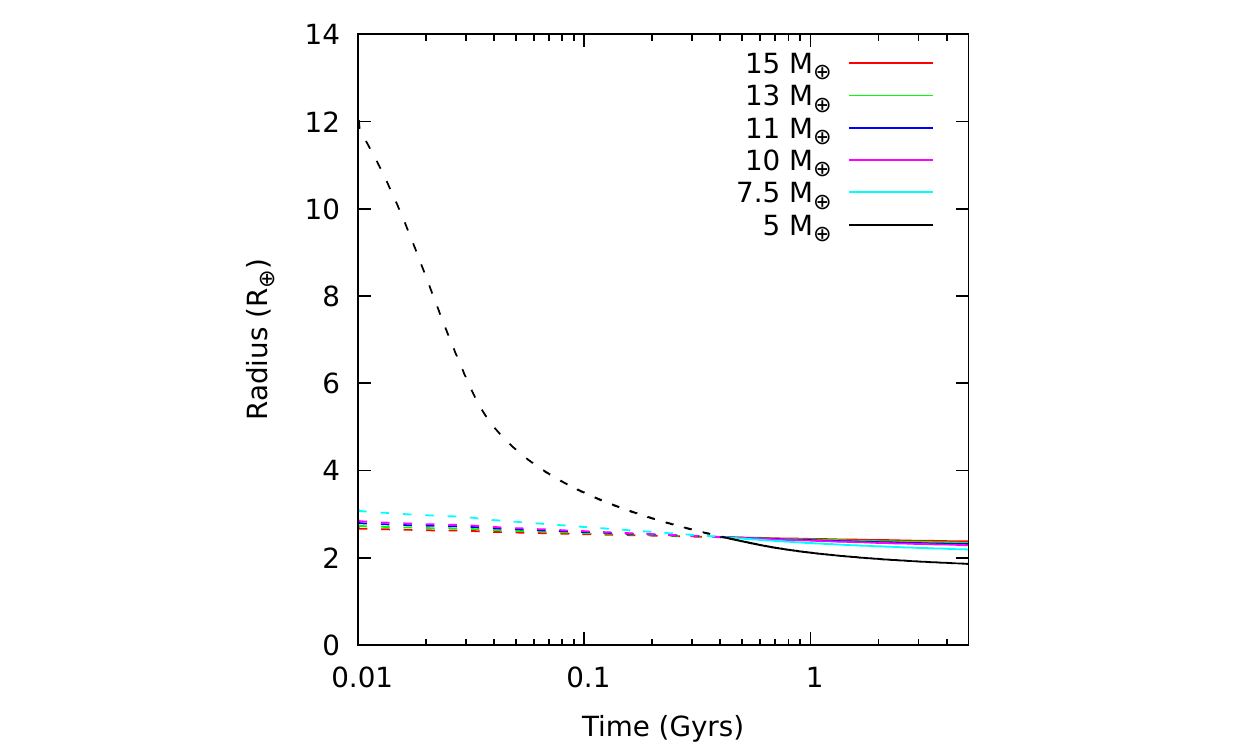}}
		\caption{Evolutionary history of planet HD 63433\,c. Panel (a):  mass-loss rate vs. time for different values of the planetary mass at the present age. Panel (b): Fraction of planetary mass lost vs. time. Panels (c) and (d): time evolution of the planetary mass and radius, respectively. Dashed lines indicate backward time evolution from the current age, solid lines forward time evolution. } 
		\label{fig:evaporation_pl_c}
	\end{figure*}
	
	\begin{table}[]
		\caption{Results of the planetary mass-loss simulations from the current age to 5\,Gyr (Sect. \ref{sec:atmphotoev}). We assumed $r_b=2.02$ and $r_c=2.44$ $\rearth$ as the current values of the planetary radii.}
		\tiny
		\centering
		\begin{tabular}{cccccc}
			\hline
			\textbf{Current} & \textbf{mass-loss rate } & \textbf{Core mass} & \textbf{Core radius} & \textbf{mass-loss } \\  
			\textbf{ planet mass} & \textbf{at 414 Myr} & & & \textbf{timescale}\\
			($\mearth$) &  (g s$^{-1}$) & ($\mearth$) & ($\rearth$) & (Myr) \\
			\hline
			\noalign{\smallskip}
			\textit{Planet b}\\
			\noalign{\smallskip}
			12 & $3.7\times10^9$ & 11.99 & 1.93 & 50 \\
			\noalign{\smallskip}
			10.5 & $6.1\times10^9$ & 10.46 & 1.86 & 103 \\
			\noalign{\smallskip}
			9 & $1.1\times10^{10}$ & 8.99 & 1.79 & 113 \\
			\noalign{\smallskip}
			7 & $2.6\times10^{10}$ & 6.99 & 1.67 & 76 \\
			\noalign{\smallskip}
			5.5\tablefootmark{a} & $1.1\times10^{11}$ & 5.47 & 1.56 & 43 \\
			\noalign{\smallskip}
			4.5 & $1.3\times10^{11}$ & 4.48 & 1.49 & 21 \\
			\noalign{\smallskip}
			2 & $2.5\times10^{12}$ & 1.99 & 1.19 & 1 \\
			\noalign{\smallskip}
			\hline
			\noalign{\smallskip}
			\textit{Planet c}\\
			\noalign{\smallskip}
			15 & 0.0 & 14.93 & 2.05 & - \\
			\noalign{\smallskip}
			13 & $1.1\times10^9$ & 12.93 & 1.97 & $> 5\times10^3$ \\
			\noalign{\smallskip}
			11 & $1.8\times10^9$ & 10.93 & 1.89 & $> 5\times10^3$ \\
			\noalign{\smallskip}
			10 & $2.4\times10^9$ & 9.93 & 1.84 & $> 5\times10^3$ \\
			\noalign{\smallskip}
			7.5 & $5.8\times10^9$ & 7.43 & 1.70 & $> 5\times10^3$ \\
			\noalign{\smallskip}
			7.3\tablefootmark{a} & $1.0 \times 10^{10}$ & 7.2 & 1.69 &  $> 5\times10^3$ \\
			\noalign{\smallskip}
			5 & $1.9\times10^{10}$ & 4.94 & 1.53 & $1.6\times10^3$ \\
			\noalign{\smallskip}
			\hline
		\end{tabular}
		\tablefoot{\tablefoottext{a}{Mass estimated by \citealt{2022AJ....163...68Z}. They assumed $r_b=2.15$ $\rearth$ and $r_c=2.67$ $\rearth$ as the current values of the planetary radii.}
		}
		\label{tab:photevaporation}
	\end{table}
	
	First, we considered the cases with the planetary parameters adopted by \cite{2022AJ....163...68Z} in their 3D hydrodynamic modelling of the hydrogen escape and of the possible Ly$\alpha$ and He\,I 10833 $\AA$ absorption features. In particular, they assumed $m_{\rm b} = 5.5$\, $\mearth$ and $R_{\rm b} = 2.15$\, $\rearth$ for planet b, and $m_{\rm c} = 7.3$\, $\mearth$ and $R_{\rm c} = 2.67$\, $\rearth$ for planet c. Using these values, we predict a current atmospheric mass fraction of $\sim 0.5$\% for planet b and $\sim 1.4$\% for planet c, which are in good agreement with the corresponding values of 0.6\% and 2\% derived by \cite{2022AJ....163...68Z}. We also checked that our X-ray to EUV luminosity scaling provides consistent fluxes at 1\,au distance, at the current age: we derived a value of 95\,erg s$^{-1}$ cm$^{-2}$, to be compared with the estimate of $91 \pm 27$\,erg s$^{-1}$ cm$^{-2}$ by \cite{2022AJ....163...68Z}, who employed a nominal stellar spectrum from 5 $\AA$ to 5 $\mu$m. Finally, we compared the predictions of the current mass-loss rates, which resulted in a factor of 1.7 higher for planet b and 2.1 lower for planet c, with respect to the values calculated by \cite{2022AJ....163...68Z}. This relatively small difference is not surprising, given the quite different modelling approaches. However, we stress that these planetary properties are just instantaneous snapshots, which do not take into account the past evolutionary history.
 
 In the following, we report the results of our backward and forward in time simulations, which include the time evolution of the XUV irradiation and of the planetary structure in response to the stellar behaviour. We are interested in investigating how the planetary masses, radii, and atmospheric mass-loss rates change with time due to photoevaporation. The results of the simulations are shown in Figures \ref{fig:evaporation_pl_b}-\ref{fig:evaporation_pl_c} and in Table \ref{tab:photevaporation}.
	
\textit{Planet b}. First, we computed atmospheric mass fractions and mass-loss rates at the present age for the grid of planet masses, and we simulated the evolution in the future. In Table \ref{tab:photevaporation} we report the e-folding mass-loss timescales. These evaporation timescales are very short (1--100\,Myr) in any case.
Hence, we confirm the conclusion of \cite{2022AJ....163...68Z}, based on observations, that the inner planet has probably lost its atmosphere entirely. Then we explored the possible backward evolutionary pathways, in order to assess the initial planetary masses and radii at the age of 10\,Myr (see Appendix \ref{app:photoevap} for details on the modelling). In practice, we found that it is always possible to find initial conditions such that the atmosphere is depleted within about 400\,Myr. 
For each assumed planetary mass at the current age, we were able to determine what could be the highest initial mass at 10\,Myr. Based on our results for the mass-loss timescales discussed above, we adopt the boundary condition that the atmospheric evaporation is completed at the current age of 414\,Myr. These evolutionary histories are shown in Fig. \ref{fig:evaporation_pl_b}. For example, if planet b has $m_{\rm b} = 12$ $\mearth$ at the present age, it could have started with a mass $m_{\rm b,\,t=10 Myr} \sim 12.02$ $\mearth$ and an atmospheric fraction $f_{\rm atm,t=10 Myr} = 0.14$\% at 10\,Myr. 
  
For planetary masses $\la 7$\,M$_{\oplus}$ we found a remarkable change in evaporation efficiency. This is due to the relative change of the Jeans escape parameter, $\Lambda$, with respect to the control parameter $e^{\Sigma}$ in the hydro-based formulation by \cite{kuby+2018b} (see Appendix \ref{app:photoevap} for details). While for relatively high planetary masses the atmospheric escape is mainly driven by the photoevaporation, for lower planetary masses the mass-loss rate is increasingly larger because the atmospheric escape is more due to the thermal energy and the lower gravity of the planet. Hence, in order to reach the current age with an atmospheric mass fraction $f_{\rm atm,\,t=414Myr} = 0 $, the initial configuration of the planet is characterised by a massive and inflated atmospheric envelope. We found that if the current mass of planet b is 7 $\mearth$, $m_{\rm b,\,t=10 Myr}$ could have been up to about 40 $\mearth$, $f_{\rm atm,t=10 Myr} = 82$\%, and the corresponding initial radius was $R_{\rm b,\,t=10 Myr} \sim 14.25$ $\rearth$. We remark that these are maximum mass values, such that complete evaporation occurs in about 400\,Myr. Lower initial masses are also possible, with shorter evaporation timescales and identical atmosphere-depleted status at the present age.

We would like to remind readers that for the backward in time modelling we are considering as a boundary condition that the current mass-loss rate of planet b is zero. That explains the difference with the non-null mass-loss rates indicated in Table \ref{tab:photevaporation}. In particular, for the cases of a planet with 2 and 4.5\,$\mearth$, the boundary condition implies the runaway values reached in the late stages before the present age.
 
A further caveat about the simulations for planets with masses $\la 7$\,M$_{\oplus}$ is that a significant decrease in the planetary mass at early ages implies an outward migration of the planet, with a rate depending on the amount of angular momentum driven away from the system by the planetary wind flow \citep{Fujita_2022}. An increase in the star-planet distance yields a decrease in XUV irradiation and equilibrium temperature, hence a decrease in mass-loss rate. Exploring this highly non-linear regime is beyond the scope of the present paper.
	
\textit{Planet c}. In this case, the evolution is more simple, with no predicted change of regime (Fig. \ref{fig:evaporation_pl_c}), due to the larger distance of the planet from the host star, a lower equilibrium temperature, and lower high-energy irradiation.
In most of the cases that we have examined, i.e. for masses in the range 7.5--15 $\mearth$ at the current age, the planet mass and radius are scarcely affected by atmospheric evaporation, although the atmospheric mass fraction can change significantly during the system lifetime. The e-folding mass-loss timescales are always $>5$\,Gyr, even for the case of a planet with $M_{\rm p}=7.3 \mearth$ considered by \cite{2022AJ....163...68Z}. The mass-loss timescale of 0.9\,Gyr predicted by these authors was significantly lower since it was computed assuming a constant XUV irradiation and mass-loss rate. Instead, in our evolutionary models, the XUV luminosity drops to about $1/3$ the current value already at 1.3\,Gyr, and it keeps decreasing at later times.
Only if a current mass of 5 $\mearth$ is considered, the mass and radius change substantially starting from an age of 10 Myr.

\section{Conclusions} \label{sec:conclusions}
In this paper, we analysed new photometric and spectroscopic data of the 400 Myr-old star HD\,63433 (TOI-1726), which hosts a multi-planet system. The presence of two planets in this relatively young system offers the opportunity to examine differences in planetary evolution pathways over the first hundreds of millions years after the system formation, under the influence of the same host star. We provided a characterisation of the system, with special focus on the measurement of the planet radii and masses, and on the investigation of the past and future planetary atmosphere evolution.
	
Thanks to new TESS photometry and follow-up with HARPS-N, we could improve the planetary ephemeris with respect to the study of \cite{mannetal2020}, with statistical evidence in favour of circular orbits for the two planets. We revised the measure and improve the precision of the planetary radii ($r_b=2.02^{+0.06}_{-0.05} \rearth$ and $r_c=2.44\pm0.07 \rearth$), and provide the first dynamic constraints on the planetary masses. After testing several approaches to mitigate the dominant stellar activity contribution in the RV time series, we can only provide upper limits for the mass $m_b$ of the innermost planet b, conservatively concluding that $m_b\lesssim$11 $\mearth$ at $95\%$ of confidence. As for the larger planet c, not all the models converge to similar mass estimates, but results from a few test models suggest a Neptune-like mass with a significance of 2.1--2.7$\sigma$, and that HD\,63433\,c could be a dense mini-Neptune, with a few known counterparts in the mass-radius diagram. The more conservative conclusion is that $m_c\lesssim$31 $\mearth$ at $95\%$ of confidence.
	
Based on the constraints to the masses derived from our analysis, we explored the evolution of the atmospheres of both planets, taking into account the decay of stellar activity and XUV irradiation with time. We estimate that the current mass-loss timescale of planet b is sufficiently short to justify the assumption that its atmosphere is already evaporated. This finding is supported by the results of \cite{2022AJ....163...68Z}, who reported no evidence of Ly$\alpha$ or He I absorption due to photoevaporation in dedicated photometric observations with HST.

 However, the past history of the planet is quite uncertain, because the evolution of the mass-loss rate is strongly dependent on the assumed mass and current atmospheric mass fraction. For masses $\ga 7 \mearth$, the atmospheric mass fraction was already small at early ages ($\sim$ 10\,Myr), hence the planetary mass and radius changed little in time. Instead, a planet with a lower mass at the present age could have started as an inflated body with a few tens of Earth masses, with a structure dominated by a heavy atmosphere. In any case, the current planetary structure is essentially depleted of a gaseous envelope.

 For HD\,63433\,c, our models indicate that the atmospheric evaporation will keep going over the next $\sim$4.5 Gyr, because the decay of the XUV irradiation determines a steady decrease of the mass-loss rate with time. Our prediction for the mass-loss timescale is different from the shorter value determined by \cite{2022AJ....163...68Z}, who did not take into account the evolution of the stellar activity. Another effect included in our modelling is the time evolution of the stellar bolometric luminosity and planetary equilibrium temperature. Neglecting this effect, the atmosphere of a $13 \mearth$ planet would stop evaporating within 0.6 Gyr, while we found that such a planet keeps losing its atmosphere with a timescale longer than 5 Gyr. Summarising, if the current mass of HD\,63433\,c is $>$ 15 ~$\mearth$, the atmosphere should be stable against evaporation; for masses in the range 7--13\,$\mearth$ the planet mass and radius do not change appreciably in time, while for the case with $m_c=5$ $\mearth$ we predict a decrease of 0.8$\%$ in mass and 25$\%$ in radius over the next $\sim$4.5 Gyr.

We acknowledge that HD\,63433 has also been followed-up with the CARMENES spectrograph \citep{mallorquin23}. The GAPS and CARMENES teams have coordinated the submission of two studies, which have been intentionally carried out in an independent way.

	
	\begin{acknowledgements}
    	This work has been supported by the PRIN-INAF 2019 "Planetary systems at young ages (PLATEA)" and ASI-INAF agreement n.2018-16-HH.0.  
		AM and DL acknowledge partial financial support from the ASI-INAF agreement n.2018-16-HH.0 (THE StellaR PAth project), and from the ARIEL ASI-INAF agreement n.2021-5-HH.0. This work has made use of data from the European Space Agency (ESA) mission
		{\it Gaia} (\url{https://www.cosmos.esa.int/gaia}), processed by the {\it Gaia}
		Data Processing and Analysis Consortium (DPAC,
		\url{https://www.cosmos.esa.int/web/gaia/dpac/consortium}). Funding for the DPAC
		has been provided by national institutions, in particular the institutions
		participating in the {\it Gaia} Multilateral Agreement.
	\end{acknowledgements}
	
	%
	
	\bibliographystyle{aa} 
	\bibliography{TOI1726.bib} 

\begin{thebibliography}{89}
\expandafter\ifx\csname natexlab\endcsname\relax\def\natexlab#1{#1}\fi

\bibitem[{{Anglada-Escud{\'e}} \& {Butler}(2012)}]{2012ApJS..200...15A}
{Anglada-Escud{\'e}}, G. \& {Butler}, R.~P. 2012, \apjs, 200, 15

\bibitem[{{Ballerini} {et~al.}(2012){Ballerini}, {Micela}, {Lanza}, \&
  {Pagano}}]{ballerini2012}
{Ballerini}, P., {Micela}, G., {Lanza}, A.~F., \& {Pagano}, I. 2012, \aap, 539,
  A140

\bibitem[{{Baratella} {et~al.}(2020{\natexlab{a}}){Baratella}, {D'Orazi},
  {Biazzo}, {Desidera}, {Gratton}, {Benatti}, {Bignamini}, {Carleo}, {Cecconi},
  {Claudi}, {Cosentino}, {Ghedina}, {Harutyunyan}, {Lanza}, {Malavolta},
  {Maldonado}, {Mallonn}, {Messina}, {Micela}, {Molinari}, {Poretti},
  {Scandariato}, \& {Sozzetti}}]{baratellaetal2020a}
{Baratella}, M., {D'Orazi}, V., {Biazzo}, K., {et~al.} 2020{\natexlab{a}},
  \aap, 640, A123

\bibitem[{{Baratella} {et~al.}(2020{\natexlab{b}}){Baratella}, {D'Orazi},
  {Carraro}, {Desidera}, {Randich}, {Magrini}, {Adibekyan}, {Smiljanic},
  {Spina}, {Tsantaki}, {Tautvai{\v{s}}ien{\.{e}}}, {Sousa}, {Jofr{\'e}},
  {Jim{\'e}nez-Esteban}, {Delgado-Mena}, {Martell}, {Van der Swaelmen},
  {Roccatagliata}, {Gilmore}, {Alfaro}, {Bayo}, {Bensby}, {Bragaglia},
  {Franciosini}, {Gonneau}, {Heiter}, {Hourihane}, {Jeffries}, {Koposov},
  {Morbidelli}, {Prisinzano}, {Sacco}, {Sbordone}, {Worley}, {Zaggia}, \&
  {Lewis}}]{baratellaetal2020b}
{Baratella}, M., {D'Orazi}, V., {Carraro}, G., {et~al.} 2020{\natexlab{b}},
  \aap, 634, A34

\bibitem[{{Barrag{\'a}n} {et~al.}(2022{\natexlab{a}}){Barrag{\'a}n}, {Aigrain},
  {Rajpaul}, \& {Zicher}}]{pyaneti2}
{Barrag{\'a}n}, O., {Aigrain}, S., {Rajpaul}, V.~M., \& {Zicher}, N.
  2022{\natexlab{a}}, \mnras, 509, 866

\bibitem[{{Barrag{\'a}n} {et~al.}(2022{\natexlab{b}}){Barrag{\'a}n},
  {Armstrong}, {Gandolfi}, {Carleo}, {Vidotto}, {Villarreal D'Angelo},
  {Oklop{\v{c}}i{\'c}}, {Isaacson}, {Oddo}, {Collins}, {Fridlund}, {Sousa},
  {Persson}, {Hellier}, {Howell}, {Howard}, {Redfield}, {Eisner}, {Georgieva},
  {Dragomir}, {Bayliss}, {Nielsen}, {Klein}, {Aigrain}, {Zhang}, {Teske},
  {Twicken}, {Jenkins}, {Esposito}, {Van Eylen}, {Rodler}, {Adibekyan},
  {Alarcon}, {Anderson}, {Akana Murphy}, {Barrado}, {Barros}, {Benneke},
  {Bouchy}, {Bryant}, {Butler}, {Burt}, {Cabrera}, {Casewell}, {Chaturvedi},
  {Cloutier}, {Cochran}, {Crane}, {Crossfield}, {Crouzet}, {Collins}, {Dai},
  {Deeg}, {Deline}, {Demangeon}, {Dumusque}, {Figueira}, {Furlan}, {Gnilka},
  {Goad}, {Goffo}, {Guti{\'e}rrez-Canales}, {Hadjigeorghiou}, {Hartman},
  {Hatzes}, {Harris}, {Henderson}, {Hirano}, {Hojjatpanah}, {Hoyer},
  {Kab{\'a}th}, {Korth}, {Lillo-Box}, {Luque}, {Marmier}, {Mo{\v{c}}nik},
  {Muresan}, {Murgas}, {Nagel}, {Osborne}, {Osborn}, {Osborn}, {Palle},
  {Raimbault}, {Ricker}, {Rubenzahl}, {Stockdale}, {Santos}, {Scott},
  {Schwarz}, {Shectman}, {Raimbault}, {Seager}, {S{\'e}gransan}, {Serrano},
  {Skarka}, {Smith}, {{\v{S}}ubjak}, {Tan}, {Udry}, {Watson}, {Wheatley},
  {West}, {Winn}, {Wang}, {Wolfgang}, \& {Ziegler}}]{barragan2022}
{Barrag{\'a}n}, O., {Armstrong}, D.~J., {Gandolfi}, D., {et~al.}
  2022{\natexlab{b}}, \mnras, 514, 1606

\bibitem[{Barrag\'an {et~al.}(2019)Barrag\'an, Gandolfi, \&
  Antoniciello}]{pyaneti}
Barrag\'an, O., Gandolfi, D., \& Antoniciello, G. 2019, \mnras, 482, 1017

\bibitem[{Barragán {et~al.}(2019)Barragán, Aigrain, Kubyshkina, Gandolfi,
  Livingston, Fridlund, Fossati, Korth, Parviainen, Malavolta, Palle, Deeg,
  Nowak, Rajpaul, Zicher, Antoniciello, Narita, Albrecht, Bedin, Cabrera,
  Cochran, de Leon, Eigmüller, Fukui, Granata, Grziwa, Guenther, Hatzes,
  Kusakabe, Latham, Libralato, Luque, Montañés-Rodríguez, Murgas, Nardiello,
  Pagano, Piotto, Persson, Redfield, \& Tamura}]{barragan2019}
Barragán, O., Aigrain, S., Kubyshkina, D., {et~al.} 2019, Monthly Notices of
  the Royal Astronomical Society, 490, 698

\bibitem[{{Benatti} {et~al.}(2021){Benatti}, {Damasso}, {Borsa}, {Locci},
  {Pillitteri}, {Desidera}, {Maggio}, {Micela}, {Wolk}, {Claudi}, {Malavolta},
  \& {Modirrousta-Galian}}]{benatti2021}
{Benatti}, S., {Damasso}, M., {Borsa}, F., {et~al.} 2021, \aap, 650, A66

\bibitem[{{Brandt}(2021)}]{brandt2021ApJS..254...42B}
{Brandt}, T.~D. 2021, \apjs, 254, 42

\bibitem[{{Brewer} {et~al.}(2016){Brewer}, {Fischer}, {Valenti}, \&
  {Piskunov}}]{Breweretal2016}
{Brewer}, J.~M., {Fischer}, D.~A., {Valenti}, J.~A., \& {Piskunov}, N. 2016,
  \apjs, 225, 32

\bibitem[{{Buchner} {et~al.}(2014){Buchner}, {Georgakakis}, {Nandra}, {Hsu},
  {Rangel}, {Brightman}, {Merloni}, {Salvato}, {Donley}, \&
  {Kocevski}}]{Buchner2014}
{Buchner}, J., {Georgakakis}, A., {Nandra}, K., {et~al.} 2014, \aap, 564, A125

\bibitem[{{Carleo} {et~al.}(2018){Carleo}, {Benatti}, {Lanza}, {Gratton},
  {Claudi}, {Desidera}, {Mace}, {Messina}, {Sanna}, {Sissa}, {Ghedina},
  {Ghinassi}, {Guerra}, {Harutyunyan}, {Micela}, {Molinari}, {Oliva}, {Tozzi},
  {Baffa}, {Baruffolo}, {Bignamini}, {Buchschacher}, {Cecconi}, {Cosentino},
  {Endl}, {Falcini}, {Fantinel}, {Fini}, {Fugazza}, {Galli}, {Giani},
  {Gonz{\'a}lez}, {Gonz{\'a}lez-{\'A}lvarez}, {Gonz{\'a}lez}, {Hernandez},
  {Hernandez Diaz}, {Iuzzolino}, {Kaplan}, {Kidder}, {Lodi}, {Malavolta},
  {Maldonado}, {Origlia}, {Perez Ventura}, {Puglisi}, {Rainer}, {Riverol},
  {Riverol}, {San Juan}, {Scuderi}, {Seemann}, {Sokal}, {Sozzetti}, \&
  {Sozzi}}]{carleo2018}
{Carleo}, I., {Benatti}, S., {Lanza}, A.~F., {et~al.} 2018, \aap, 613, A50

\bibitem[{{Carleo} {et~al.}(2021){Carleo}, {Desidera}, {Nardiello},
  {Malavolta}, {Lanza}, {Livingston}, {Locci}, {Marzari}, {Messina}, {Turrini},
  {Baratella}, {Borsa}, {D'Orazi}, {Nascimbeni}, {Pinamonti}, {Rainer}, {Alei},
  {Bignamini}, {Gratton}, {Micela}, {Montalto}, {Sozzetti}, {Squicciarini},
  {Affer}, {Benatti}, {Biazzo}, {Bonomo}, {Claudi}, {Cosentino}, {Covino},
  {Damasso}, {Esposito}, {Fiorenzano}, {Frustagli}, {Giacobbe}, {Harutyunyan},
  {Leto}, {Magazz{\`u}}, {Maggio}, {Mainella}, {Maldonado}, {Mallonn},
  {Mancini}, {Molinari}, {Molinaro}, {Pagano}, {Pedani}, {Piotto}, {Poretti},
  {Redfield}, \& {Scandariato}}]{carleo2021}
{Carleo}, I., {Desidera}, S., {Nardiello}, D., {et~al.} 2021, \aap, 645, A71

\bibitem[{{Carleo} {et~al.}(2020){Carleo}, {Malavolta}, {Lanza}, {Damasso},
  {Desidera}, {Borsa}, {Mallonn}, {Pinamonti}, {Gratton}, {Alei}, {Benatti},
  {Mancini}, {Maldonado}, {Biazzo}, {Esposito}, {Frustagli},
  {Gonz{\'a}lez-{\'A}lvarez}, {Micela}, {Scandariato}, {Sozzetti}, {Affer},
  {Bignamini}, {Bonomo}, {Claudi}, {Cosentino}, {Covino}, {Fiorenzano},
  {Giacobbe}, {Harutyunyan}, {Leto}, {Maggio}, {Molinari}, {Nascimbeni},
  {Pagano}, {Pedani}, {Piotto}, {Poretti}, {Rainer}, {Redfield}, {Baffa},
  {Baruffolo}, {Buchschacher}, {Billotti}, {Cecconi}, {Falcini}, {Fantinel},
  {Fini}, {Galli}, {Ghedina}, {Ghinassi}, {Giani}, {Gonzalez}, {Gonzalez},
  {Guerra}, {Hernandez Diaz}, {Hernandez}, {Iuzzolino}, {Lodi}, {Oliva},
  {Origlia}, {Perez Ventura}, {Puglisi}, {Riverol}, {Riverol}, {San Juan},
  {Sanna}, {Scuderi}, {Seemann}, {Sozzi}, \& {Tozzi}}]{carleo2020}
{Carleo}, I., {Malavolta}, L., {Lanza}, A.~F., {et~al.} 2020, \aap, 638, A5

\bibitem[{{Casagrande} {et~al.}(2010){Casagrande}, {Ram{\'\i}rez},
  {Mel{\'e}ndez}, {Bessell}, \& {Asplund}}]{Casagrandeetal2010}
{Casagrande}, L., {Ram{\'\i}rez}, I., {Mel{\'e}ndez}, J., {Bessell}, M., \&
  {Asplund}, M. 2010, \aap, 512, A54

\bibitem[{{Castelli} \& {Kurucz}(2003)}]{castellikurucz2003}
{Castelli}, F. \& {Kurucz}, R.~L. 2003, in Modelling of Stellar Atmospheres,
  ed. N.~{Piskunov}, W.~W. {Weiss}, \& D.~F. {Gray}, Vol. 210, A20

\bibitem[{{Choi} {et~al.}(2016){Choi}, {Dotter}, {Conroy}, {Cantiello},
  {Paxton}, \& {Johnson}}]{choi+2016}
{Choi}, J., {Dotter}, A., {Conroy}, C., {et~al.} 2016, \apj, 823, 102

\bibitem[{{Cosentino} {et~al.}(2012){Cosentino}, {Lovis}, {Pepe}, {Collier
  Cameron}, {Latham}, {Molinari}, {Udry}, {Bezawada}, {Black}, {Born},
  {Buchschacher}, {Charbonneau}, {Figueira}, {Fleury}, {Galli}, {Gallie},
  {Gao}, {Ghedina}, {Gonzalez}, {Gonzalez}, {Guerra}, {Henry}, {Horne},
  {Hughes}, {Kelly}, {Lodi}, {Lunney}, {Maire}, {Mayor}, {Micela}, {Ordway},
  {Peacock}, {Phillips}, {Piotto}, {Pollacco}, {Queloz}, {Rice}, {Riverol},
  {Riverol}, {San Juan}, {Sasselov}, {Segransan}, {Sozzetti}, {Sosnowska},
  {Stobie}, {Szentgyorgyi}, {Vick}, \& {Weber}}]{Cosentino2012}
{Cosentino}, R., {Lovis}, C., {Pepe}, F., {et~al.} 2012, Society of
  Photo-Optical Instrumentation Engineers (SPIE) Conference Series, Vol. 8446,
  {Harps-N: the new planet hunter at TNG}, 84461V

\bibitem[{{Covino} {et~al.}(2013){Covino}, {Esposito}, {Barbieri}, {Mancini},
  {Nascimbeni}, {Claudi}, {Desidera}, {Gratton}, {Lanza}, {Sozzetti}, {Biazzo},
  {Affer}, {Gandolfi}, {Munari}, {Pagano}, {Bonomo}, {Collier Cameron},
  {H{\'e}brard}, {Maggio}, {Messina}, {Micela}, {Molinari}, {Pepe}, {Piotto},
  {Ribas}, {Santos}, {Southworth}, {Shkolnik}, {Triaud}, {Bedin}, {Benatti},
  {Boccato}, {Bonavita}, {Borsa}, {Borsato}, {Brown}, {Carolo}, {Ciceri},
  {Cosentino}, {Damasso}, {Faedi}, {Mart{\'\i}nez Fiorenzano}, {Latham},
  {Lovis}, {Mordasini}, {Nikolov}, {Poretti}, {Rainer}, {Rebolo L{\'o}pez},
  {Scandariato}, {Silvotti}, {Smareglia}, {Alcal{\'a}}, {Cunial}, {Di
  Fabrizio}, {Di Mauro}, {Giacobbe}, {Granata}, {Harutyunyan}, {Knapic},
  {Lattanzi}, {Leto}, {Lodato}, {Malavolta}, {Marzari}, {Molinaro},
  {Nardiello}, {Pedani}, {Prisinzano}, \& {Turrini}}]{Covino2013}
{Covino}, E., {Esposito}, M., {Barbieri}, M., {et~al.} 2013, \aap, 554, A28

\bibitem[{{Cutri} {et~al.}(2003){Cutri}, {Skrutskie}, {van Dyk}, {Beichman},
  {Carpenter}, {Chester}, {Cambresy}, {Evans}, {Fowler}, {Gizis}, {Howard},
  {Huchra}, {Jarrett}, {Kopan}, {Kirkpatrick}, {Light}, {Marsh}, {McCallon},
  {Schneider}, {Stiening}, {Sykes}, {Weinberg}, {Wheaton}, {Wheelock}, \&
  {Zacarias}}]{cutri2003}
{Cutri}, R.~M., {Skrutskie}, M.~F., {van Dyk}, S., {et~al.} 2003, VizieR Online
  Data Catalog, II/246

\bibitem[{{Cutri} {et~al.}(2021){Cutri}, {Wright}, {Conrow}, {Fowler},
  {Eisenhardt}, {Grillmair}, {Kirkpatrick}, {Masci}, {McCallon}, {Wheelock},
  {Fajardo-Acosta}, {Yan}, {Benford}, {Harbut}, {Jarrett}, {Lake}, {Leisawitz},
  {Ressler}, {Stanford}, {Tsai}, {Liu}, {Helou}, {Mainzer}, {Gettngs},
  {Gonzalez}, {Hoffman}, {Marsh}, {Padgett}, {Skrutskie}, {Beck}, {Papin}, \&
  {Wittman}}]{cutri2013}
{Cutri}, R.~M., {Wright}, E.~L., {Conrow}, T., {et~al.} 2021, VizieR Online
  Data Catalog, II/328

\bibitem[{{Dai} {et~al.}(2020){Dai}, {Roy}, {Fulton}, {Robertson}, {Hirsch},
  {Isaacson}, {Albrecht}, {Mann}, {Kristiansen}, {Batalha}, {Beard}, {Behmard},
  {Chontos}, {Crossfield}, {Dalba}, {Dressing}, {Giacalone}, {Hill}, {Howard},
  {Huber}, {Kane}, {Kosiarek}, {Lubin}, {Mayo}, {Mocnik}, {Akana Murphy},
  {Petigura}, {Rosenthal}, {Rubenzahl}, {Scarsdale}, {Weiss}, {Van Zandt},
  {Ricker}, {Vanderspek}, {Latham}, {Seager}, {Winn}, {Jenkins}, {Caldwell},
  {Charbonneau}, {Daylan}, {G{\"u}nther}, {Morgan}, {Quinn}, {Rose}, \&
  {Smith}}]{2020AJ....160..193D}
{Dai}, F., {Roy}, A., {Fulton}, B., {et~al.} 2020, \aj, 160, 193

\bibitem[{{Damasso} {et~al.}(2020){Damasso}, {Lanza}, {Benatti}, {Rajpaul},
  {Mallonn}, {Desidera}, {Biazzo}, {D'Orazi}, {Malavolta}, {Nardiello},
  {Rainer}, {Borsa}, {Affer}, {Bignamini}, {Bonomo}, {Carleo}, {Claudi},
  {Cosentino}, {Covino}, {Giacobbe}, {Gratton}, {Harutyunyan}, {Knapic},
  {Leto}, {Maggio}, {Maldonado}, {Mancini}, {Micela}, {Molinari}, {Nascimbeni},
  {Pagano}, {Piotto}, {Poretti}, {Scandariato}, {Sozzetti}, {Capuzzo Dolcetta},
  {Di Mauro}, {Carosati}, {Fiorenzano}, {Frustagli}, {Pedani}, {Pinamonti},
  {Stoev}, \& {Turrini}}]{damasso_v830tau}
{Damasso}, M., {Lanza}, A.~F., {Benatti}, S., {et~al.} 2020, \aap, 642, A133

\bibitem[{{Desidera} {et~al.}(2022){Desidera}, {Damasso}, {Gratton}, {Benatti},
  {Nardiello}, {D'Orazi}, {Lanza}, {Locci}, {Marzari}, {Mesa}, {Messina},
  {Pillitteri}, {Sozzetti}, {Girard}, {Maggio}, {Micela}, {Malavolta},
  {Nascimbeni}, {Pinamonti}, {Squicciarini}, {Alcala}, {Biazzo}, {Bohn},
  {Bonavita}, {Brooks}, {Chauvin}, {Covino}, {Delorme}, {Hagelberg}, {Janson},
  {Lagrange}, \& {Lazzoni}}]{2022arXiv221007933D}
{Desidera}, S., {Damasso}, M., {Gratton}, R., {et~al.} 2022, arXiv e-prints,
  arXiv:2210.07933

\bibitem[{{Dotter}(2016)}]{Dotter2016}
{Dotter}, A. 2016, \apjs, 222, 8

\bibitem[{{Dutra-Ferreira} {et~al.}(2016){Dutra-Ferreira}, {Pasquini},
  {Smiljanic}, {Porto de Mello}, \& {Steffen}}]{DutraFerreiraetal2016}
{Dutra-Ferreira}, L., {Pasquini}, L., {Smiljanic}, R., {Porto de Mello}, G.~F.,
  \& {Steffen}, M. 2016, \aap, 585, A75

\bibitem[{{Eastman} {et~al.}(2019){Eastman}, {Rodriguez}, {Agol}, {Stassun},
  {Beatty}, {Vanderburg}, {Gaudi}, {Collins}, \& {Luger}}]{Eastmanetal2019}
{Eastman}, J.~D., {Rodriguez}, J.~E., {Agol}, E., {et~al.} 2019, arXiv
  e-prints, arXiv:1907.09480

\bibitem[{{Feroz} {et~al.}(2011){Feroz}, {Balan}, \& {Hobson}}]{feroz2011}
{Feroz}, F., {Balan}, S.~T., \& {Hobson}, M.~P. 2011, \mnras, 415, 3462

\bibitem[{{Feroz} {et~al.}(2019){Feroz}, {Hobson}, {Cameron}, \&
  {Pettitt}}]{Feroz2019}
{Feroz}, F., {Hobson}, M.~P., {Cameron}, E., \& {Pettitt}, A.~N. 2019, The Open
  Journal of Astrophysics, 2, 10

\bibitem[{{Foreman-Mackey}(2018)}]{2018RNAAS...2...31F}
{Foreman-Mackey}, D. 2018, Research Notes of the American Astronomical Society,
  2, 31

\bibitem[{{Foreman-Mackey} {et~al.}(2017){Foreman-Mackey}, {Agol},
  {Ambikasaran}, \& {Angus}}]{2017AJ....154..220F}
{Foreman-Mackey}, D., {Agol}, E., {Ambikasaran}, S., \& {Angus}, R. 2017, \aj,
  154, 220

\bibitem[{{Fossati} {et~al.}(2017){Fossati}, {Erkaev}, {Lammer}, {Cubillos},
  {Odert}, {Juvan}, {Kislyakova}, {Lendl}, {Kubyshkina}, \&
  {Bauer}}]{Fossati+2017}
{Fossati}, L., {Erkaev}, N.~V., {Lammer}, H., {et~al.} 2017, \aap, 598, A90

\bibitem[{Fujita {et~al.}(2022)Fujita, Hori, \& Sasaki}]{Fujita_2022}
Fujita, N., Hori, Y., \& Sasaki, T. 2022, The Astrophysical Journal, 928, 105

\bibitem[{{Gaia Collaboration}(2022)}]{Gaia2022}
{Gaia Collaboration}. 2022, VizieR Online Data Catalog, I/355

\bibitem[{{Gaia Collaboration} {et~al.}(2021){Gaia Collaboration}, {Brown},
  {Vallenari}, {Prusti}, {de Bruijne}, {Babusiaux}, {Biermann}, {Creevey},
  {Evans}, {Eyer}, {Hutton}, {Jansen}, {Jordi}, {Klioner}, {Lammers},
  {Lindegren}, {Luri}, {Mignard}, {Panem}, {Pourbaix}, {Randich}, {Sartoretti},
  {Soubiran}, {Walton}, {Arenou}, {Bailer-Jones}, {Bastian}, {Cropper},
  {Drimmel}, {Katz}, {Lattanzi}, {van Leeuwen}, {Bakker}, {Cacciari},
  {Casta{\~n}eda}, {De Angeli}, {Ducourant}, {Fabricius}, {Fouesneau},
  {Fr{\'e}mat}, {Guerra}, {Guerrier}, {Guiraud}, {Jean-Antoine Piccolo},
  {Masana}, {Messineo}, {Mowlavi}, {Nicolas}, {Nienartowicz}, {Pailler},
  {Panuzzo}, {Riclet}, {Roux}, {Seabroke}, {Sordo}, {Tanga}, {Th{\'e}venin},
  {Gracia-Abril}, {Portell}, {Teyssier}, {Altmann}, {Andrae}, {Bellas-Velidis},
  {Benson}, {Berthier}, {Blomme}, {Brugaletta}, {Burgess}, {Busso}, {Carry},
  {Cellino}, {Cheek}, {Clementini}, {Damerdji}, {Davidson}, {Delchambre},
  {Dell'Oro}, {Fern{\'a}ndez-Hern{\'a}ndez}, {Galluccio}, {Garc{\'\i}a-Lario},
  {Garcia-Reinaldos}, {Gonz{\'a}lez-N{\'u}{\~n}ez}, {Gosset}, {Haigron},
  {Halbwachs}, {Hambly}, {Harrison}, {Hatzidimitriou}, {Heiter},
  {Hern{\'a}ndez}, {Hestroffer}, {Hodgkin}, {Holl}, {Jan{\ss}en}, {Jevardat de
  Fombelle}, {Jordan}, {Krone-Martins}, {Lanzafame}, {L{\"o}ffler}, {Lorca},
  {Manteiga}, {Marchal}, {Marrese}, {Moitinho}, {Mora}, {Muinonen}, {Osborne},
  {Pancino}, {Pauwels}, {Petit}, {Recio-Blanco}, {Richards}, {Riello},
  {Rimoldini}, {Robin}, {Roegiers}, {Rybizki}, {Sarro}, {Siopis}, {Smith},
  {Sozzetti}, {Ulla}, {Utrilla}, {van Leeuwen}, {van Reeven}, {Abbas}, {Abreu
  Aramburu}, {Accart}, {Aerts}, {Aguado}, {Ajaj}, {Altavilla}, {{\'A}lvarez},
  {{\'A}lvarez Cid-Fuentes}, {Alves}, {Anderson}, {Anglada Varela}, {Antoja},
  {Audard}, {Baines}, {Baker}, {Balaguer-N{\'u}{\~n}ez}, {Balbinot}, {Balog},
  {Barache}, {Barbato}, {Barros}, {Barstow}, {Bartolom{\'e}}, {Bassilana},
  {Bauchet}, {Baudesson-Stella}, {Becciani}, {Bellazzini}, {Bernet}, {Bertone},
  {Bianchi}, {Blanco-Cuaresma}, {Boch}, {Bombrun}, {Bossini}, {Bouquillon},
  {Bragaglia}, {Bramante}, {Breedt}, {Bressan}, {Brouillet}, {Bucciarelli},
  {Burlacu}, {Busonero}, {Butkevich}, {Buzzi}, {Caffau}, {Cancelliere},
  {C{\'a}novas}, {Cantat-Gaudin}, {Carballo}, {Carlucci}, {Carnerero},
  {Carrasco}, {Casamiquela}, {Castellani}, {Castro-Ginard}, {Castro Sampol},
  {Chaoul}, {Charlot}, {Chemin}, {Chiavassa}, {Cioni}, {Comoretto}, {Cooper},
  {Cornez}, {Cowell}, {Crifo}, {Crosta}, {Crowley}, {Dafonte}, {Dapergolas},
  {David}, {David}, {de Laverny}, {De Luise}, {De March}, {De Ridder}, {de
  Souza}, {de Teodoro}, {de Torres}, {del Peloso}, {del Pozo}, {Delbo},
  {Delgado}, {Delgado}, {Delisle}, {Di Matteo}, {Diakite}, {Diener},
  {Distefano}, {Dolding}, {Eappachen}, {Edvardsson}, {Enke}, {Esquej}, {Fabre},
  {Fabrizio}, {Faigler}, {Fedorets}, {Fernique}, {Fienga}, {Figueras},
  {Fouron}, {Fragkoudi}, {Fraile}, {Franke}, {Gai}, {Garabato},
  {Garcia-Gutierrez}, {Garc{\'\i}a-Torres}, {Garofalo}, {Gavras}, {Gerlach},
  {Geyer}, {Giacobbe}, {Gilmore}, {Girona}, {Giuffrida}, {Gomel}, {Gomez},
  {Gonzalez-Santamaria}, {Gonz{\'a}lez-Vidal}, {Granvik},
  {Guti{\'e}rrez-S{\'a}nchez}, {Guy}, {Hauser}, {Haywood}, {Helmi}, {Hidalgo},
  {Hilger}, {H{\l}adczuk}, {Hobbs}, {Holland}, {Huckle}, {Jasniewicz},
  {Jonker}, {Juaristi Campillo}, {Julbe}, {Karbevska}, {Kervella}, {Khanna},
  {Kochoska}, {Kontizas}, {Kordopatis}, {Korn}, {Kostrzewa-Rutkowska},
  {Kruszy{\'n}ska}, {Lambert}, {Lanza}, {Lasne}, {Le Campion}, {Le Fustec},
  {Lebreton}, {Lebzelter}, {Leccia}, {Leclerc}, {Lecoeur-Taibi}, {Liao},
  {Licata}, {Lindstr{\o}m}, {Lister}, {Livanou}, {Lobel}, {Madrero Pardo},
  {Managau}, {Mann}, {Marchant}, {Marconi}, {Marcos Santos}, {Marinoni},
  {Marocco}, {Marshall}, {Martin Polo}, {Mart{\'\i}n-Fleitas}, {Masip},
  {Massari}, {Mastrobuono-Battisti}, {Mazeh}, {McMillan}, {Messina},
  {Michalik}, {Millar}, {Mints}, {Molina}, {Molinaro}, {Moln{\'a}r},
  {Montegriffo}, {Mor}, {Morbidelli}, {Morel}, {Morris}, {Mulone}, {Munoz},
  {Muraveva}, {Murphy}, {Musella}, {Noval}, {Ord{\'e}novic}, {Orr{\`u}},
  {Osinde}, {Pagani}, {Pagano}, {Palaversa}, {Palicio}, {Panahi}, {Pawlak},
  {Pe{\~n}alosa Esteller}, {Penttil{\"a}}, {Piersimoni}, {Pineau}, {Plachy},
  {Plum}, {Poggio}, {Poretti}, {Poujoulet}, {Pr{\v{s}}a}, {Pulone}, {Racero},
  {Ragaini}, {Rainer}, {Raiteri}, {Rambaux}, {Ramos}, {Ramos-Lerate}, {Re
  Fiorentin}, {Regibo}, {Reyl{\'e}}, {Ripepi}, {Riva}, {Rixon}, {Robichon},
  {Robin}, {Roelens}, {Rohrbasser}, {Romero-G{\'o}mez}, {Rowell}, {Royer},
  {Rybicki}, {Sadowski}, {Sagrist{\`a} Sell{\'e}s}, {Sahlmann}, {Salgado},
  {Salguero}, {Samaras}, {Sanchez Gimenez}, {Sanna}, {Santove{\~n}a},
  {Sarasso}, {Schultheis}, {Sciacca}, {Segol}, {Segovia}, {S{\'e}gransan},
  {Semeux}, {Shahaf}, {Siddiqui}, {Siebert}, {Siltala}, {Slezak}, {Smart},
  {Solano}, {Solitro}, {Souami}, {Souchay}, {Spagna}, {Spoto}, {Steele},
  {Steidelm{\"u}ller}, {Stephenson}, {S{\"u}veges}, {Szabados}, {Szegedi-Elek},
  {Taris}, {Tauran}, {Taylor}, {Teixeira}, {Thuillot}, {Tonello}, {Torra},
  {Torra}, {Turon}, {Unger}, {Vaillant}, {van Dillen}, {Vanel}, {Vecchiato},
  {Viala}, {Vicente}, {Voutsinas}, {Weiler}, {Wevers}, {Wyrzykowski}, {Yoldas},
  {Yvard}, {Zhao}, {Zorec}, {Zucker}, {Zurbach}, \& {Zwitter}}]{gaia2021}
{Gaia Collaboration}, {Brown}, A.~G.~A., {Vallenari}, A., {et~al.} 2021, \aap,
  649, A1

\bibitem[{{Gaia Collaboration} {et~al.}(2016){Gaia Collaboration}, {Prusti},
  {de Bruijne}, {Brown}, {Vallenari}, {Babusiaux}, {Bailer-Jones}, {Bastian},
  {Biermann}, {Evans}, {Eyer}, {Jansen}, {Jordi}, {Klioner}, {Lammers},
  {Lindegren}, {Luri}, {Mignard}, {Milligan}, {Panem}, {Poinsignon},
  {Pourbaix}, {Randich}, {Sarri}, {Sartoretti}, {Siddiqui}, {Soubiran},
  {Valette}, {van Leeuwen}, {Walton}, {Aerts}, {Arenou}, {Cropper}, {Drimmel},
  {H{\o}g}, {Katz}, {Lattanzi}, {O'Mullane}, {Grebel}, {Holland}, {Huc},
  {Passot}, {Bramante}, {Cacciari}, {Casta{\~n}eda}, {Chaoul}, {Cheek}, {De
  Angeli}, {Fabricius}, {Guerra}, {Hern{\'a}ndez}, {Jean-Antoine-Piccolo},
  {Masana}, {Messineo}, {Mowlavi}, {Nienartowicz}, {Ord{\'o}{\~n}ez-Blanco},
  {Panuzzo}, {Portell}, {Richards}, {Riello}, {Seabroke}, {Tanga},
  {Th{\'e}venin}, {Torra}, {Els}, {Gracia-Abril}, {Comoretto},
  {Garcia-Reinaldos}, {Lock}, {Mercier}, {Altmann}, {Andrae}, {Astraatmadja},
  {Bellas-Velidis}, {Benson}, {Berthier}, {Blomme}, {Busso}, {Carry},
  {Cellino}, {Clementini}, {Cowell}, {Creevey}, {Cuypers}, {Davidson}, {De
  Ridder}, {de Torres}, {Delchambre}, {Dell'Oro}, {Ducourant}, {Fr{\'e}mat},
  {Garc{\'\i}a-Torres}, {Gosset}, {Halbwachs}, {Hambly}, {Harrison}, {Hauser},
  {Hestroffer}, {Hodgkin}, {Huckle}, {Hutton}, {Jasniewicz}, {Jordan},
  {Kontizas}, {Korn}, {Lanzafame}, {Manteiga}, {Moitinho}, {Muinonen},
  {Osinde}, {Pancino}, {Pauwels}, {Petit}, {Recio-Blanco}, {Robin}, {Sarro},
  {Siopis}, {Smith}, {Smith}, {Sozzetti}, {Thuillot}, {van Reeven}, {Viala},
  {Abbas}, {Abreu Aramburu}, {Accart}, {Aguado}, {Allan}, {Allasia},
  {Altavilla}, {{\'A}lvarez}, {Alves}, {Anderson}, {Andrei}, {Anglada Varela},
  {Antiche}, {Antoja}, {Ant{\'o}n}, {Arcay}, {Atzei}, {Ayache}, {Bach},
  {Baker}, {Balaguer-N{\'u}{\~n}ez}, {Barache}, {Barata}, {Barbier}, {Barblan},
  {Baroni}, {Barrado y Navascu{\'e}s}, {Barros}, {Barstow}, {Becciani},
  {Bellazzini}, {Bellei}, {Bello Garc{\'\i}a}, {Belokurov}, {Bendjoya},
  {Berihuete}, {Bianchi}, {Bienaym{\'e}}, {Billebaud}, {Blagorodnova},
  {Blanco-Cuaresma}, {Boch}, {Bombrun}, {Borrachero}, {Bouquillon}, {Bourda},
  {Bouy}, {Bragaglia}, {Breddels}, {Brouillet}, {Br{\"u}semeister},
  {Bucciarelli}, {Budnik}, {Burgess}, {Burgon}, {Burlacu}, {Busonero}, {Buzzi},
  {Caffau}, {Cambras}, {Campbell}, {Cancelliere}, {Cantat-Gaudin}, {Carlucci},
  {Carrasco}, {Castellani}, {Charlot}, {Charnas}, {Charvet}, {Chassat},
  {Chiavassa}, {Clotet}, {Cocozza}, {Collins}, {Collins}, {Costigan}, {Crifo},
  {Cross}, {Crosta}, {Crowley}, {Dafonte}, {Damerdji}, {Dapergolas}, {David},
  {David}, {De Cat}, {de Felice}, {de Laverny}, {De Luise}, {De March}, {de
  Martino}, {de Souza}, {Debosscher}, {del Pozo}, {Delbo}, {Delgado},
  {Delgado}, {di Marco}, {Di Matteo}, {Diakite}, {Distefano}, {Dolding}, {Dos
  Anjos}, {Drazinos}, {Dur{\'a}n}, {Dzigan}, {Ecale}, {Edvardsson}, {Enke},
  {Erdmann}, {Escolar}, {Espina}, {Evans}, {Eynard Bontemps}, {Fabre},
  {Fabrizio}, {Faigler}, {Falc{\~a}o}, {Farr{\`a}s Casas}, {Faye}, {Federici},
  {Fedorets}, {Fern{\'a}ndez-Hern{\'a}ndez}, {Fernique}, {Fienga}, {Figueras},
  {Filippi}, {Findeisen}, {Fonti}, {Fouesneau}, {Fraile}, {Fraser}, {Fuchs},
  {Furnell}, {Gai}, {Galleti}, {Galluccio}, {Garabato}, {Garc{\'\i}a-Sedano},
  {Gar{\'e}}, {Garofalo}, {Garralda}, {Gavras}, {Gerssen}, {Geyer}, {Gilmore},
  {Girona}, {Giuffrida}, {Gomes}, {Gonz{\'a}lez-Marcos},
  {Gonz{\'a}lez-N{\'u}{\~n}ez}, {Gonz{\'a}lez-Vidal}, {Granvik}, {Guerrier},
  {Guillout}, {Guiraud}, {G{\'u}rpide}, {Guti{\'e}rrez-S{\'a}nchez}, {Guy},
  {Haigron}, {Hatzidimitriou}, {Haywood}, {Heiter}, {Helmi}, {Hobbs},
  {Hofmann}, {Holl}, {Holland}, {Hunt}, {Hypki}, {Icardi}, {Irwin}, {Jevardat
  de Fombelle}, {Jofr{\'e}}, {Jonker}, {Jorissen}, {Julbe}, {Karampelas},
  {Kochoska}, {Kohley}, {Kolenberg}, {Kontizas}, {Koposov}, {Kordopatis},
  {Koubsky}, {Kowalczyk}, {Krone-Martins}, {Kudryashova}, {Kull}, {Bachchan},
  {Lacoste-Seris}, {Lanza}, {Lavigne}, {Le Poncin-Lafitte}, {Lebreton},
  {Lebzelter}, {Leccia}, {Leclerc}, {Lecoeur-Taibi}, {Lemaitre}, {Lenhardt},
  {Leroux}, {Liao}, {Licata}, {Lindstr{\o}m}, {Lister}, {Livanou}, {Lobel},
  {L{\"o}ffler}, {L{\'o}pez}, {Lopez-Lozano}, {Lorenz}, {Loureiro},
  {MacDonald}, {Magalh{\~a}es Fernandes}, {Managau}, {Mann}, {Mantelet},
  {Marchal}, {Marchant}, {Marconi}, {Marie}, {Marinoni}, {Marrese},
  {Marschalk{\'o}}, {Marshall}, {Mart{\'\i}n-Fleitas}, {Martino}, {Mary},
  {Matijevi{\v{c}}}, {Mazeh}, {McMillan}, {Messina}, {Mestre}, {Michalik},
  {Millar}, {Miranda}, {Molina}, {Molinaro}, {Molinaro}, {Moln{\'a}r},
  {Moniez}, {Montegriffo}, {Monteiro}, {Mor}, {Mora}, {Morbidelli}, {Morel},
  {Morgenthaler}, {Morley}, {Morris}, {Mulone}, {Muraveva}, {Musella},
  {Narbonne}, {Nelemans}, {Nicastro}, {Noval}, {Ord{\'e}novic},
  {Ordieres-Mer{\'e}}, {Osborne}, {Pagani}, {Pagano}, {Pailler}, {Palacin},
  {Palaversa}, {Parsons}, {Paulsen}, {Pecoraro}, {Pedrosa}, {Pentik{\"a}inen},
  {Pereira}, {Pichon}, {Piersimoni}, {Pineau}, {Plachy}, {Plum}, {Poujoulet},
  {Pr{\v{s}}a}, {Pulone}, {Ragaini}, {Rago}, {Rambaux}, {Ramos-Lerate},
  {Ranalli}, {Rauw}, {Read}, {Regibo}, {Renk}, {Reyl{\'e}}, {Ribeiro},
  {Rimoldini}, {Ripepi}, {Riva}, {Rixon}, {Roelens}, {Romero-G{\'o}mez},
  {Rowell}, {Royer}, {Rudolph}, {Ruiz-Dern}, {Sadowski}, {Sagrist{\`a}
  Sell{\'e}s}, {Sahlmann}, {Salgado}, {Salguero}, {Sarasso}, {Savietto},
  {Schnorhk}, {Schultheis}, {Sciacca}, {Segol}, {Segovia}, {Segransan},
  {Serpell}, {Shih}, {Smareglia}, {Smart}, {Smith}, {Solano}, {Solitro},
  {Sordo}, {Soria Nieto}, {Souchay}, {Spagna}, {Spoto}, {Stampa}, {Steele},
  {Steidelm{\"u}ller}, {Stephenson}, {Stoev}, {Suess}, {S{\"u}veges}, {Surdej},
  {Szabados}, {Szegedi-Elek}, {Tapiador}, {Taris}, {Tauran}, {Taylor},
  {Teixeira}, {Terrett}, {Tingley}, {Trager}, {Turon}, {Ulla}, {Utrilla},
  {Valentini}, {van Elteren}, {Van Hemelryck}, {van Leeuwen}, {Varadi},
  {Vecchiato}, {Veljanoski}, {Via}, {Vicente}, {Vogt}, {Voss}, {Votruba},
  {Voutsinas}, {Walmsley}, {Weiler}, {Weingrill}, {Werner}, {Wevers},
  {Whitehead}, {Wyrzykowski}, {Yoldas}, {{\v{Z}}erjal}, {Zucker}, {Zurbach},
  {Zwitter}, {Alecu}, {Allen}, {Allende Prieto}, {Amorim},
  {Anglada-Escud{\'e}}, {Arsenijevic}, {Azaz}, {Balm}, {Beck}, {Bernstein},
  {Bigot}, {Bijaoui}, {Blasco}, {Bonfigli}, {Bono}, {Boudreault}, {Bressan},
  {Brown}, {Brunet}, {Bunclark}, {Buonanno}, {Butkevich}, {Carret}, {Carrion},
  {Chemin}, {Ch{\'e}reau}, {Corcione}, {Darmigny}, {de Boer}, {de Teodoro}, {de
  Zeeuw}, {Delle Luche}, {Domingues}, {Dubath}, {Fodor}, {Fr{\'e}zouls},
  {Fries}, {Fustes}, {Fyfe}, {Gallardo}, {Gallegos}, {Gardiol}, {Gebran},
  {Gomboc}, {G{\'o}mez}, {Grux}, {Gueguen}, {Heyrovsky}, {Hoar}, {Iannicola},
  {Isasi Parache}, {Janotto}, {Joliet}, {Jonckheere}, {Keil}, {Kim},
  {Klagyivik}, {Klar}, {Knude}, {Kochukhov}, {Kolka}, {Kos}, {Kutka}, {Lainey},
  {LeBouquin}, {Liu}, {Loreggia}, {Makarov}, {Marseille}, {Martayan},
  {Martinez-Rubi}, {Massart}, {Meynadier}, {Mignot}, {Munari}, {Nguyen},
  {Nordlander}, {Ocvirk}, {O'Flaherty}, {Olias Sanz}, {Ortiz}, {Osorio},
  {Oszkiewicz}, {Ouzounis}, {Palmer}, {Park}, {Pasquato}, {Peltzer}, {Peralta},
  {P{\'e}turaud}, {Pieniluoma}, {Pigozzi}, {Poels}, {Prat}, {Prod'homme},
  {Raison}, {Rebordao}, {Risquez}, {Rocca-Volmerange}, {Rosen}, {Ruiz-Fuertes},
  {Russo}, {Sembay}, {Serraller Vizcaino}, {Short}, {Siebert}, {Silva},
  {Sinachopoulos}, {Slezak}, {Soffel}, {Sosnowska}, {Strai{\v{z}}ys}, {ter
  Linden}, {Terrell}, {Theil}, {Tiede}, {Troisi}, {Tsalmantza}, {Tur},
  {Vaccari}, {Vachier}, {Valles}, {Van Hamme}, {Veltz}, {Virtanen}, {Wallut},
  {Wichmann}, {Wilkinson}, {Ziaeepour}, \& {Zschocke}}]{gaia2016}
{Gaia Collaboration}, {Prusti}, T., {de Bruijne}, J.~H.~J., {et~al.} 2016,
  \aap, 595, A1

\bibitem[{{Gaudi} {et~al.}(2021){Gaudi}, {Meyer}, \&
  {Christiansen}}]{gaudi2021}
{Gaudi}, B.~S., {Meyer}, M., \& {Christiansen}, J. 2021, in ExoFrontiers; Big
  Questions in Exoplanetary Science, ed. N.~{Madhusudhan}, 2--1

\bibitem[{{Georgieva} {et~al.}(2021){Georgieva}, {Persson}, {Barrag{\'a}n},
  {Nowak}, {Fridlund}, {Locci}, {Palle}, {Luque}, {Carleo}, {Gandolfi}, {Kane},
  {Korth}, {Stassun}, {Livingston}, {Matthews}, {Collins}, {Howell}, {Serrano},
  {Albrecht}, {Bieryla}, {Brasseur}, {Ciardi}, {Cochran}, {Colon},
  {Crossfield}, {Csizmadia}, {Deeg}, {Esposito}, {Furlan}, {Gan}, {Goffo},
  {Gonzales}, {Grziwa}, {Guenther}, {Guerra}, {Hirano}, {Jenkins}, {Jensen},
  {Kab{\'a}th}, {Knudstrup}, {Lam}, {Latham}, {Levine}, {Matson}, {McDermott},
  {Osborne}, {Paegert}, {Quinn}, {Redfield}, {Ricker}, {Schlieder}, {Scott},
  {Seager}, {Smith}, {Tenenbaum}, {Twicken}, {Vanderspek}, {Van Eylen}, \&
  {Winn}}]{Georgieva2021}
{Georgieva}, I.~Y., {Persson}, C.~M., {Barrag{\'a}n}, O., {et~al.} 2021,
  \mnras, 505, 4684

\bibitem[{{Gomes da Silva} {et~al.}(2018){Gomes da Silva}, {Figueira},
  {Santos}, \& {Faria}}]{gomesdasilva18}
{Gomes da Silva}, J., {Figueira}, P., {Santos}, N., \& {Faria}, J. 2018, The
  Journal of Open Source Software, 3, 667

\bibitem[{Haywood {et~al.}(2014)Haywood, Collier~Cameron, Queloz, Barros,
  Deleuil, Fares, Gillon, Lanza, Lovis, Moutou, Pepe, Pollacco, Santerne,
  Ségransan, \& Unruh}]{haywood2014}
Haywood, R.~D., Collier~Cameron, A., Queloz, D., {et~al.} 2014, \mnras, 443,
  2517

\bibitem[{{Hirsch} {et~al.}(2021){Hirsch}, {Rosenthal}, {Fulton}, {Howard},
  {Ciardi}, {Marcy}, {Nielsen}, {Petigura}, {de Rosa}, {Isaacson}, {Weiss},
  {Sinukoff}, \& {Macintosh}}]{hirsch2021}
{Hirsch}, L.~A., {Rosenthal}, L., {Fulton}, B.~J., {et~al.} 2021, \aj, 161, 134

\bibitem[{{H{\o}g} {et~al.}(2000){H{\o}g}, {Fabricius}, {Makarov}, {Urban},
  {Corbin}, {Wycoff}, {Bastian}, {Schwekendiek}, \& {Wicenec}}]{hog2000}
{H{\o}g}, E., {Fabricius}, C., {Makarov}, V.~V., {et~al.} 2000, \aap, 355, L27

\bibitem[{{Hunter} {et~al.}(2012){Hunter}, {Macgregor}, {Szabo}, {Wellington},
  \& {Bellgard}}]{YABI}
{Hunter}, A., {Macgregor}, A.~B., {Szabo}, T., {Wellington}, C., \& {Bellgard},
  M.~I. 2012, Source Code for Biology and Medicine, 7, 1

\bibitem[{{Johnstone} {et~al.}(2021){Johnstone}, {Bartel}, \&
  {G{\"u}del}}]{Johnstone+2021}
{Johnstone}, C.~P., {Bartel}, M., \& {G{\"u}del}, M. 2021, \aap, 649, A96

\bibitem[{{Jones} {et~al.}(2015){Jones}, {White}, {Boyajian}, {Schaefer},
  {Baines}, {Ireland}, {Patience}, {ten Brummelaar}, {McAlister}, {Ridgway},
  {Sturmann}, {Sturmann}, {Turner}, {Farrington}, \& {Goldfinger}}]{jones2015}
{Jones}, J., {White}, R.~J., {Boyajian}, T., {et~al.} 2015, \apj, 813, 58

\bibitem[{{Kervella} {et~al.}(2022){Kervella}, {Arenou}, \&
  {Th{\'e}venin}}]{kervella2022A&A...657A...7K}
{Kervella}, P., {Arenou}, F., \& {Th{\'e}venin}, F. 2022, \aap, 657, A7

\bibitem[{{King} \& {Wheatley}(2021)}]{King+Wheatley21}
{King}, G.~W. \& {Wheatley}, P.~J. 2021, \mnras, 501, L28

\bibitem[{{Kipping}(2013)}]{kipping2013}
{Kipping}, D.~M. 2013, \mnras, 435, 2152

\bibitem[{{Kreidberg}(2015)}]{Kreidberg2015}
{Kreidberg}, L. 2015, \pasp, 127, 1161

\bibitem[{{Kubyshkina} {et~al.}(2018{\natexlab{a}}){Kubyshkina}, {Fossati},
  {Erkaev}, {Cubillos}, {Johnstone}, {Kislyakova}, {Lammer}, {Lendl}, \&
  {Odert}}]{kuby+2018a}
{Kubyshkina}, D., {Fossati}, L., {Erkaev}, N.~V., {et~al.} 2018{\natexlab{a}},
  \apjl, 866, L18

\bibitem[{{Kubyshkina} {et~al.}(2018{\natexlab{b}}){Kubyshkina}, {Fossati},
  {Erkaev}, {Johnstone}, {Cubillos}, {Kislyakova}, {Lammer}, {Lendl}, \&
  {Odert}}]{kuby+2018b}
{Kubyshkina}, D., {Fossati}, L., {Erkaev}, N.~V., {et~al.} 2018{\natexlab{b}},
  \aap, 619, A151

\bibitem[{{Lind} {et~al.}(2009){Lind}, {Asplund}, \& {Barklem}}]{lindetal2009}
{Lind}, K., {Asplund}, M., \& {Barklem}, P.~S. 2009, \aap, 503, 541

\bibitem[{{Lopez} \& {Fortney}(2014)}]{LopFor14}
{Lopez}, E.~D. \& {Fortney}, J.~J. 2014, \apj, 792, 1

\bibitem[{{Lopez} {et~al.}(2012){Lopez}, {Fortney}, \& {Miller}}]{LopForMil12}
{Lopez}, E.~D., {Fortney}, J.~J., \& {Miller}, N. 2012, \apj, 761, 59

\bibitem[{{Maggio} {et~al.}(2022){Maggio}, {Locci}, {Pillitteri}, {Benatti},
  {Claudi}, {Desidera}, {Micela}, {Damasso}, {Sozzetti}, \& {Suarez
  Mascare{\~n}o}}]{maggio2022}
{Maggio}, A., {Locci}, D., {Pillitteri}, I., {et~al.} 2022, \apj, 925, 172

\bibitem[{{Malavolta}(2016)}]{2016ascl.soft12008M}
{Malavolta}, L. 2016, {PyORBIT: Exoplanet orbital parameters and stellar
  activity}, Astrophysics Source Code Library, record ascl:1612.008

\bibitem[{{Malavolta} {et~al.}(2018){Malavolta}, {Mayo}, {Louden}, {Rajpaul},
  {Bonomo}, {Buchhave}, {Kreidberg}, {Kristiansen}, {Lopez-Morales}, {Mortier},
  {Vanderburg}, {Coffinet}, {Ehrenreich}, {Lovis}, {Bouchy}, {Charbonneau},
  {Ciardi}, {Collier Cameron}, {Cosentino}, {Crossfield}, {Damasso},
  {Dressing}, {Dumusque}, {Everett}, {Figueira}, {Fiorenzano}, {Gonzales},
  {Haywood}, {Harutyunyan}, {Hirsch}, {Howell}, {Johnson}, {Latham}, {Lopez},
  {Mayor}, {Micela}, {Molinari}, {Nascimbeni}, {Pepe}, {Phillips}, {Piotto},
  {Rice}, {Sasselov}, {S{\'e}gransan}, {Sozzetti}, {Udry}, \&
  {Watson}}]{2018AJ....155..107M}
{Malavolta}, L., {Mayo}, A.~W., {Louden}, T., {et~al.} 2018, \aj, 155, 107

\bibitem[{{Malavolta} {et~al.}(2016){Malavolta}, {Nascimbeni}, {Piotto},
  {Quinn}, {Borsato}, {Granata}, {Bonomo}, {Marzari}, {Bedin}, {Rainer},
  {Desidera}, {Lanza}, {Poretti}, {Sozzetti}, {White}, {Latham}, {Cunial},
  {Libralato}, {Nardiello}, {Boccato}, {Claudi}, {Cosentino}, {Covino},
  {Gratton}, {Maggio}, {Micela}, {Molinari}, {Pagano}, {Smareglia}, {Affer},
  {Andreuzzi}, {Aparicio}, {Benatti}, {Bignamini}, {Borsa}, {Damasso}, {Di
  Fabrizio}, {Harutyunyan}, {Esposito}, {Fiorenzano}, {Gandolfi}, {Giacobbe},
  {Gonz{\'a}lez Hern{\'a}ndez}, {Maldonado}, {Masiero}, {Molinaro}, {Pedani},
  \& {Scandariato}}]{2016A&A...588A.118M}
{Malavolta}, L., {Nascimbeni}, V., {Piotto}, G., {et~al.} 2016, \aap, 588, A118

\bibitem[{{Mallonn} \& {Strassmeier}(2016)}]{mallonn2016}
{Mallonn}, M. \& {Strassmeier}, K.~G. 2016, \aap, 590, A100

\bibitem[{{Mallorqu\'in D\'iaz, M., B\'ejar, V~J~S, Lodieu, N. et al.}(2023,
  accepted)}]{mallorquin23}
{Mallorqu\'in D\'iaz, M., B\'ejar, V~J~S, Lodieu, N. et al.} 2023, accepted,
  \aap

\bibitem[{{Mann} {et~al.}(2020){Mann}, {Johnson}, {Vanderburg}, {Kraus},
  {Rizzuto}, {Wood}, {Bush}, {Rockcliffe}, {Newton}, {Latham}, {Mamajek},
  {Zhou}, {Quinn}, {Thao}, {Benatti}, {Cosentino}, {Desidera}, {Harutyunyan},
  {Lovis}, {Mortier}, {Pepe}, {Poretti}, {Wilson}, {Kristiansen}, {Gagliano},
  {Jacobs}, {LaCourse}, {Omohundro}, {Schwengeler}, {Terentev}, {Kane}, {Hill},
  {Rabus}, {Esquerdo}, {Berlind}, {Collins}, {Murawski}, {Sallam}, {Aitken},
  {Massey}, {Ricker}, {Vanderspek}, {Seager}, {Winn}, {Jenkins}, {Barclay},
  {Caldwell}, {Dragomir}, {Doty}, {Glidden}, {Tenenbaum}, {Torres}, {Twicken},
  \& {Villanueva}}]{mannetal2020}
{Mann}, A.~W., {Johnson}, M.~C., {Vanderburg}, A., {et~al.} 2020, \aj, 160, 179

\bibitem[{{Mucciarelli} {et~al.}(2021){Mucciarelli}, {Bellazzini}, \&
  {Massari}}]{Mucciarellietal2021}
{Mucciarelli}, A., {Bellazzini}, M., \& {Massari}, D. 2021, \aap, 653, A90

\bibitem[{{Nardiello}(2020)}]{2020MNRAS.498.5972N}
{Nardiello}, D. 2020, \mnras, 498, 5972

\bibitem[{{Nardiello} {et~al.}(2021){Nardiello}, {Deleuil}, {Mantovan},
  {Malavolta}, {Lacedelli}, {Libralato}, {Bedin}, {Borsato}, {Granata}, \&
  {Piotto}}]{2021MNRAS.505.3767N}
{Nardiello}, D., {Deleuil}, M., {Mantovan}, G., {et~al.} 2021, \mnras, 505,
  3767

\bibitem[{{Nardiello} {et~al.}(2022){Nardiello}, {Malavolta}, {Desidera},
  {Baratella}, {D'Orazi}, {Messina}, {Biazzo}, {Benatti}, {Damasso}, {Rajpaul},
  {Bonomo}, {Capuzzo Dolcetta}, {Mallonn}, {Cale}, {Plavchan}, {El Mufti},
  {Bignamini}, {Borsa}, {Carleo}, {Claudi}, {Covino}, {Lanza}, {Maldonado},
  {Mancini}, {Micela}, {Molinari}, {Pinamonti}, {Piotto}, {Poretti},
  {Scandariato}, {Sozzetti}, {Andreuzzi}, {Boschin}, {Cosentino}, {Fiorenzano},
  {Harutyunyan}, {Knapic}, {Pedani}, {Affer}, {Maggio}, \&
  {Rainer}}]{Nardiello2022}
{Nardiello}, D., {Malavolta}, L., {Desidera}, S., {et~al.} 2022, \aap, 664,
  A163

\bibitem[{{Nardiello} {et~al.}(2020){Nardiello}, {Piotto}, {Deleuil},
  {Malavolta}, {Montalto}, {Bedin}, {Borsato}, {Granata}, {Libralato}, \&
  {Manthopoulou}}]{Nardielloetal2020}
{Nardiello}, D., {Piotto}, G., {Deleuil}, M., {et~al.} 2020, \mnras, 495, 4924

\bibitem[{{Pecaut} \& {Mamajek}(2013)}]{PecautMamajek2013}
{Pecaut}, M.~J. \& {Mamajek}, E.~E. 2013, \apjs, 208, 9

\bibitem[{{Penz} {et~al.}(2008){Penz}, {Micela}, \& {Lammer}}]{Penz08a}
{Penz}, T., {Micela}, G., \& {Lammer}, H. 2008, \aap, 477, 309

\bibitem[{{Perger} {et~al.}(2021){Perger}, {Anglada-Escud{\'e}}, {Ribas},
  {Rosich}, {Herrero}, \& {Morales}}]{perger2021}
{Perger}, M., {Anglada-Escud{\'e}}, G., {Ribas}, I., {et~al.} 2021, \aap, 645,
  A58

\bibitem[{{Poppenhaeger} {et~al.}(2021){Poppenhaeger}, {Ketzer}, \&
  {Mallonn}}]{poppen2021}
{Poppenhaeger}, K., {Ketzer}, L., \& {Mallonn}, M. 2021, \mnras, 500, 4560

\bibitem[{{Poretti} {et~al.}(2016){Poretti}, {Boccato}, {Claudi}, {Cosentino},
  {Covino}, {Desidera}, {Gratton}, {Lanza}, {Maggio}, {Micela}, {Molinari},
  {Pagano}, {Piotto}, {Smareglia}, {Sozzetti}, \& {GAPS
  Collaboration}}]{Poretti2016}
{Poretti}, E., {Boccato}, C., {Claudi}, R., {et~al.} 2016, \memsai, 87, 141

\bibitem[{{Rajpaul} {et~al.}(2015){Rajpaul}, {Aigrain}, {Osborne}, {Reece}, \&
  {Roberts}}]{rajpaul2015}
{Rajpaul}, V., {Aigrain}, S., {Osborne}, M.~A., {Reece}, S., \& {Roberts}, S.
  2015, \mnras, 452, 2269

\bibitem[{{Ricker} {et~al.}(2016){Ricker}, {Vanderspek}, {Winn}, {Seager},
  {Berta-Thompson}, {Levine}, {Villasenor}, {Latham}, {Charbonneau}, {Holman},
  {Johnson}, {Sasselov}, {Szentgyorgyi}, {Torres}, {Bakos}, {Brown},
  {Christensen-Dalsgaard}, {Kjeldsen}, {Clampin}, {Rinehart}, {Deming}, {Doty},
  {Dunham}, {Ida}, {Kawai}, {Sato}, {Jenkins}, {Lissauer}, {Jernigan},
  {Kaltenegger}, {Laughlin}, {Lin}, {McCullough}, {Narita}, {Pepper},
  {Stassun}, \& {Udry}}]{ricker2016}
{Ricker}, G.~R., {Vanderspek}, R., {Winn}, J., {et~al.} 2016, in Society of
  Photo-Optical Instrumentation Engineers (SPIE) Conference Series, Vol. 9904,
  Space Telescopes and Instrumentation 2016: Optical, Infrared, and Millimeter
  Wave, ed. H.~A. {MacEwen}, G.~G. {Fazio}, M.~{Lystrup}, N.~{Batalha},
  N.~{Siegler}, \& E.~C. {Tong}, 99042B

\bibitem[{{Sanz-Forcada} {et~al.}(2022){Sanz-Forcada}, {L{\'o}pez-Puertas},
  {Nortmann}, \& {Lamp{\'o}n}}]{SF22}
{Sanz-Forcada}, J., {L{\'o}pez-Puertas}, M., {Nortmann}, L., \& {Lamp{\'o}n},
  M. 2022, 21st Cambridge Workshop on Cool Stars, Stellar Systems, and the Sun,
  \url{https://backoffice.inviteo.com/upload/compte153/Base/inscriptions_projets/fichier/108345-posterjsanzcs21.pdf}

\bibitem[{{Sanz-Forcada} {et~al.}(2011){Sanz-Forcada}, {Micela}, {Ribas},
  {Pollock}, {Eiroa}, {Velasco}, {Solano}, \& {Garc{\'\i}a-{\'A}lvarez}}]{SF11}
{Sanz-Forcada}, J., {Micela}, G., {Ribas}, I., {et~al.} 2011, \aap, 532, A6

\bibitem[{{Smith} {et~al.}(2012){Smith}, {Stumpe}, {Van Cleve}, {Jenkins},
  {Barclay}, {Fanelli}, {Girouard}, {Kolodziejczak}, {McCauliff}, {Morris}, \&
  {Twicken}}]{2012PASP..124.1000S}
{Smith}, J.~C., {Stumpe}, M.~C., {Van Cleve}, J.~E., {et~al.} 2012, \pasp, 124,
  1000

\bibitem[{{Sneden}(1973)}]{sneden1973}
{Sneden}, C. 1973, \apj, 184, 839

\bibitem[{{Sousa} {et~al.}(2015){Sousa}, {Santos}, {Adibekyan}, {Delgado-Mena},
  \& {Israelian}}]{sousaetal2015}
{Sousa}, S.~G., {Santos}, N.~C., {Adibekyan}, V., {Delgado-Mena}, E., \&
  {Israelian}, G. 2015, \aap, 577, A67

\bibitem[{{Southworth}(2011)}]{2011MNRAS.417.2166S}
{Southworth}, J. 2011, \mnras, 417, 2166

\bibitem[{{Strassmeier} {et~al.}(2004){Strassmeier}, {Granzer}, {Weber},
  {Woche}, {Andersen}, {Bartus}, {Bauer}, {Dionies}, {Popow}, {Fechner},
  {Hildebrandt}, {Washuettl}, {Ritter}, {Schwope}, {Staude}, {Paschke},
  {Stolz}, {Serre-Ricart}, {de la Rosa}, \& {Arnay}}]{strassmeier2004}
{Strassmeier}, K.~G., {Granzer}, T., {Weber}, M., {et~al.} 2004, Astronomische
  Nachrichten, 325, 527

\bibitem[{{Stumpe} {et~al.}(2014){Stumpe}, {Smith}, {Catanzarite}, {Van Cleve},
  {Jenkins}, {Twicken}, \& {Girouard}}]{2014PASP..126..100S}
{Stumpe}, M.~C., {Smith}, J.~C., {Catanzarite}, J.~H., {et~al.} 2014, \pasp,
  126, 100

\bibitem[{{Stumpe} {et~al.}(2012){Stumpe}, {Smith}, {Van Cleve}, {Twicken},
  {Barclay}, {Fanelli}, {Girouard}, {Jenkins}, {Kolodziejczak}, {McCauliff}, \&
  {Morris}}]{2012PASP..124..985S}
{Stumpe}, M.~C., {Smith}, J.~C., {Van Cleve}, J.~E., {et~al.} 2012, \pasp, 124,
  985

\bibitem[{{Su{\'a}rez Mascare{\~n}o} {et~al.}(2021){Su{\'a}rez Mascare{\~n}o},
  {Damasso}, {Lodieu}, {Sozzetti}, {B{\'e}jar}, {Benatti}, {Zapatero Osorio},
  {Micela}, {Rebolo}, {Desidera}, {Murgas}, {Claudi}, {Gonz{\'a}lez
  Hern{\'a}ndez}, {Malavolta}, {del Burgo}, {D'Orazi}, {Amado}, {Locci},
  {Tabernero}, {Marzari}, {Aguado}, {Turrini}, {Cardona Guill{\'e}n},
  {Toledo-Padr{\'o}n}, {Maggio}, {Aceituno}, {Bauer}, {Caballero},
  {Chinchilla}, {Esparza-Borges}, {Gonz{\'a}lez-{\'A}lvarez}, {Granzer},
  {Luque}, {Mart{\'\i}n}, {Nowak}, {Oshagh}, {Pall{\'e}}, {Parviainen},
  {Quirrenbach}, {Reiners}, {Ribas}, {Strassmeier}, {Weber}, \&
  {Mallonn}}]{Mascareno2021}
{Su{\'a}rez Mascare{\~n}o}, A., {Damasso}, M., {Lodieu}, N., {et~al.} 2021,
  Nature Astronomy, 6, 232

\bibitem[{{Sun} {et~al.}(2019){Sun}, {Ioannidis}, {Gu}, {Schmitt}, {Wang}, \&
  {Kouwenhoven}}]{sun2019}
{Sun}, L., {Ioannidis}, P., {Gu}, S., {et~al.} 2019, \aap, 624, A15

\bibitem[{{Vines} {et~al.}(2023){Vines}, {Jenkins}, {Berdi{\~n}as}, {Soto},
  {D{\'\i}az}, {Alves}, {Tuomi}, {Wittenmyer}, {de Leon}, {Pe{\~n}a},
  {Lissauer}, {Ballard}, {Bedding}, {Bowler}, {Horner}, {Jones}, {Kane},
  {Kielkopf}, {Plavchan}, {Shporer}, {Tinney}, {Zhang}, {Wright}, {Addison},
  {Mengel}, {Okumura}, \& {Samadi-Ghadim}}]{vines2023}
{Vines}, J.~I., {Jenkins}, J.~S., {Berdi{\~n}as}, Z., {et~al.} 2023, \mnras,
  518, 2627

\bibitem[{{Zechmeister} \& {K{\"u}rster}(2009)}]{zechmeister2009}
{Zechmeister}, M. \& {K{\"u}rster}, M. 2009, \aap, 496, 577

\bibitem[{{Zeng} {et~al.}(2019){Zeng}, {Jacobsen}, {Sasselov}, {Petaev},
  {Vanderburg}, {Lopez-Morales}, {Perez-Mercader}, {Mattsson}, {Li}, {Heising},
  {Bonomo}, {Damasso}, {Berger}, {Cao}, {Levi}, \&
  {Wordsworth}}]{2019PNAS..116.9723Z}
{Zeng}, L., {Jacobsen}, S.~B., {Sasselov}, D.~D., {et~al.} 2019, Proceedings of
  the National Academy of Science, 116, 9723

\bibitem[{{Zhang} {et~al.}(2022){Zhang}, {Knutson}, {Wang}, {Dai}, {dos
  Santos}, {Fossati}, {Henry}, {Ehrenreich}, {Alibert}, {Hoyer}, {Wilson}, \&
  {Bonfanti}}]{2022AJ....163...68Z}
{Zhang}, M., {Knutson}, H.~A., {Wang}, L., {et~al.} 2022, \aj, 163, 68

\end{thebibliography}
	
	\begin{appendix}
		
		\section{Description of the GP models and model priors} \label{app:modeldetails}
		Hereafter we provide details about the different GP kernels tested in this work to filter out the stellar activity contribution to the observed RV variability.
		
		\textit{GP quasi-periodic (QP) kernel.} An element of the QP covariance matrix (e.g. \citealt{haywood2014}) used in our work is defined as follows:
		\begin{gather} 
			\label{eq:eqgpqpkernel}
			k_{QP}(t, t^{\prime}) = h^2\cdot\exp\Bigg[-\frac{(t-t^{\prime})^2}{2\lambda_{\rm QP}^2} - \frac{\sin^{2}\Bigg(\pi(t-t^{\prime})/\theta\Bigg)}{2w^2}\Bigg] + \nonumber \\
			+\, (\sigma^{2}_{\rm RV}(t)+\sigma^{2}_{\rm jit})\cdot\delta_{t, t^{\prime}}
		\end{gather}
		Here, $t$ and $t^{\prime}$ represent two different epochs of observations, $\sigma_{\rm RV}$ is the radial velocity uncertainty, and $\delta_{t, t^{\prime}}$ is the Kronecker delta. Our analysis takes into account other sources of uncorrelated noise -- instrumental and/or astrophysical -- by including a constant jitter term $\sigma_{\rm jit}$ which is added in quadrature to the formal uncertainties $\sigma_{\rm RV}$. The GP hyper-parameters are $h$, which denotes the scale amplitude of the correlated signal; $\theta$, which represents the periodic timescale of the correlated signal, and corresponds to the stellar rotation period; $w$, which describes the "weight" of the rotation period harmonic content within a complete stellar rotation (i.e. a low value of $w$ indicates that the periodic variations contain a significant contribution from the harmonics of the rotation periods); and $\lambda_{\rm QP}$, which represents the decay timescale of the correlations, and is related to the temporal evolution of the magnetically active regions responsible for the correlated signal observed in the RVs.\\
		
		\textit{GP quasi-periodic with cosine (QPC) kernel.} The QPC kernel has been introduced by \cite{perger2021}. The covariance matrix element implemented in our work is described by the equation 
		
		\begin{gather}
			\label{eq:eqgpqpckernel}
			k_{QPC}(t, t^{\prime}) = \exp\Big(-2\frac{(t-t^{\prime})^2}{\lambda_{QPC}^2}\Big)\cdot \Bigg[h_1^2\exp\Big(-\frac{1}{2w^2}\sin^2\Big(\frac{\pi(t-t^{\prime})}{\theta}\Big)\Big)+ \nonumber \\ 
			+h_2^2\cos\Big(\frac{4\pi(t-t^{\prime})}{\theta}\Big)\Bigg]+(\sigma^{2}_{\rm RV}(t)+\sigma^{2}_{\rm jit})\cdot\delta_{t, t^{\prime}}\,
		\end{gather}
		
		Again, $t$ and $t^{\prime}$ represent two different epochs of observations; $h_1$ and $h_2$ are scale amplitudes; $\theta$ still represents the periodic time-scale of the modelled signal, and corresponds to the stellar rotation period; $w$ still describes the weight of the rotation period harmonic content within a complete stellar rotation; $\lambda_{\rm QPC}$ is defined as $2\cdot \lambda_{QP}$, better representing the average lifetime of the activity-related features responsible for the stellar correlated signal in the RVs according to \cite{perger2021}; $\sigma_{\rm RV}(t)$ and $\sigma_{\rm jit}$ are the radial velocity uncertainty at epoch \textit{t} and the uncorrelated jitter, respectively, and $\delta_{t, t^{\prime}}$ is the Kronecker delta.\\
		
		\textit{GP rotational or double simple harmonic oscillator (dSHO) kernel.} 
		The ``rotational'' kernel is defined by a mixture of two stochastically driven, damped simple harmonic oscillators (SHOs) with undamped periods of $P_{\star,\, rot}$ and $P_{\star,\, rot}$/2. This can be obtained by combining two \texttt{SHOTerm} kernels included in the package \texttt{celerite} \citep{2017AJ....154..220F}\footnote{\url{https://github.com/dfm/celerite/blob/main/celerite/terms.py}}. 
		The power spectral density corresponding to this kernel is 
		\begin{eqnarray}
			S(\omega) = \sqrt{\frac{2}{\pi}} \frac{S_1\omega_1^4}{(\omega^2-\omega_1^2)^2 +  2\omega_1^2\omega^2} + \nonumber\\
			+\sqrt{\frac{2}{\pi}} \frac{S_2\omega_2^4}{(\omega^2-\omega_2^2)^2 + 2\omega_2^2\omega^2/Q^2},
		\end{eqnarray}
		
		where
		\begin{gather} 
			\label{eqn:sho1}
			S_{\rm 1}=\frac{A^2}{\omega_{\rm 1}Q_{\rm 1}(1+f)}, \\ 
			S_{\rm 2}=\frac{A^2}{\omega_{\rm 2}Q_{\rm 2}(1+f)}\cdot f, \\
			\omega_{\rm 1}=\frac{4 \pi Q_{\rm 1}}{P_{\rm rot}\sqrt{4Q_{\rm 1}^{2}-1}}, \\
			\omega_{\rm 2}=\frac{8 \pi Q_{\rm 2}}{P_{\rm rot}\sqrt{4Q_{\rm 2}^{2}-1}}, \\
			Q_{\rm 1}= \frac{1}{2}+Q_{\rm 0}+\Delta Q,\\
			\label{eqn:sho2}
			Q_{\rm 2}= \frac{1}{2}+Q_{\rm 0}. 
		\end{gather}
		
		The parameters in [\ref{eqn:sho1}-\ref{eqn:sho2}], where the subscripts 1 and 2 refer to the primary ($P_{\star,\, rot}$) and secondary ($P_{\star,\, rot}$/2) modes, represent the inputs to the \texttt{SHOTerm} kernels. However, instead of using them directly, we adopt a different parametrization using the following variables as free hyper-parameters in the MC analysis, from which $S_{\rm i}$, $Q_{\rm i}$, and $\omega_{\rm i}$ are derived through Eq. [\ref{eqn:sho1}-\ref{eqn:sho2}]: the variability amplitude $A$, the stellar rotation period $P_{\star,\, rot}$, the quality factor $Q_{0}$, the difference $\Delta$Q between the quality factors of the first and second terms, and the fractional amplitude $f$ of the secondary mode relative to the primary.\\
		
		\textit{Multi-dimensional GP framework}. This framework has been first introduced by \cite{rajpaul2015}. A \texttt{python} version of the code is implemented in the package \texttt{pyaneti} \citep{pyaneti}\footnote{\url{https://github.com/oscaribv/pyaneti}}. In our work, we included the module \texttt{pyaneti} in our own code to make it work with the nested sampler \texttt{MultiNest}. In brief, the multi-dimensional GP framework uses ancillary data sensitive to stellar activity in a combined fit with the RVs (in our case, we used the activity diagnostics \textit{BIS} and $\log R^{\prime}_{\rm HK}$), with each dataset represented in the time domain by linear combinations of the same GP-drawn function $G(t)$ and its derivative $\dot{G}(t)$. For all the details of the multi-dimensional GP implementation see \cite{pyaneti2}, especially Sect. 3.1.2 and Eq. (8) therein. For our test, we adopted a QP kernel, as described above, and we fixed to zero the coefficient of $\dot{G}(t)$ for the case of $\log R^{\prime}_{\rm HK}$. 
		
		The prior distributions for the free parameters of the models tested in this work are provided in Table \ref{table:priors}.
		
		\begin{table}
			\caption{Priors used for the models listed in Table \ref{tab:models}}.         
			\label{table:priors}      
			\centering      
			\tiny
			\begin{tabular}{lc}       
				\hline             
				\textbf{Parameter} & \textbf{Prior}\tablefootmark{a} \\    
				\hline              
				\noalign{\smallskip}
				\textbf{Stellar Activity - QP kernel} \\
				\noalign{\smallskip}
				$h$ [\ms] & $\mathcal{U}(0,500)$ \\
				\noalign{\smallskip}
				$\theta$ [d] & $\mathcal{U}(0,10)$  \\
				\noalign{\smallskip}
				$\lambda_{\rm QP}$ [d] & $\mathcal{U}(0,1000)$  \\
				\noalign{\smallskip}
				$w$ & $\mathcal{U}(0,1)$  \\
				\noalign{\smallskip}
				\textbf{Stellar Activity - QPC kernel} \\
				\noalign{\smallskip}
				$h_1$ [\ms] & $\mathcal{U}(0,500)$ \\
				\noalign{\smallskip}
				$h_2$ [\ms] & $\mathcal{U}(0,500)$ \\
				\noalign{\smallskip}
				$\theta$ [d] & $\mathcal{U}(0,10)$  \\
				\noalign{\smallskip}
				$\lambda_{\rm QPC}$ [d] & $\mathcal{U}(0,1000)$  \\
				\noalign{\smallskip}
				$w$ & $\mathcal{U}(0,1)$  \\
				\noalign{\smallskip}
				\textbf{Stellar Activity - dSHO kernel} \\
				\noalign{\smallskip}
				$\log A$ & $\mathcal{U}$(0.05,10) \\
				\noalign{\smallskip}
				$\theta$ [d] & $\mathcal{U}$(0,10)  \\
				\noalign{\smallskip}
				$\log Q_0$ & $\mathcal{U}$(-10,10) \\
				\noalign{\smallskip}
				$\log \Delta Q$ & $\mathcal{U}$(-10,10) \\
				\noalign{\smallskip}
				$f$ & $\mathcal{U}$(0,10) \\
				\noalign{\smallskip}
				\textbf{Stellar Activity - Multi-dim. QP kernel\tablefootmark{b}} \\
				\noalign{\smallskip}
				$A_{\rm RV}$, $B_{\rm RV}$ [km s$^{-1}$] & $\mathcal{U}$(-0.1,0.1), $\mathcal{U}$(-0.1,0.1) \\
				\noalign{\smallskip}
				$A_{\rm BIS}$, $B_{\rm BIS}$ [km s$^{-1}$] & $\mathcal{U}$(-0.1,0.1), $\mathcal{U}$(-0.1,0.1) \\
				\noalign{\smallskip}
				$A_{\rm R^\prime HK}$, $B_{\rm R^\prime HK}$ [dex] & $\mathcal{U}$(-0.1,0.1), fixed to 0 \\
				\noalign{\smallskip}
				\textbf{Keplerian of planet b} \\
				\noalign{\smallskip}
				$K_b$ [\ms] & $\mathcal{U}$(0,30)\\
				\noalign{\smallskip}
				$P_b$ [d] & $\mathcal{U}$(6.9,7.3)\\ 
				\noalign{\smallskip}
				$T_{conj,\,b}$ [BJD-2\,450\,000] & $\mathcal{U}$(9584.4,9584.7) \\
				\noalign{\smallskip}
				$\sqrt{e_{\rm b}}\cos\omega_{\rm \star\:,b}$ & $\mathcal{U}$(-1,1)\\
				\noalign{\smallskip}
				$\sqrt{e_{\rm b}}\sin\omega_{\rm \star\:,b}$ & $\mathcal{U}$(-1,1)\\
				\noalign{\smallskip}
				\textbf{Keplerian of planet c} \\
				\noalign{\smallskip}
				$K_c$ [\ms] & $\mathcal{U}$(0,30)\\
				\noalign{\smallskip}
				$P_c$ [d] & $\mathcal{U}$(19.5,21.5)  \\
				\noalign{\smallskip}
				$T_{conj,\,c}$ [BJD-2\,450\,000] & $\mathcal{U}$(9583.5,9583.8) \\
				\noalign{\smallskip}
				$\sqrt{e_{\rm c}}\cos\omega_{\rm \star\:,c}$ & $\mathcal{U}$(-1,1)\\
				\noalign{\smallskip}
				$\sqrt{e_{\rm c}}\sin\omega_{\rm \star\:,c}$ & $\mathcal{U}$(-1,1)\\
				\noalign{\smallskip}
				RV acceleration, $\dot{\gamma}$ [$\ms d^{-1}$] & $\mathcal{U}$(-1,1) \\
				\noalign{\smallskip}
				\textbf{Transit parameters for planet b} \\
				\noalign{\smallskip}
				$R_{\rm b}/R_{\star}$ & $\mathcal{U}$(0,1)\\
				\noalign{\smallskip}
				orbital plane inclination, $i_b$ [deg] &  $\mathcal{U}$(80,90)  \\
				\noalign{\smallskip}
				\textbf{Transit parameters for planet c} \\
				\noalign{\smallskip}
				$R_{\rm c}/R_{\star}$ & $\mathcal{U}$(0,1)\\
				\noalign{\smallskip}
				orbital plane inclination, $i_c$ [deg] &  $\mathcal{U}$(80,90)  \\
				\noalign{\smallskip}
				\textbf{Instrument-related} \\
				\noalign{\smallskip}
				$\gamma_{\rm RV}$ [\ms] & \texttt{DRS}: $\mathcal{U}$(-16000,-15500) \\
				\noalign{\smallskip}
				& \texttt{TERRA}: $\mathcal{U}$(-100,100) \\
				\noalign{\smallskip}
				$\sigma_{\rm jit,\, RV}$ [\ms] & $\mathcal{U}$(0,50) \\
				\noalign{\smallskip}
				$\sigma_{\rm jit, \,TESS\,sect.\,20}$ & $\mathcal{U}$(0,300e-5) \\
				\noalign{\smallskip}
				$\sigma_{\rm jit, \,TESS\,sect.\,44-47}$ & $\mathcal{U}$(0,300e-5) \\
				\noalign{\smallskip}
				Limb darkening param.\tablefootmark{c} $q_1$ & $\mathcal{U}$(0,1)  \\
				\noalign{\smallskip}
				Limb darkening param.\tablefootmark{c} $q_2$ & $\mathcal{U}$(0,1) \\
				\noalign{\smallskip}
				\hline
			\end{tabular}
			\tablefoot{
				\tablefoottext{a}{$\mathcal{U}(a,b)$ denotes a prior drawn from an uninformative distribution in the range (a,b). $\mathcal{N}(a,b)$ denotes a prior drawn from a Gaussian distribution with mean \textit{a} and sigma \textit{b}. }
				\tablefoottext{b}{We follow the notation of Eq. (8) in \cite{pyaneti2}}
				\tablefoottext{c}{Following \cite{kipping2013}, $q_1$ and $q_2$ are related to the limb darkening coefficients LD$_{\rm c1}$ and LD$_{\rm c2}$ by the relations LD$_{\rm c1}=2q_2\sqrt{q_1}$ and LD$_{\rm c2}=\sqrt{q_1}(1-2q_2)$. }
			}
		\end{table}
		
		\section{Details on atmospheric evaporation modelling} \label{app:photoevap}
		
		The hydro-based approximation developed by \cite{kuby+2018a,kuby+2018b} provides an analytic expression of the planetary mass-loss rate as a function of planetary mass and radius, star-planet distance, and stellar high-energy flux. We employed this approximation to model the time evolution of the planet characteristics, taking into account both the stellar evolutionary track and the long-term change of the XUV irradiation.    
		
		For the X-ray evolution as a function of time, we used the description by \cite{Penz08a}, based on open cluster and field G-type stars. It describes the saturation and decay phases of stellar activity with simple analytic power laws. \cite{maggio2022} showed that they represent a good approximation of the more complex behaviour modelled by \cite{Johnstone+2021} for late-type stars. In order to evaluate the stellar irradiation in an extreme ultraviolet band we adopted a scaling law between EUV (100--920\,\AA) and X-ray (5--100\,\AA) luminosities, calibrated on stars observed in both bands, derived by \cite{SF22}, namely
		\begin{equation}
			\log L_{\rm EUV} = (0.793 \pm 0.058) \log L_{\rm x} + (6.53 \pm 1.61)
		\end{equation}
		which is an updated version of the better-known and widely used relationship by \cite{SF11}. The predicted evolution of the X-ray and EUV stellar luminosities is shown in Fig.\ \ref{fig:xuvevol}.
		
		We also took into account the evolution of the planetary radius according to \cite{LopFor14}.
		This relation was developed for H-He dominated atmospheres, and provides the envelope radius $R_{\rm env}$, as a function of the planetary mass, the atmospheric mass fraction $f_{\rm atm}$, the bolometric flux received, and the age of the system.
  It takes into account also the cooling and contraction of the envelope as a consequence of its  thermal evolution, based on the grid of models presented in \cite{LopForMil12}. 
        
        The track followed by HD 63433 in the theoretical temperature-luminosity diagram is shown in Fig. \ref{fig:evoltrack}. It was computed for a star with a mass $0.994 \msun$ and metallicity $[\rm Fe/H] = +0.02$ (Sect. \ref{sec:stellarparameters}) using the web-based interpolator\footnote{\url{https://waps.cfa.harvard.edu/MIST/interp_tracks.html}} of the MESA Isochrones and Stellar Tracks (MIST, \citealt{choi+2016}).
		
		The \cite{LopFor14} relation can be inverted to estimate $f_{\rm atm}$ at the current age, starting with the known (measured) planetary radius and with a core radius, $R_{\rm core}$, computed as in \cite{benatti2021}, assuming that all the planetary mass is concentrated in the core. Different assumptions (or future measurements) of the planetary mass, yield different $R_{\rm core}$ and $f_{\rm atm}$ values.
		
		We then proceeded with the time evolution. For each time step of the simulation, we computed the mass-loss rate and we updated $f_{\rm atm}$ and the planetary mass, thus obtaining a new value of $R_{\rm env}$ with the relation by \cite{LopFor14}. The latter quantity added to the core radius (assumed constant) provides the updated planetary radius. For each starting planetary mass, we let the system evolve from the current age (414 Myr for HD 63433) until 5 Gyr. In order to check if the atmosphere is hydrodynamically stable or not, we use the Jeans escape parameter
		\begin{equation}
			\Lambda = \frac{G m_{\rm H} M_{\rm p}}{k_{\rm B} T_{\rm eq} R_{\rm p}}
			\label{eq:J}
		\end{equation}
		where $G$ is the gravitational constant, $m_{\rm H}$ is the hydrogen mass, and $k_{\rm B}$ is the Boltzman constant, while $M_{\rm p}$ and $R_{\rm p}$ are the planet mass and radius, and $T_{\rm eq}$ is the planet equilibrium temperature, computed as
		\begin{equation}
			T_{\rm eq}=T_{\rm eff}\biggl[f_{\rm p}(1-A_{\rm B})\biggr]^{1/4}\biggl(\frac{R_{\rm *}}{2 d}\biggr)^{1/2}
		\end{equation}
		where $T_{\rm eff}$ and $R_{\rm *}$ are the stellar effective temperature and radius, respectively, $A_{\rm B}$ is the Bond albedo, $d$ the star-planet distance, $f_{\rm p}$ is a parameter that takes into account if the planet is tidally locked or not. We assumed $A_{\rm B} = 0.5$, and $f_{\rm p}=\frac{2}{3}$, because tidal locking is assumed as in \citep{2022AJ....163...68Z}. 
		
		Note that $\Lambda$ depends on the stellar $T_{\rm eff}$ through $T_{\rm eq}$, and  $\Lambda$ enters also in the analytic expression for the atmospheric mass-loss rate derived by \cite{kuby+2018b}, which is also a function of the XUV flux, $F_{\rm XUV}$, at the planetary orbital position. According to \cite{Fossati+2017}, photoevaporation of planetary atmospheres occurs when the Jeans escape parameter is smaller than a critical value ($\Lambda < 80$), but two different evaporation regimes are possible, depending on how $\Lambda$ compares with a control parameter $e^{\Sigma}$, determined by $d$ and by time-dependent values of $R_{\rm p}$ and $F_{\rm XUV}$ \citep{kuby+2018b}: if $\Lambda > e^{\Sigma}$, the evaporation is controlled by the XUV flux, while for smaller values of $\Lambda$ the atmospheric escape is mainly driven by the planetary intrinsic thermal energy and low gravity.
		
		In this work, assuming an X-ray luminosity of $5 \times 10^{28}$\,erg/s at the current age of the system, and allowing for the evolution of the bolometric luminosity and the effective temperature of the star as described above (Fig. \ref{fig:evoltrack}), we investigated the hydrodynamic atmospheric evolution of the two planets of HD 63433 in the future.
		Furthermore, we studied the system’s evolution back in time, as described by \cite{Georgieva2021}. According to the aforementioned scenario and assuming a core which does not change in size or mass, we created a synthetic population of planets with different initial masses, and we assigned to them the core radius and core mass fixed by the total mass assumed for the planet at its present age. This framework leaves $f_{\rm atm}$ to dictate the total initial mass, while the total radius is again based on the analytic fit by \cite{LopFor14}. Then we searched for the planet in the synthetic set which ends up with mass, radius and $f_{\rm atm}$  most similar to the values at the present age, and we selected its predicted evolutionary history. We started all the simulations in the past at an age of 10\,Myr, when we suppose that the circumstellar disk is already disappeared and all planets are in their final, stable orbit. These simulations always end at an age of 414\,Myr.
		
		\section{Additional plots} \label{app:addplots}
		
		
		\begin{figure*}
			\centering
			\includegraphics[width=0.8\textwidth]{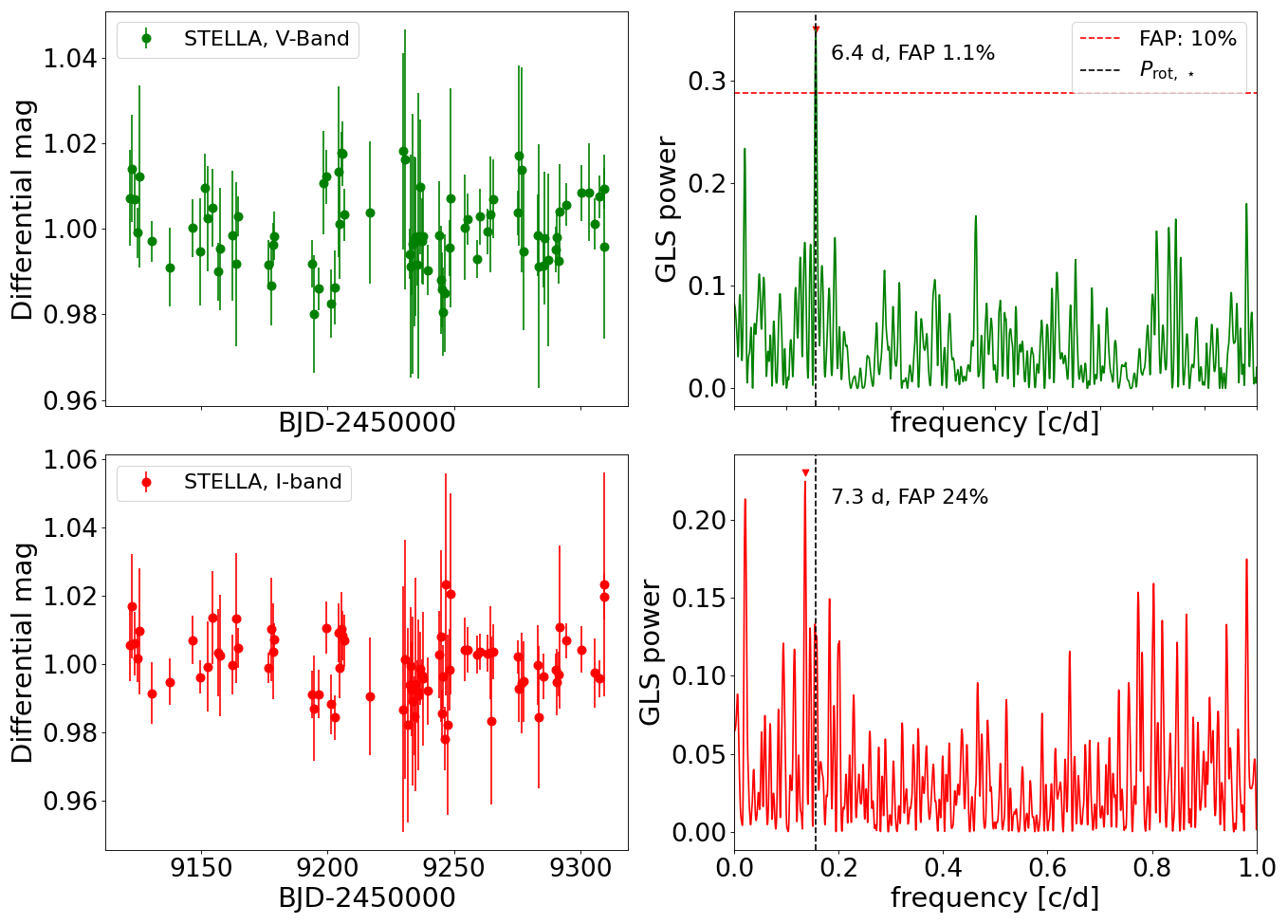}
			\caption{Time series and \texttt{GLS} periodograms of the light curves collected by STELLA in \textit{V} and \textit{I} bands. The horizontal dashed line denotes the FAP level of 10$\%$, the vertical dashed line identifies the stellar rotation period.}
			\label{fig:stelladata}
		\end{figure*}

            \begin{figure}
			\centering
			\includegraphics[width=0.5\textwidth]{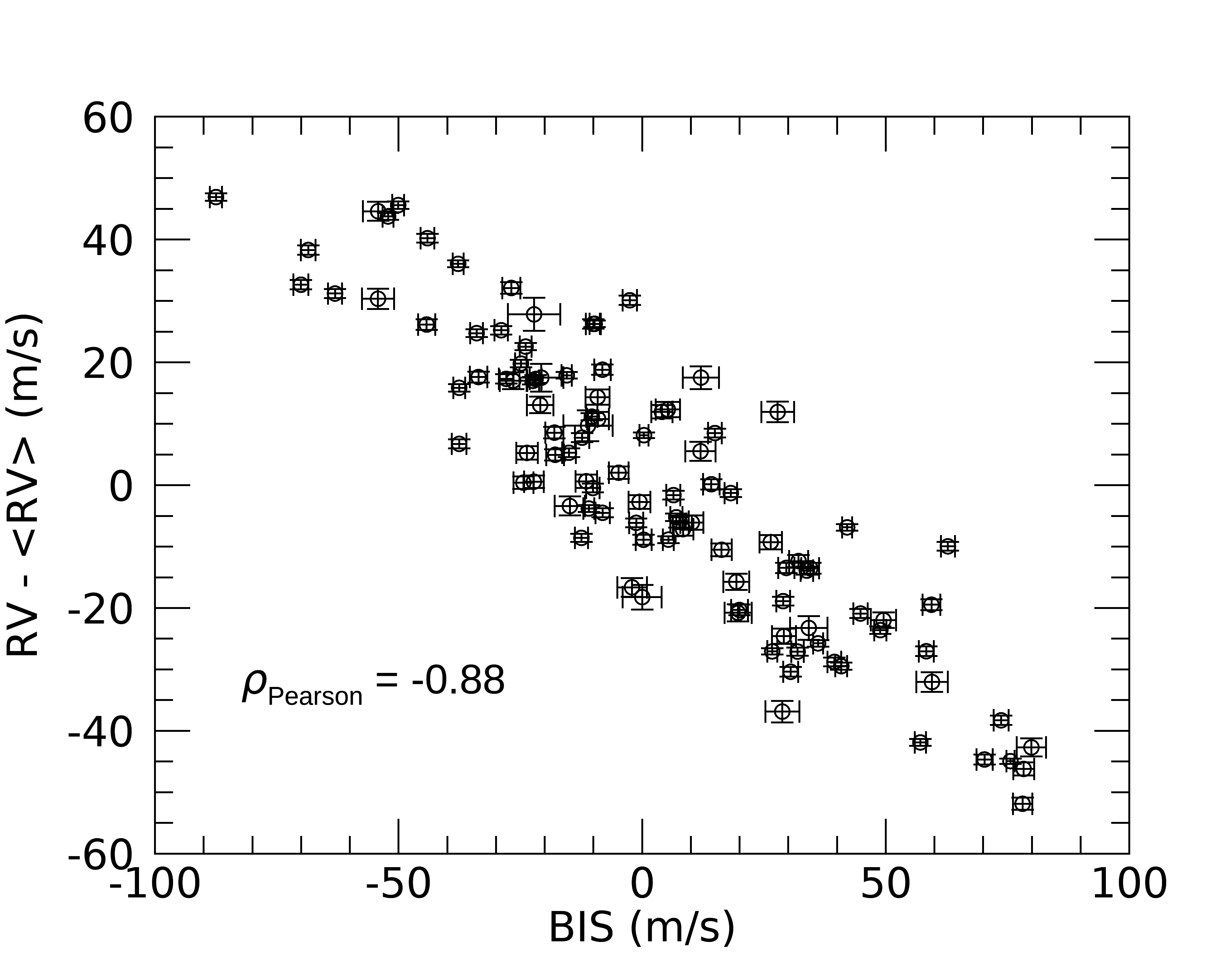}
			\caption{RVs as a function of the CCF-BIS, showing the significant correlation between the two dataset.}
			\label{fig:rvbiscorr}
		\end{figure}

		\begin{figure}
			\centering
			\includegraphics[width=0.5\textwidth, trim = 1cm 13cm 2cm 3cm]{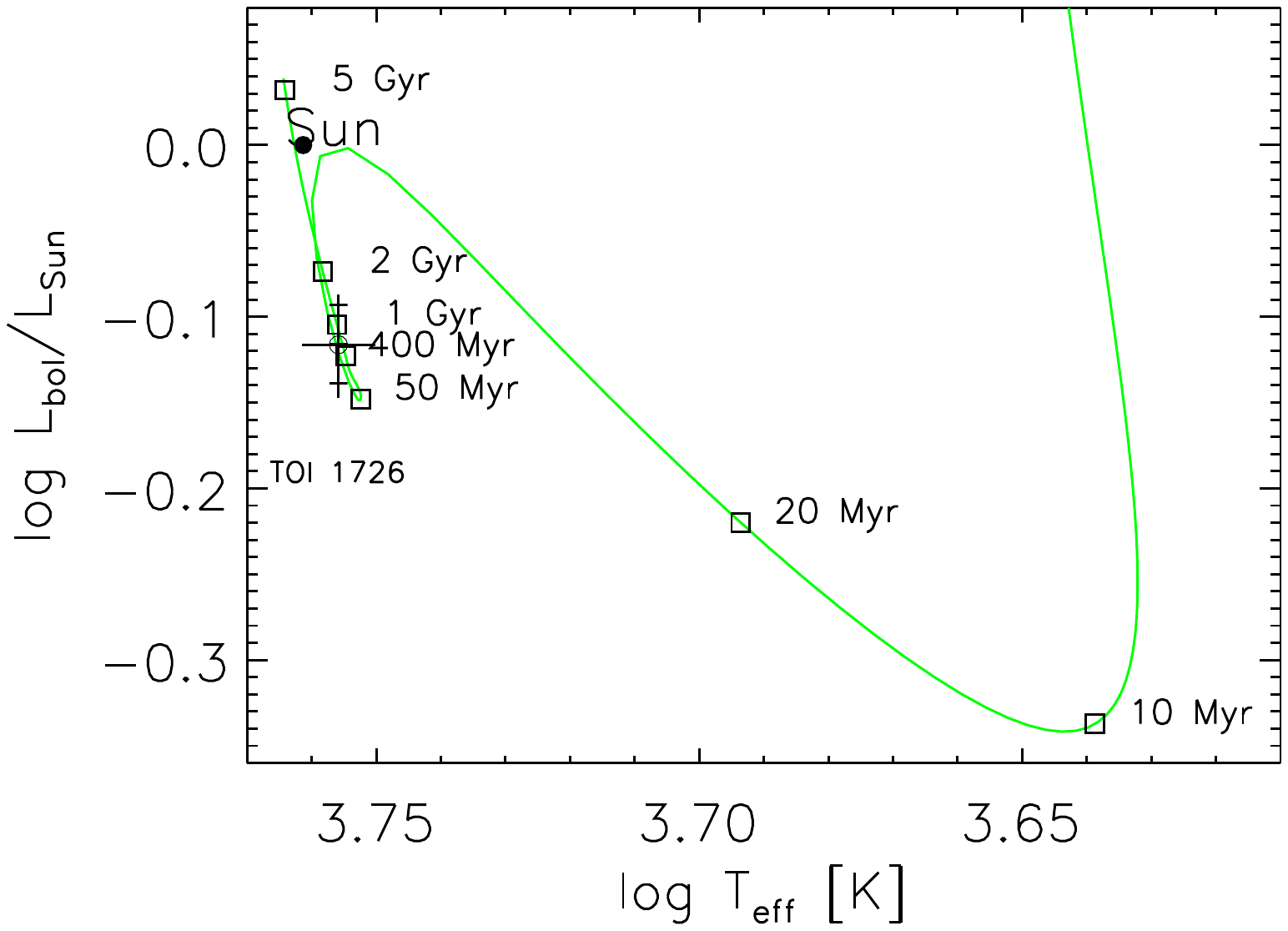}
			\caption{Evolutionary track of HD\,63433 up to an age of 5 Gyr in the effective temperature-bolometric luminosity plane. The current location of the star on the track is marked by a black cross. }
			\label{fig:evoltrack}
		\end{figure}
  
  \begin{figure}
			\centering
			\includegraphics[width=0.5\textwidth, trim = 1cm 13cm 2cm 3cm]{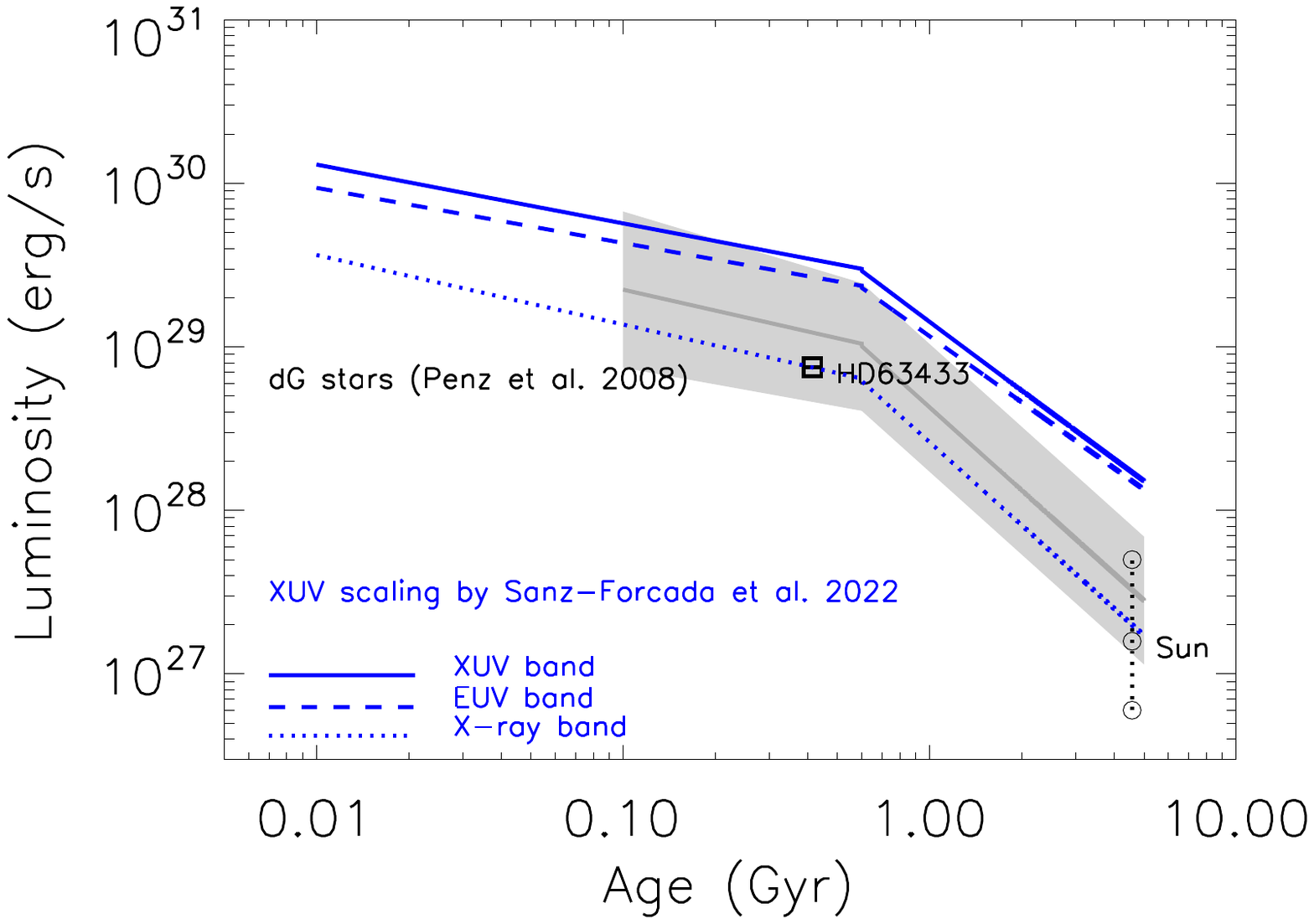}
			\caption{Time evolution of X-ray (5--100\,\AA), EUV (100--920\,\AA), and total XUV luminosity of HD\,63433, according to \cite{Penz08a} and the X-ray/EUV scaling by \citet{SF22}. The grey area is the observed $1\sigma$ spread around the median (dark grey line) of the X-ray luminosity distributions for dG stars in the Hyades and Pleiades open clusters. Uncertainties on the age and X-ray luminosity of HD\,63433 are within the size of the square symbol.}
			\label{fig:xuvevol}
		\end{figure}
		
		
		\section{Dataset} \label{app:datatables}

\longtab[1]{ 
\tiny
\begin{longtable}{lcccc}
       \caption{RVs derived from HARPS-N spectra and used in this work.} 
       \label{tab:dataset1}\\
       \hline\hline
		Time [BJD] & $RV_{\rm DRS}$ [\ms] & $\sigma_{\rm RV,\, DRS}$ [\ms] & $RV_{\rm TERRA}$ [\ms] & $\sigma_{\rm RV,\, TERRA}$ [\ms] \\
		\hline
		\endfirsthead
		\caption{continued.}\\
		\hline\hline
		Time [BJD] & $RV_{\rm DRS}$ [\ms] & $\sigma_{\rm RV,\, DRS}$ [\ms] & $RV_{\rm TERRA}$ [\ms] & $\sigma_{\rm RV,\, TERRA}$ [\ms] \\
		\hline
		\endhead
		\hline
		\endfoot
		2458906.346145 & -15775.70 & 0.74 &  32.74 & 1.24 \\ 
2458908.346263 & -15834.58 & 0.70 & -29.73 & 1.10 \\ 
... & ... & ... & ... & ...  \\ 
\end{longtable}
}

\longtab[2]{ 
\tiny
\begin{longtable}{lcccccccc}
       \caption{Activity diagnostics derived from HARPS-N spectra and used in this work.}
       \label{tab:dataset2}\\
       \hline\hline
		Time [BJD] & $BIS_{\rm DRS}$ [\ms] & $\sigma_{\rm BIS}$ [\ms] & $FWHM_{\rm DRS}$ [\ms] & $\sigma_{\rm FWHM}$ [\ms] & $\log R^{\prime}_{\rm HK}$ [dex] & $\sigma_{\rm \log R^{\prime}_{\rm HK}}$ [dex] & $H\alpha$ & $\sigma_{H\alpha}$ \\
		\hline
		\endfirsthead
		\caption{continued.}\\
		\hline\hline
		Time [BJD] & $BIS_{\rm DRS}$ [\ms] & $\sigma_{\rm BIS}$ [\ms] & $FWHM_{\rm DRS}$ [\ms]  & $\sigma_{\rm FWHM}$ [\ms] & $\log R^{\prime}_{\rm HK}$ [dex] & $\sigma_{\rm \log R^{\prime}_{\rm HK}}$ [dex] & $H\alpha$ & $\sigma_{H\alpha}$ \\
		\hline
		\endhead
		\hline
		\endfoot
		2458906.346145 &  -2.54 & 1.48 & 11463.30 & 1.48 & -4.378 & 0.0010 & 0.1580 & 0.0003 \\ 
2458908.346263 &  39.46 & 1.41 & 11399.39 & 1.41 & -4.407 & 0.0009 & 0.1545 & 0.0003 \\ 
... & ... & ... & ... & ... & ... & ... & ... & ... \\ 
\end{longtable}
}       
	\end{appendix}
	
\end{document}